\documentclass[10pt]{iopart} 

\usepackage{iopams}
\usepackage[usenames,dvipsnames]{xcolor}
\usepackage{graphicx}

\newcommand{\Tc}{\ensuremath{T_{\rm c}}} 
\newcommand{\xA}{\ensuremath{x_{\rm A}}} 
\newcommand{\xB}{\ensuremath{x_{\rm B}}} 
\newcommand{\NA}{\ensuremath{N_{\rm A}}} 
\newcommand{\NB}{\ensuremath{N_{\rm B}}} 
\newcommand{\RA}{\ensuremath{R_{\rm A}}} 
\newcommand{\RB}{\ensuremath{R_{\rm B}}} 
\newcommand{\iAA}{\ensuremath{{\rm AA}}} 
\newcommand{\iBB}{\ensuremath{{\rm BB}}} 
\newcommand{\iAB}{\ensuremath{{\rm AB}}} 
\newcommand{\lA}{\ensuremath{{\rm A}}}   
\newcommand{\lB}{\ensuremath{{\rm B}}}  
  
\begin{document}

\title[Glassy dynamics of a binary Voronoi fluid]%
{Glassy dynamics of a binary Voronoi fluid: A mode-coupling analysis}

\author{C Ruscher$^{1,2}$, S Ciarella$^3$, C Luo$^3$\footnote{S Ciarella and C Luo have contributed equally to this work.}, L M C Janssen$^3$, J Farago$^1$, J Baschnagel$^1$}

\address{$^1$ Universit\'e de Strasbourg, Institut Charles Sadron, CNRS--UPR22, 23 rue du Loess, BP 84047, 67034 Strasbourg Cedex 2, France}
\address{$^2$ Department of Physics and Astronomy and Quantum Matter Institute, University of British Columbia, Vancouver BC V6T 1Z1, Canada}
\address{$^3$ Theory of Polymers and Soft Matter, Department of Applied Physics, Eindhoven University of Technology, P.O. Box 513, 5600MB Eindhoven, The Netherlands}
\ead{celine.ruscher@ics-cnrs.unistra.fr}

%

\begin{abstract}
The binary Voronoi mixture is a fluid model whose interactions are derived from the Voronoi-Laguerre tessellation of the configurations of the system. The resulting interactions are local and many-body. Here we perform  molecular-dynamics (MD) simulations of an equimolar mixture that is weakly polydisperse and additive. For the first time we study the structural relaxation of this mixture in the supercooled-liquid regime. From the simulations we determine the time- and temperature-dependent coherent and incoherent scattering functions for a large range of wave vectors, as well as the mean-square displacements of both particle species. We perform a detailed analysis of the dynamics by comparing the MD results with the first-principles-based idealized mode-coupling theory (MCT). To this end, we employ two approaches: fits to the asymptotic predictions of the theory, and fit-parameter-free binary MCT calculations based on static-structure-factor input from the simulations. We find that many-body interactions of the Voronoi mixture do not lead to strong qualitative differences relative to similar analyses carried out for simple liquids with pair-wise interactions. For instance, the fits give an exponent parameter $\lambda \approx 0.746$ comparable to typical values found for simple liquids, the wavevector dependence of the Kohlrausch relaxation time is in good qualitative agreement with literature results for polydisperse hard spheres, and the MCT calculations based on static input overestimate the critical temperature, albeit only by a factor of about 1.2. This overestimation appears to be weak relative to other well-studied supercooled-liquid models such as the binary Kob--Andersen Lennard-Jones mixture. 
Overall, the agreement between MCT and simulation suggests that it is possible to predict several microscopic dynamic properties with qualitative, and in some cases near-quantitative, accuracy based solely on static two-point structural correlations, even though the system itself is inherently governed by many-body interactions. 
\end{abstract}

%
\vspace{2pc}

%
\submitto{\JPCM}
%
%
\ioptwocol
%

\section{Introduction}
\label{sec:intro}
Disordered materials, such as polymers, metallic alloys, and polydisperse colloidal suspensions, are of huge practical interest as they can be designed with specific mechanical, optical, or thermal properties. At low density or high temperature these materials are in the liquid state. Provided the liquid can be supercooled without undergoing structural ordering, the dynamics strongly slows down upon cooling or increasing the density, shifting the time scale for viscous flow to ever longer times. Ultimately, the glass transition is reached, beyond which structural relaxation can no longer occur on experimental time scales. Such systems are then in a nonequilibrium solid-like state where they exhibit mechanical rigidity but, contrary to crystalline materials, they lack any long-range order. Developing a microscopic understanding of the nature of the glassy state and the glass transition is one of the challenging problems in condensed matter physics \cite{Cavagna:PhysRep2009,BerthierBiroliRMP}. 

In the dense liquid phase, relaxation takes place through cooperative rearrangements of (groups of) neighboring particles. That is, in order for a particle to move, its neighbors must also move, and hence the local particle environment plays an important role in the dynamics. A way to access information on the neighborhood of a particle is to apply a Voronoi tessellation \cite{OkabeEtal:Voronoi}. Voronoi tessellation is a geometric partitioning of space into contiguous cells whose volume can be thought of as the zone of influence of a given particle. Such tessellations have been extensively used for glass-forming \cite{StarrEtal:PRL2002,FaragoEtal:EPJE2014} or granular systems \cite{Morse_2014,Morse_2016,Rieser_2016}, mostly as a tool to define free volume or to obtain local geometric information. For instance, Morse and Corwin \cite{Morse_2014,Morse_2016} emphasize the geometric nature of the jamming transition in granular systems by showing that a large set of geometrical observables (surface area, aspect ratio, standard deviation of the volume, etc.) extracted from Voronoi tessellation shows a marked signature at the jamming point. A similar observation was made by Rieser \etal \cite{Rieser_2016} who found a strong signature of jamming in a quantity related to the relative free volume of the particles. These observations highlight the importance of Voronoi tessellation to get a deeper level of structural information which is either not contained or too strongly averaged in the usual two-point correlation functions, like the radial distribution function or static structure factor that are known to vary only weakly on approach to the glass transition.

Voronoi tessellation offers more than only a diagnostic tool, however; it also provides the basis for a new class of complex liquid models. During the past five years, two new models have emerged whose interactions are intrinsically many-body and derived from the inherent geometrical properties of the Voronoi tessellation: the ``Voronoi liquid'' introduced by some of us \cite{RuscherEtal:EPL2015} and the ``self-propelled Voronoi (SPV) model'' proposed by Bi \etal \cite{Bi_2016}. The SPV model aims at describing cell motility and cell-cell interactions in confluent tissues. One major achievement of the SPV model has been the identification of a structural order parameter, the shape index, which depends only the perimeter and area of the Voronoi cell, and which identifies, for given single-cell motility and persistence time, a liquid-to-solid transition reminiscent of the glass transition. The SPV model has also found use in understanding collective cell phenomena such as the epithelial-to-mesenchymal transition---a key step in the propagation of cancer cells \cite{Yang_2017}---and in the design of a new generation of bioinspired materials, such as tunable photonic fluids \cite{Li_2018}. Moreover, the model can also shed light on the glass transition in active matter \cite{Janssen2019}. Recently, Sussman \etal \cite{Sussman_2018} studied a passive version of the SPV model. Their findings differ from the usual phenomenology of glass formers as they observed a sub-Arrhenius behavior of the relaxation time---i.e.\ an anomalous fragility which thus far has been found in only a very limited number of systems \cite{Ciarella2019}---and a high density of collective low-frequency vibrational modes associated with low-temperature energy minima. The specific many-body nature of the interactions is at the core of this anomalous dynamics, meaning that going beyond the usual pairwise potentials could broaden the view on the ``stylist facts'' \cite{BiroliGarrahan:JCP2013} commonly associated with the glass transition phenomenology. This has also been a main motivation for the introduction of the Voronoi liquid. 
From its conception the Voronoi liquid is a passive fluid model. In many ways, perhaps surprisingly, the model behaves like an ordinary simple liquid regarding traditional structural and dynamical correlation functions \cite{RuscherEtal:EPL2015}. However, the Voronoi fluid also has some unique specificities. The most striking one is arguably that the potential energy of an $N+1$ particle system, $E_{\rm p}(\bi{r}_1, \cdots, \bi{r}_N,\bi{r}_{N+1})$, becomes equal to that of $N$ particle system $E_{\rm p}(\bi{r}_1, \cdots, \bi{r}_N)$, if $\bi{r}_{N+1} \rightarrow \bi{r}_{N}$ \cite{RuscherEtal:EPL2015}. In this sense, the potential is ``hypersoft''. This property does not compromise the stability of the liquid because the interactions are locally repulsive and the superposition of two particles has a finite energy cost. Hypersoftness has, however, an impact in situations where the dynamics is slow, e.g.\ for the crystalline phase. At low temperature the monodisperse Voronoi liquid forms a bcc crystal that is ``plastic'' in that the particles can diffuse freely in the solid without destroying the crystalline structure \cite{PhDthesis:Celine}. A further striking property of the Voronoi liquid is an anomalous scaling of the sound attenuation rate ($\propto q$ instead of $\propto q^2$ with $q$ being the modulus of the wave vector) at mesoscopic scales and a shift of the hydrodynamic limit to very small $q$-values with respect to a standard Lennard-Jones (LJ) system \cite{RuscherEtal:JCP2017}. This specific behavior can be attributed to a very weak resistance to shear deformations at high frequency. For the Voronoi liquid the product of the infinite-frequency shear modulus ($G_\infty$) and the isothermal compressibility ($\chi_T$) is much smaller than 1, whereas $G_\infty \chi_T \sim 1$ for the LJ fluid at the triple point \cite{RuscherEtal:JCP2017}. 

To suppress the tendency for crystallization, the model has recently been generalized to binary mixtures \cite{RuscherEtal:PRE2018}. Reference~\cite{RuscherEtal:PRE2018} discusses this generalization and explores numerically and theoretically the thermodynamic and structural properties of an equimolar mixture. It was shown that the system is thermodynamically stable against demixing and can be supercooled to low temperature while keeping a liquid-like structure. The present work extends the characterization of the model to dynamic properties. In doing so, we present results from molecular-dynamics (MD) simulations which we analyze in terms of the idealized mode-coupling theory (MCT).

The layout of the paper is as follows. We first review the definition of the model in \sref{sec:voronoiliquid} and then describe the MD simulations. In \sref{sec:statics}, we discuss various static structure factors. The contents of this section overlaps with \cite{RuscherEtal:PRE2018} but also extends the analysis to structure factors related to number and composition fluctuations. Next, we provide an overview of the idealized MCT (\sref{sec:mct}). Two approaches are considered: fully microscopic, fit-parameter-free MCT calculations based on static input from the simulations, and fits to the asymptotic predictions of MCT. Both approaches will be compared with the MD results and with each other. \Sref{sec:results} discusses this comparison. A summary of the main results and an outlook on possible future research directions are given in \sref{sec:sum}.

\section{Model and simulation method}
\label{sec:voronoiliquid}

\subsection{Monodisperse and binary Voronoi liquid}
\label{subsec:voronoiliquid}
Consider a system of $N$ point particles at positions $\bi{r}_j$ ($j = 1, \ldots, N$) in a three-dimensional volume $V$. To each configuration $\{\bi{r}_j\}_{i=j,\ldots,N}$ one can associate a Voronoi tessellation, a space-filling partitioning of $V$ into $N$ cells assigning one cell to each particle. The cell of particle $j$ is defined as the region of space being closer to $j$ than to any other particle in the system. The cell has a volume $v_j$ and a centroid at position $\bi{g}_j$. Since $\bi{g}_j$ does in general not coincide with $\bi{r}_j$, we can introduce the ``geometric polarization'' $\btau_j$ of a cell by $\btau_j = v_j (\bi{g}_j - \bi{r}_j)$. Analysis of a supercooled liquid of short polymer chains revealed that the geometric polarization $\btau_j(t)$ at time $t$ is correlated to the total interaction force $\bi{F}_j(t)$ on particle $j$ and $\btau_j$ obeys the conservation law $\sum_{j=1}^N \btau_j = \boldsymbol{0}$, analogous to $\bi{F}_j$ \cite{FaragoEtal:EPJE2014}. The equilibrium properties of the geometric polarization are thus reminiscent of those of a force. This observation motivated us to introduce a new model for a liquid---the Voronoi liquid---where the force on particle $j$ is taken proportional to $\btau_j$ \cite{RuscherEtal:EPL2015}: 
\begin{eqnarray}
  \bi{F}_j = \gamma {\btau}_{j} = \gamma \int_{v_j} \rmd^3 \bi{r} \,\bi{r} .
  \label{voronoi_force}
\end{eqnarray}
Here the constant $\gamma$ is a parameter of the model and $\bi{r}$ denotes the vector from the particle position $\bi{r}_j$ to the boundary of its Voronoi cell. The force $\bi{F}_j$ can be written as $\bi{F}_j = - \boldsymbol{\nabla}_j E_{\rm p}$ with \cite{RuscherEtal:EPL2015}
\begin{eqnarray}
  E_{\rm p}(\bi{r}_1, \cdots, \bi{r}_N)=  \sum_{j=1}^N \left [ \frac{\gamma}{2}
  \int_{v_j} \rmd^3 \bi{r} \, r^2 \right ] .
  \label{voronoi_energy_mono}
\end{eqnarray}
This defines the potential energy of the monodisperse Voronoi liquid as the sum of the interaction energies of all particles. The interaction energy of a particle is positive and determined by its nearest-neighbor shell: it is local, many-body, and soft in the sense that the energy cost for particle overlap is finite. 

Thermodynamic, structural and dynamic properties \cite{RuscherEtal:EPL2015,RuscherEtal:JCP2017} have been studied for the monodisperse system. Upon cooling the liquid becomes metastable and eventually crystallizes in a bcc structure \cite{PhDthesis:Celine}. For the study of glasses the tendency of structural ordering has to be suppressed. This can be achieved by using systems with multiple components of different sizes (and interaction energies) \cite{IngebrigtsenEtal:PRX2019,NinarelloEtal:PRX2017}. Therefore, we introduce size dispersity into our model by choosing the Voronoi-Laguerre generalization of the Voronoi tessalation \cite{RuscherEtal:PRE2018}. The Voronoi-Laguerre tessellation assigns a ``natural radius'' $R_j$ ($>0$) to every particle $j$, which enters the construction of its cell, and has the advantage of preserving the defining features of the Voronoi liquid. For the Voronoi-Laguerre tessellation we still have $\sum_j \btau_j = \boldsymbol{0}$ and $\bi{F}_j = \gamma \btau_j = - \boldsymbol{\nabla}_j E_{\rm p}$ with the following generalization of the potential energy \cite{RuscherEtal:PRE2018}:  
\begin{eqnarray}
  E_{\rm p}(\bi{r}_1, \cdots, \bi{r}_N)=  \sum_{j=1}^N \left [\frac{\gamma}{2} \int_{v_j} \rmd^3 \bi{r}  \left ( r^2 - R_j^2 + R^2 \right ) \right ].
  \label{voronoi_energy_poly}
\end{eqnarray}
Here $R^2 = \sum_{j=1}^{N} R^2_j/N$ is mean-square natural radius averaged over the polydispersity. \Eref{voronoi_energy_poly} reduces to \eref{voronoi_energy_mono} in the monodisperse case. 

Since $E_{\rm p}$ is defined in terms of $v_j$, the relevant length scale of the Voronoi liquid is given by $v^{1/3}$, where $v = \sum_{j=1}^N v_j / N = V/N$ is the average volume per particle, and the temperature scale by $\gamma v^{5/3}/ k_{\rm B}$ with $k_{\rm B}$ being the Boltzmann constant. As in previous work \cite{RuscherEtal:EPL2015,RuscherEtal:JCP2017,RuscherEtal:PRE2018} we choose the density $v^{-1}=1$ and take $\gamma = 1000$ so that the temperatures of the liquid phase are in the range $T \sim 1$ (with $k_{\rm B} = 1$).

Here we examine the simplest representative of a polydisperse system, a binary mixture of $\NA$ large particles of radius $\RA$ and $\NB = N - \NA$ small particles of radius $\RB < \RA$. The mixture is characterized by the number concentration of small particles $\xB = \NB/ N$ and the size ratio $\RB/\RA$. The choice of these parameters is motivated by theoretical work on binary hard-sphere mixtures \cite{GoetzeVoigtmann:PRE2003}, suggesting that the propensity to form a glass is enhanced, relative to the monodisperse system, for size ratios $\sim 0.8$ and $\xB \sim 0.5$ (cf Fig.~1 in \cite{GoetzeVoigtmann:PRE2003}). Therefore, we take $\RB/\RA = 0.83$ and $\xB = \xA = 0.5$. 

A particular feature of the binary Voronoi mixture is that the natural radii determine the potential energy only by the (dimensionless) ``polydispersity parameter'' $\xi = \RA^2 - \RB^2$ (recall that the length scale $v^{1/3}=1$). As pointed out in \cite{RuscherEtal:PRE2018}, $ \xi$ needs to be smaller than 1 to avoid unphysical situations where small particles are situated outside their Voronoi cell. On the other hand, $\xi$ has to be large enough to suppress crystallization. From continuous cooling runs at finite rates, it was found that the binary mixture forms glasses for $0.06 \lesssim \xi \lesssim 0.36$ \cite{PhDthesis:Celine}. Here, to probe the glassy regime, we choose $\xi = (0.375)^2 \simeq 0.141$, finally leading to $\RA = 0.6729$ and $\RB = 0.5585$.   

A priori, $\xi$ is the relevant parameter. The physical properties of the mixture are not changed when varying $\RA$ and $\RB$ but keeping $\xi$ the same. Our choices for the natural radii, however, turn out to be physically meaningful. The partial pair-distribution functions of the $\lA$ particles, $g_\iAA(r)$, and the $\lB$ particles, $g_\iBB(r)$, show a first maximum at $r_\iAA = 1.225 \approx 2 \RA$ and $r_\iBB = 0.975 \approx 2 \RB$ \cite{RuscherEtal:PRE2018}. Therefore, $\RA$ and $\RB$ can be thought of as the radii of soft repulsive particles. Moreover, the cross pair-distribution function of $\lA$ and $\lB$, $g_\iAB(r)$, peaks at $r_\iAB = 1.125 \approx (r_\iAA + r_\iBB)/2 \approx (\RA + \RB)$, suggesting that the studied binary Voronoi mixture is intrinsically additive \cite{RuscherEtal:PRE2018}.

\subsection{Molecular-dynamics simulations}
\label{subsec:MD}
We performed molecular-dynamics (MD) simulations with a modified version of the LAMMPS code \cite{Plimpton:LAMMPS1995}, enabling the computation of the geometric polarization $\btau_j$, and therefore of $\bi{F}_j$, by using the Voro++ library \cite{voropp}. The system contained $N=1000$ particles of mass $m =1$ in a cubic box of linear dimension $L$ with periodic boundary conditions. Since $v = V/N = 1$, this implies $L=10$. Thus, the smallest accessible wave vector ($\bi{q}$) has the modulus $q_{\rm min} = 2 \pi / L = 0.628$. With the energy scale $\gamma v^{5/3}$, the mass $m$ and the length scale $v^{1/3}$, the characteristic time scale of the Voronoi liquid is $\tau_{\rm voro} = \sqrt{m / \gamma v} = \sqrt{1/1000} \approx 0.03$ with $\gamma = 1000$, $m=1$ and $v=1$. The time step ($\delta t$) of the MD simulation has to be smaller than $\tau_{\rm voro}$. We used $\delta t=0.001 \approx 0.03 \tau_{\rm voro}$ when integrating the classical equations of motion by the velocity-Verlet algorithm. In the following, all times are measured in units of $\tau_{\rm voro}$.

The simulations were carried out in the canonical ensemble with the Nos\'e-Hoover thermostat (using a damping parameter of $T_{\rm damp}=0.1$). We investigated equilibrium properties for $0.83 \leq T \leq 2$. This interval ranges from the regime of the normal to the moderately supercooled liquid (above the critical temperature of MCT $\Tc = 0.798$, cf \tref{tablefitparam}). 

Equilibration was done as follows. Starting from an equilibrated configuration at $T_1$, the temperature was instantaneously decreased by a small step to $T_2 < T_1$. The system was allowed to evolve in the canonical ensemble until the potential energy fluctuated around an average value. Then, the isothermal simulation was continued over a time interval $\Delta t$ (adapted to $T_2$) before starting the production run for data analysis. Control of equilibrium was carried out by dividing the production trajectory in half and checking that dynamic observables gave the same results on both portions of the trajectory. As an example, $\Delta t = 10^5$ for $T=0.84$, corresponding to about 10 times the $\alpha$ relaxation time at that temperature.

\section{Static structure factors}
\label{sec:statics}
\begin{figure}
\begin{center}
  \includegraphics*[scale=0.38]{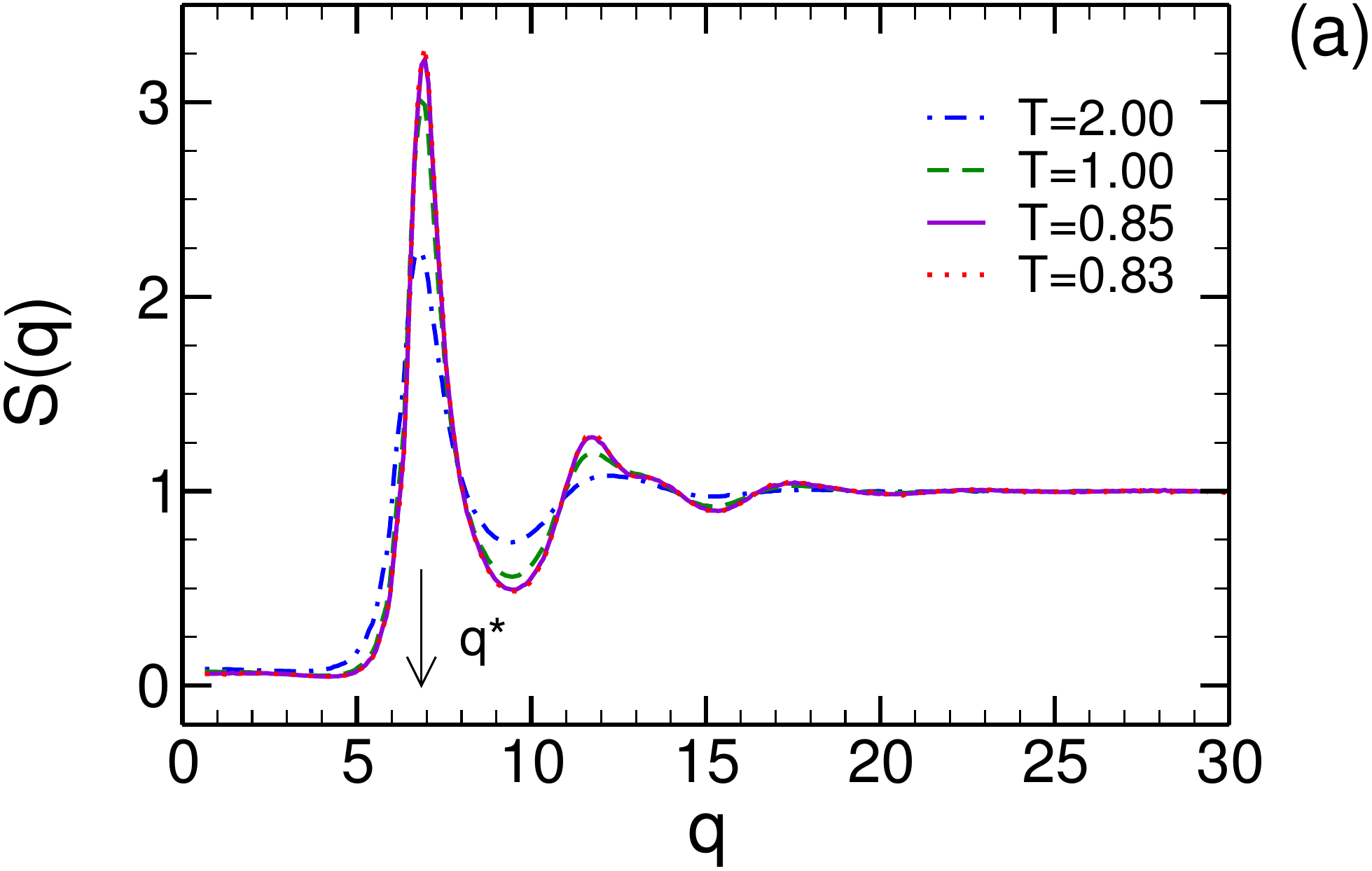}\\[2mm]
  \includegraphics*[scale=0.38]{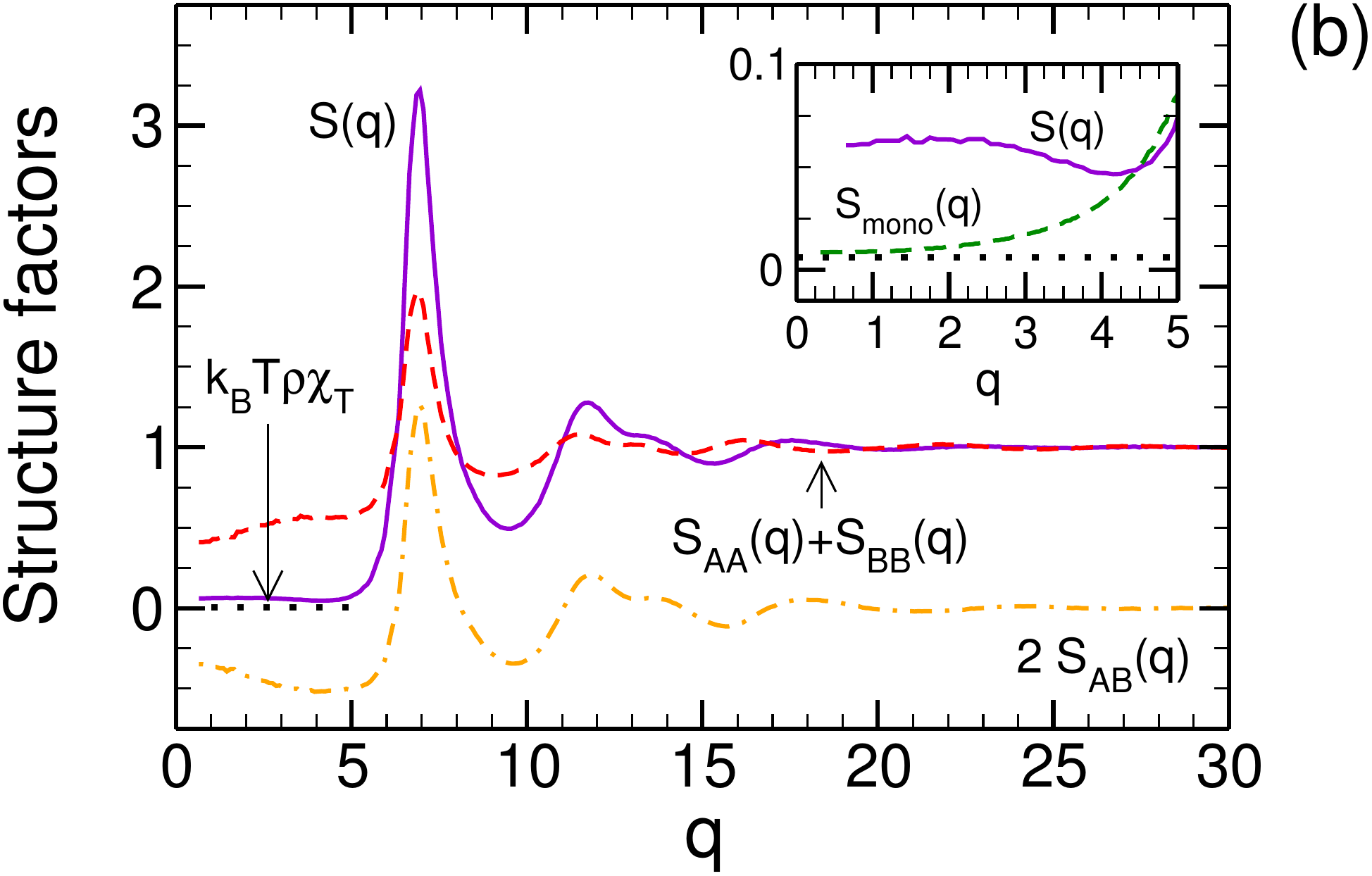}\\[2mm]
  \includegraphics*[scale=0.38]{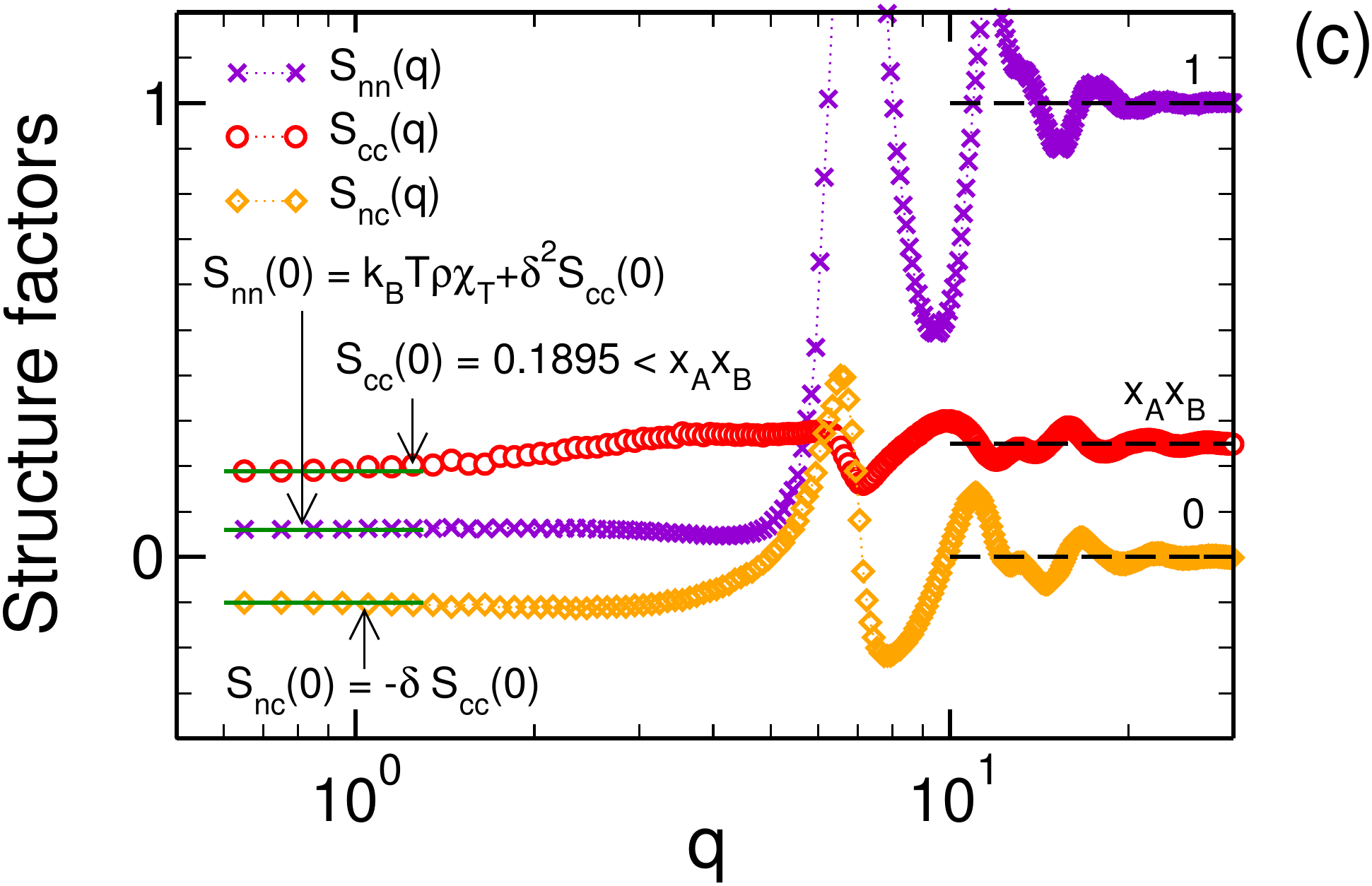}
\end{center}
\caption{(a) Collective static structure factor $S(q)$ as a function of the modulus of the wave vector $q$ for $T = 0.83$ (\dotted), 0.85 (\full), 1 (\broken) and 2 (\chain). $S(q)$ has a maximum near $q^* = 6.85$ which is indicated by an arrow.
(b) $S(q)$ versus $q$ at $T=0.85$ and its decomposition into partial structure factors according to \eref{eq:defSnn}: $S_\iAA(q) + S_\iBB(q)$ (\broken) and $2 S_\iAB(q)$ (\chain). The horizontal dotted line indicates the value of $k_{\rm B}T\rho \chi_T$ (= $0.006$ \cite{RuscherEtal:PRE2018}) with $\rho = 1/v$ the particle density and $\chi_T$ the isothermal compressibility. Inset: Zoom for small $q$ focusing on $S(q)$ (\full). The dashed line presents the collective structure factor $S_{\rm mono}(q)$ of the monodisperse Voronoi liquid at $T=1.05$ which has a comparable compressibility as the binary mixture. $k_{\rm B}T\rho \chi_T$ of the mixture from the main figure is shown as a horizontal dotted line.
(c) Bhatia--Thornton structure factors versus $q$ at $T=0.85$: $S_{\rm nn}(q) = S(q)$ ($\times$), $S_{\rm cc}(q)$ (\opencircle) and $S_{\rm nc}(q)$ (\opendiamond). The horizontal dashed lines represent the large-$q$ limits: $S_{\rm nn}(q \rightarrow \infty) = 1$, $S_{\rm cc}(q \rightarrow \infty) = \xA\xB$ and $S_{\rm nc}(q \rightarrow \infty) = 0$. The horizontal full lines indicate the limits for $q \rightarrow 0$ given by \eref{eq:Snnq0}, \eref{eq:Sccq0} and \eref{eq:Sncq0}, cf text for details.}
\label{sdek}
\end{figure}

Previous work studied the thermodynamics, the stress tensor, and structural properties of the binary Voronoi liquid \cite{RuscherEtal:PRE2018}. Due to their importance for mode-coupling theory we here revisit the discussion of the static structure factors. The collective static structure factor
\begin{equation}
  S(q) = \frac{1}{N} \left \langle \rho(\bi{q}) \rho(-\bi{q}) \right \rangle 
  \label{eq:defsq}
\end{equation}
is defined in terms of the coherent density fluctuations for wave vector $\bi{q}$,
\begin{equation}
  \rho(\bi{q}) = \sum_{j=1}^N 
  \exp \left ( \rmi \bi{q} \cdot \bi{r}_j \right )
  \quad \mbox{(for $q \neq 0$)},
  \label{eq:defrhoq}
\end{equation}
where $\langle \ldots \rangle$ denotes the canonical average and $\bi{r}_j$ is the position of particle $j$. For a spatially homogeneous and isotropic system, the structure factor depends only on the modulus of the wave vector, $q = | \bi{q} |$. \Fref{sdek}(a) presents $S(q)$ for four temperatures in the investigated interval $0.83 \leq T \leq 2$. We see that the collective structure of the Voronoi mixture is typical of a dense disordered system. In the limit $q \rightarrow 0$, $S(q)$ is small because the fluctuations of the particle number relative to the average $\langle N \rangle$ ($ = N$) are weak in a dense system,
\begin{equation}
  S(q \rightarrow 0) = \frac{\langle N^2 \rangle - \langle N \rangle^2}{\langle N \rangle} \ll 1.
  \label{eq:sqlimit0}
\end{equation} 
With increasing $q$, $S(q)$ increases toward a maximum that occurs around $q^* = 6.85$. The corresponding length scale $2\pi/q^* \sim 1$ is on the order of the particle diameters. Thus, the dominant contribution to $S(q^*)$ comes from the amorphous packing in neighbor shells around a particle. Upon cooling the packing becomes tighter, which is reflected by the increase of the height and the decrease of the width of $S(q)$ near $q^*$.

Further insight can be obtained from the partial static structure factors
\begin{equation}
  S_{\alpha\beta}(q)
  = \frac{1}{N} \left \langle \rho_\alpha(\bi{q}) 
  \rho_\beta(-\bi{q}) \right \rangle 
  \quad \mbox{($\alpha,\beta = \lA, \lB$)}
  \label{eq:defsqpartial}
\end{equation}
defined by the partial density fluctuations
\begin{equation}
  \rho_\alpha(\bi{q}) 
  = \sum_{j_\alpha = 1}^{N_\alpha}
  \exp \left ( \rmi \bi{q} \cdot \bi{r}_{j_\alpha} \right ),
  \label{eq:defpartialdensfluc}
\end{equation}
where $\bi{r}_{j_\alpha}$ is the position of particle $j_\alpha$ of species $\alpha$. As $\rho_\lA(\bi{q}) + \rho_\lB(\bi{q}) = \rho(\bi{q}) \equiv \rho_{\rm n}(\bi{q})$, the collective structure factor can be expressed as
\begin{equation}
  S(q) = S_{\rm nn}(q) = S_\iAA(q) + S_\iBB(q) + 2 S_\iAB(q) . 
  \label{eq:defSnn}
\end{equation}
While $S_{\alpha\beta}(q)$ characterize spatial correlations between like and unlike particles, $S(q)$ describes number-number (nn) correlations (whence the notation $S = S_{\rm nn}$). \Eref{eq:defSnn} is not the only physically significant linear combination of the partial structure factors. Since composition (or concentration) fluctuations $\rho_{\rm c}(\bi{q})$ are defined by $\rho_{\rm c}(\bi{q}) = \rho_\lA(\bi{q}) - \xA \rho(\bi{q}) = \xB \rho_\lA(\bi{q}) - \xA \rho_\lB(\bi{q})$, the structure factor 
\begin{eqnarray}
  S_{\rm cc}(q)  
  & = & \frac{1}{N} \left \langle \rho_{\rm c}(\bi{q}) 
  \rho_{\rm c}(-\bi{q}) \right \rangle \nonumber \\
  & = & \xB^2 S_\iAA(q)  + \xA^2 S_\iBB(q)  - 2 \xA\xB S_\iAB(q) 
  \label{eq:defScc} 
\end{eqnarray}
represents composition-composition (cc) correlations, and the structure factor between $\rho_{\rm n}$ and $\rho_{\rm c}$, 
\begin{eqnarray}
  \fl S_{\rm nc} & (q) 
   = \frac{1}{N} \left \langle \rho_{\rm n}(\bi{q}) 
  \rho_{\rm c}(-\bi{q}) \right \rangle \nonumber \\ 
  & = \xB S_\iAA(q)- \xA S_\iBB(q) + (\xB - \xA) S_\iAB(q) , 
  \label{eq:defSnc} 
\end{eqnarray}
describes number-composition (nc) correlations. The structure factors $S_{\rm nn}(q)$, $S_{\rm cc}(q)$ and $S_{\rm nc}(q)$ are often referred to as Bhatia--Thornton structure factors \cite{BhatiaThornton:PRB1970}. They have been studied extensively for metallic alloys \cite{HorbachEtal:PRB2007,DasEtal:PRB2008} or colloidal suspensions \cite{ThorneyworkEtal:MolPhys2018}. 

\Fref{sdek}(b) compares $S(q)$ with $S_\iAA(q) + S_\iBB(q)$ and $2S_\iAB(q)$ at $T=0.85$. In the limit $q \rightarrow \infty$, the system behaves like an ideal mixture with vanishing correlations. This implies $S_\iAB(q\rightarrow \infty) = 0$ as well as $S_\iAA(q\rightarrow \infty) = \xA$ and $S_\iBB(q\rightarrow \infty) = \xB$. For large $q$, say $q \gtrsim 20$, the behavior of $S(q)$ is therefore dominated by correlations between like particles. The sum $S_\iAA(q) + S_\iBB(q)$ is positive for all $q$, whereas $S_\iAB(q)$ oscillates around 0 and remains negative for $q < q^*$, displaying a minimum at $q \approx 4$. These negative values indicate that long-range $\iAB$ correlations are suppressed in the Voronoi mixture, a feature also found in other binary systems \cite{DasEtal:PRB2008,NaurothKob1997,MorenoColmenero:PRE2006}. The minimum of $S_\iAB(q)$ at $q \approx 4$ outweighs the positive contribution of $S_\iAA(q) + S_\iBB(q)$, leading to a dip in $S(q)$ at $q \approx 4$ before $S(q)$ increases again toward a plateau as $q \rightarrow 0$. Such a dip is not observed for the monodisperse Voronoi liquid. Here $S_{\rm mono}(q)$ continuously decreases toward the compressibility plateau $S_{\rm mono}(q \rightarrow 0) = k_{\rm B}T \rho \chi_T$ with $\chi_T$ being the isothermal compressibility (cf inset in \fref{sdek}(b)). For the mixture, however, we see that $S(q\rightarrow 0)$ adopts a value larger than $k_{\rm B}T \rho \chi_T$. 

For binary mixtures this deviation between $S(q\rightarrow 0)$ and $k_{\rm B}T \rho \chi_T$ is expected from the work of Bhatia and Thornton \cite{BhatiaThornton:PRB1970} and also from the Kirkwood--Buff theory for multicomponent solutions \cite{BenNaim}. For $q \rightarrow 0$ the Bhatia--Thornton structure factors are related to the thermodynamic properties of the binary mixture:
\begin{eqnarray}
  S(q\rightarrow 0)
  & = & k_{\rm B}T\rho \chi_T + \delta^2 S_{\rm cc} (q \rightarrow 0) ,
  \label{eq:Snnq0} \\
  S_{\rm cc} (q \rightarrow 0) 
  & = & \frac{Nk_{\rm B} T}{(\partial^2 G/\partial \xA^2)_{p,T,N}} ,
  \label{eq:Sccq0} \\
  S_{\rm nc} (q \rightarrow 0) 
  & = & - \delta \, S_{\rm cc} (q \rightarrow 0) ,\label{eq:Sncq0}
\end{eqnarray}
where $G$ is the Gibbs free energy, $p$ the pressure and
\begin{equation}
  \delta = \rho(v_\lA - v_\lB) 
  \label{eq:defdelta}
\end{equation}
is a dilatation factor given by the partial molar volumes $v_\lA = (\partial V / \partial N_\lA)_{p,T,N_\lB}$ and $v_\lB = (\partial V / \partial N_\lB)_{p,T,N_\lA}$. \Eref{eq:Snnq0} shows that in a mixture fluctuations of the total particle number, $S(q\rightarrow 0)$, do not only stem from compressibility effects---that is, from the volume response of the system to a pressure fluctuation---but also from composition fluctuations and their coupling to the number density. Since thermodynamic stability requires $(\partial^2 G/\partial \xA^2)_{p,T,N} > 0$, we have in general $S(q\rightarrow 0) > k_{\rm B}T\rho \chi_T$, as seen in the inset of \fref{sdek}(b). This implies that the second term in the right-hand-side of \eref{eq:Snnq0} does not vanish, in particular $\delta \neq 0$ or $v_\lA \neq v_\lB$. The molar volumes of the two species can be calculated via the Kirkwood--Buff theory from the partial structure factors in the limit $q \rightarrow 0$ \cite{BenNaim},
\begin{eqnarray}
  v_\lA & = & v \, \frac{\xA S_\iBB(0) - \xB S_\iAB(0)}{\xA^2 S_\iBB(0) + 
  \xB^2 S_\iAA(0) - 2 \xA\xB S_{\iAB}(0)} , 
  \label{eq:defvA} \\
  v_\lB & = & v \, \frac{\xB S_\iAA(0) - \xA S_\iAB(0)}{\xA^2 S_\iBB(0) + 
  \xB^2 S_\iAA(0) - 2 \xA\xB S_{\iAB}(0)} , 
  \label{eq:defvB}
\end{eqnarray}
where $S_{\alpha\beta}(0)$ is an abbreviation for $S_{\alpha\beta}(q \rightarrow 0)$. 

We compare these theoretical predictions to the simulation data in \fref{sdek}(c). The figure shows $S_{\rm cc} (q)$ and $S_{\rm nc} (q)$, as obtained from \eref{eq:defScc} and \eref{eq:defSnc}, together with $S(q)$ at $T=0.85$. We find that $S_{\rm cc}(q)$ is positive for all $q$. For large $q$, $S_{\rm cc} (q)$ oscillates around $\xA \xB$ ($=0.25$)---the value expected for an ideal (equimolar) mixture---and decreases toward $S_{\rm cc} (q\rightarrow 0) = 0.1895$ in the small-$q$ limit. The ratio $\Phi(\xA,T)=\xA(1-\xA)/S_{\rm cc}(q \rightarrow 0)$ enters the definition of the interdiffusion coefficient of the mixture (cf \eref{eq:Darken}) \cite{HorbachEtal:PRB2007}. In systems that favor mixing, as the binary Voronoi mixture \cite{RuscherEtal:PRE2018}, one has $\Phi > 1$ \cite{HorbachEtal:PRB2007,ThorneyworkEtal:MolPhys2018,KuhnEtal:PRB2014}. Here we find $\Phi(\xA=1/2,T=0.85) \simeq 1.319$. Using finally \eref{eq:defvA} and \eref{eq:defvB} we can determine the partial molar volumes, $v_\lA = 1.266$ and $v_\lB = 0.734$, and so the dilatation factor $\delta = 0.532$ at $T=0.85$. If we also take $k_{\rm B}T\rho \chi_T=0.006$ from \cite{RuscherEtal:PRE2018} and read off $S_{\rm cc} (q\rightarrow 0)$ from \fref{sdek}(c), the values of $S(q\rightarrow 0)$ and $S_{\rm nc}(q\rightarrow 0)$ can be computed. These results are shown as horizontal full lines in \fref{sdek}(c). As can be seen, we find excellent agreement between simulation and theoretical expectation. 

\section{Mode-coupling theory}
\label{sec:mct}
The idealized mode-coupling theory (MCT) and its application to simple and molecular glass formers are described in detail in a monograph \cite{GoetzeBook2009}, as well as in several review papers \cite{ReichmanCharbonneau:JSMTE2005,Janssen2018}. Specifically for binary mixtures, MCT is also discussed in several publications, see e.g.\ \cite{GoetzeVoigtmann:PRE2003,DasEtal:PRB2008,NaurothKob1997,%
KuhnEtal:PRB2014,FoffiEtal:PRE2004,FlennerSzamel:PRE2005_2,weysser2011}. Below we first recapitulate the main equations and general predictions of binary MCT, and subsequently apply the theory to our Voronoi mixture.

\subsection{MCT equations for coherent density fluctuations}
\label{subsec:mct_coherent}
MCT assumes that the slow dynamics of glass-forming liquids results from the relaxation of collective density fluctuations. For binary mixtures the central dynamic correlation functions are therefore the partial dynamic structure factors
\begin{equation}
  S_{\alpha\beta}(q,t)
  = \frac{1}{N} \left \langle \rho_\alpha(\bi{q},t) 
  \rho_\beta(-\bi{q}) \right \rangle 
  \quad \mbox{($\alpha,\beta = \lA, \lB$)}
  \label{eq:defsqtpartial}
\end{equation}
where 
\begin{equation}
  \rho_\alpha(\bi{q},t) 
  = \sum_{j_\alpha = 1}^{N_\alpha}
  \exp \left [ \rmi \bi{q} \cdot \bi{r}_{j_\alpha}(t) \right ]
  \label{eq:defpartialdensfluc_of_t}
\end{equation}
and $\bi{r}_{j_\alpha}(t)$ is the position of particle $j_\alpha$ of species $\alpha$ at time $t$. Let us combine these functions in a $2\times 2$ matrix $\mathbf{S}(q,t)$ with $(\mathbf{S}(q,t))_{\alpha\beta} = S_{\alpha\beta}(q,t)$. By means of the Zwanzig--Mori projection operator formalism, an exact equation of motion for $\mathbf{S}(q,t)$ is derived \cite{FoffiEtal:PRE2004,weysser2011}
\begin{eqnarray}
  \partial^2_t \mathbf{S}(q,t)
  & + \mathbf{J}(q)\mathbf{S}^{-1}(q) \mathbf{S}(q,t) \nonumber \\
  & + \mathbf{J}(q) \int_0^t \rmd t' \, \mathbf{M}(q,t-t')
  \partial_{t'} \mathbf{S}(q,t')= \mathbf{0} .
  \label{eq:eomMCT} 
\end{eqnarray}
The matrix $J_{\alpha\beta}(q) = q^2 (k_{\rm B}T/m_\alpha) \delta_{\alpha\beta}$ is given in terms of the square of the thermal velocities, $k_{\rm B}T/m_\alpha$, where $m_\alpha$ is the mass of a particle of species $\alpha$; this matrix describes inertial effects. Outside the initial inertial regime the dynamics is determined by the memory kernels $M_{\alpha\beta}(q,t)$. These kernels are fluctuating force-correlation functions, reflecting many-body interaction effects. MCT writes $\mathbf{M}(q,t)$ as a sum of two terms: $\mathbf{M}(q,t) = \mathbf{M}_{\rm reg}(q,t) + \mathbf{M}_{\rm MCT}(q,t)$. The ``regular'' term $\mathbf{M}_{\rm reg}(q,t)$ is supposed to describe the normal liquid-state dynamics; it decays on short time scales and is not responsible for slow glassy dynamics. We model the regular term as a Markovian process with friction constant $\nu$: $\mathbf{M}_{\rm reg}(q,t) = \mathbf{J}^{-1}(q)\nu \delta(t)$ \cite{WeysserEtal:PRE2010}. The slow dynamics is encapsulated in the second term. For $\mathbf{M}_{\rm MCT}(q,t)$ the theory assumes that the dominant contribution to the fluctuating forces stems from pairs of density fluctuations and factorizes the resulting four-point correlation function into a product of two two-point correlation functions:
\begin{equation}
  \mathbf{M}_{\rm MCT}(q,t) = \boldsymbol{\mathcal{F}}
  [\mathbf{S}(t),\mathbf{S}(t)](q) ,
\end{equation}
where the components of the mode-coupling functional $\boldsymbol{\mathcal{F}}$ are given by
\begin{eqnarray}
  \fl \mathcal{F}_{\alpha\beta}
  & [\mathbf{S}(t),\mathbf{S}(t)](q) \nonumber \\
  & = \frac{1}{2q^2} \frac{\rho}{\xA\xB} \int \frac{\rmd^3 k}{(2\pi)^3}
  \sum_{\alpha'\beta'\alpha''\beta''} V_{\alpha\alpha'\alpha''}(\bi{q},\bi{k})
  \nonumber \\
  & \times V_{\beta\beta'\beta''}(\bi{q},\bi{k}) S_{\alpha'\beta'}(k,t)
  S_{\alpha''\beta''}(|\bi{q}-\bi{k}|,t) .
  \label{eq:defMCTMF}
\end{eqnarray}
Here $V_{\alpha\alpha'\alpha''}$ are the coupling vertices
\begin{eqnarray}
  \fl V_{\alpha\alpha'\alpha''}(\bi{q},\bi{k}) 
  = &
  \frac{\bi{q}\cdot\bi{k}}{q} c_{\alpha\alpha'}(k)\delta_{\alpha\alpha''} 
  \nonumber \\
  & +
  \frac{\bi{q}\cdot(\bi{q}-\bi{k})}{q} c_{\alpha\alpha''}(|\bi{q}-\bi{k}|)
  \delta_{\alpha\alpha'} ,
  \label{eq:defcouplingvertices}
\end{eqnarray}
which only depend on the equilibrium structure of the system via the matrix of direct correlation functions, $c_{\alpha\beta}(q)$, defined in terms of $\mathbf{S}(q)$ by the Ornstein--Zernike equation
\begin{equation}
  \rho c_{\alpha\beta}(q) = \frac{\delta_{\alpha\beta}}{\xA} - 
  \left (\mathbf{S}^{-1}(q) \right )_{\alpha\beta} .
  \label{eq:dcf}
\end{equation}
In writing \eref{eq:defcouplingvertices} we assume that static triple correlations can be treated by the convolution approximation. This approximation has been justified for the Kob--Andersen mixture \cite{SciortinoKob2001} and we suppose that it also holds for the Voronoi mixture.

\Eref{eq:defsqtpartial} to \eref{eq:dcf} establish a link between the equilibrium structure and dynamics of a glass former. This opens the possibility to predict the temperature dependence of the dynamics based on static input from simulations and to compare these predictions against the simulated relaxation behavior. Here we will carry out such a comparison for the Voronoi mixture. Similar comparisons have been performed for a variety of different models, including binary \cite{FoffiEtal:PRE2004,weysser2011} and polydisperse hard-sphere systems \cite{WeysserEtal:PRE2010}, the Kob--Andersen Lennard-Jones mixture \cite{FlennerSzamel:PRE2005_2,KobNaurothSciortino2002,FlennerSzamel:PRE2005_1}, metallic glasses \cite{DasEtal:PRB2008,KuhnEtal:PRB2014}, strong liquids \cite{SciortinoKob2001,VoigtmannHorbach:EPL2006}, orthoterphenyl \cite{RinaldiSciortinoTartaglia2001,ChongSciortino:PRE2004}, and polymer melts \cite{Ciarella2019,FreyEtal:dscf2013,ChongEtal:Polymer1,Colmenero:JPCM2015}.

\subsection{MCT equations for single-particle dynamics}
\label{subsec:mct_incoherent}
To describe the single-particle dynamics, MCT considers the correlation function of the tagged-particle density, i.e.\ the incoherent intermediate scattering function
\begin{equation}
  \phi^{{\rm s},\alpha}(q,t) = \frac{1}{N_\alpha} \sum_{j =1}^{N_\alpha}
  \left \langle \exp \left \{ \rmi \bi{q} \cdot [\bi{r}_{j}(t) 
  - \bi{r}_{j}(0 ) ] \right \} \right \rangle ,
  \label{eq:defiisf}
\end{equation}
of species $\alpha$ ($= \lA, \lB$). By the Zwanzig--Mori projection operator formalism the following equation of motion is obtained
\begin{eqnarray}
  \frac{m_\alpha}{q^2 k_{\rm B}T} & \partial^2_t \phi^{{\rm s},\alpha}(q,t)
  + \phi^{{\rm s},\alpha}(q,t)\nonumber \\
  & + \int_0^t \rmd t' \, M^{{\rm s},\alpha} (q,t-t') 
  \partial_{t'} \phi^{{\rm s},\alpha}(q,t') = 0 .
  \label{eq:eomiisf} 
\end{eqnarray}
As for the coherent density fluctuations, the memory kernel is approximated by a sum of a regular part, modeled as a damped Markovian process $M_{\rm reg}^{{\rm s},\alpha}(q,t) = m_\alpha\nu/(q^2 k_{\rm B}T)\delta(t)$ \cite{WeysserEtal:PRE2010}, and an MCT contribution. The expression for the latter reads \cite{KuhnEtal:PRB2014,WeysserEtal:PRE2010}
\begin{eqnarray}
  \fl M_{\rm MCT}^{{\rm s},\alpha} (q,t)
  & = \mathcal{F}^{{\rm s},\alpha}[\mathbf{S}(t),\phi^{{\rm s},\alpha}(t)](q) 
    \nonumber \\
  & = \frac{\rho}{q^2} \int \frac{\rmd^3 k}{(2\pi)^3}
  \sum_{\alpha'\beta'} \left (\frac{\bi{q}\cdot\bi{k}}{q} \right )^2
  c_{\alpha\alpha'}(k)c_{\alpha\beta'}(k)  
  \nonumber \\
  & \times S_{\alpha'\beta'}(k,t)
  \phi^{{\rm s},\alpha}(|\bi{q}-\bi{k}|,t) .
  \label{eq:defMCTMfiisf}
\end{eqnarray}
The solution of \eref{eq:eomiisf} requires not only static input, but also the collective $\mathbf{S}(q,t)$ which needs to be determined from \eref{eq:defMCTMF}.

\subsection{Numerical solution of the MCT equations}
\label{subsec:numerics}
Using bipolar coordinates and the rotational symmetry of the system, the three-dimensional integral over $\bi{k}$ in \eref{eq:defMCTMF} and \eref{eq:defMCTMfiisf} is written as a double integral over $k = |\bi{k}|$ and $p = |\bi{q} - \bi{k}|$. Then, $q$ is discretized by introducing a finite, equally spaced grid of $M$ points $q = q_0 + \hat{q} \Delta q$ with $\hat{q} = 0, 1, \ldots, M-1$. This allows us to replace the double integral by Riemann sums
\begin{equation}
  \int_0^\infty \rmd k  \int_{|q-k|}^{q+k} \rmd p
  \; \rightarrow \;
  (\Delta q)^2 \sum_{\hat{k}=0}^{M-1} \sum_{\hat{p} = |\hat{q}-\hat{k}|}^{\min[M-1,\hat{q}+\hat{k}]} .
  \label{eq:discreteqintegral}
\end{equation}
Following commonly made choices \cite{WeysserEtal:PRE2010} we took $M = 300$, $\Delta q = 0.1998333(\ldots)$ and $q_0 = 0.09991666(\ldots)$ so that $0.0999 \lesssim q \lesssim 59.8501$. The partial static structure factors that serve as input in \eref{eq:defcouplingvertices} were obtained from the simulations. Upon insertion of \eref{eq:discreteqintegral} into \eref{eq:eomMCT} one gets a finite number of coupled nonlinear integro-differential equations. For the solution of these equations we employ the algorithm of \cite{FuchsGoetzeHofackerLatz1991} in which the first $64$ time points were calculated with a step size of $\Delta t = 10^{-6}$, and $\Delta t$ was subsequently doubled for every $32$ new points. The friction constant of the regular kernel was set to $\nu = 1$.

\subsection{Universal MCT predictions}
\label{sec:mct_universal}
MCT makes a number of ``universal'' predictions. They are universal in the sense that they do not depend on the details of the static input, but are mathematical consequences of the form of the MCT equations \cite{GoetzeBook2009}. Here we summarize those results which will be important for the analysis of the MD simulations.

Let us denote the long-time limits of the solutions of \eref{eq:eomMCT} and \eref{eq:eomiisf} by 
\begin{equation}
  \eqalign{
  \mathbf{F}(q) = \lim_{t \rightarrow \infty} \mathbf{S}(q,t) , \\
  f^{{\rm s},\alpha}(q) = \lim_{t \rightarrow \infty} 
  \phi^{{\rm s},\alpha}(q,t) .}
  \label{eq:deflongtlimits}
\end{equation}
By means of the Laplace transform and the final value theorem, one can show that $\mathbf{F}(q)$ and $f^{{\rm s},\alpha}(q)$ obey the equations
\begin{eqnarray}
  \mathbf{F}(q) = \mathbf{S}(q) - \left (\mathbf{S}^{-1}(q) + 
  \boldsymbol{\mathcal{F}}[\mathbf{F},\mathbf{F}](q) \right )^{-1} ,
  \label{eq:NEP} \\
  \frac{f^{{\rm s},\alpha}(q)}{1 - f^{{\rm s},\alpha}(q)} = 
  \mathcal{F}^{{\rm s},\alpha}[\mathbf{F},f^{{\rm s},\alpha}](q) .
  \label{eq:incNEP}
\end{eqnarray}
These equations are defined by the static structure; neither the inertia matrix $\mathbf{J}$ nor the regular kernel $\mathbf{M}_{\rm reg}$ enter. Therefore, the solutions are independent of the microscopic dynamics. Equations~\eref{eq:NEP} and \eref{eq:incNEP} can be solved by an iteration procedure \cite{FranoschVoigtmann:JSP2002}.

The solutions of \eref{eq:NEP} display bifurcations. For structural glasses usually the $\mathcal{A}_2$ bifurcation is relevant \cite{GoetzeBook2009}. If $T$ is the control variable, the bifurcation occurs at a critical temperature $\Tc$ (depending on composition and particle size ratio \cite{GoetzeVoigtmann:PRE2003}). For $T > \Tc$, one has $\mathbf{F}(q) = \mathbf{0}$. This behavior corresponds to an ergodic liquid where density correlations decay to 0 for $t \rightarrow \infty$. For $T \leq \Tc$, the long-time limits are given by a (nondegenerate symmetric) positive-definite matrix $\mathbf{F}(q)$. Since density correlations no longer decay to zero, MCT describes an amorphous solid, i.e.\ a nonergodic ideal glass. Accordingly, the corresponding $\mathbf{F}(q)$ are called ``nonergodicity parameters''. The glass transition point $\Tc$ can be identified with the highest temperature at which the system is a glass, i.e.\ at which $\mathbf{F}$ jumps from $\mathbf{0}$ to some finite $\mathbf{F}^{\rm c}$. In the generic case, the tagged-particle dynamics is strongly coupled to the collective dynamics and undergoes a glass transition. This implies that the solution of \eref{eq:incNEP} also jumps from zero to a finite value $f^{{\rm sc},\alpha}(q)$ at $\Tc$. Along with the finite value $\mathbf{F}^{\rm c}$ of the nonergodicity parameter, the corresponding stability matrix $\boldsymbol{\mathcal{C}}^{\rm c}$ of \eref{eq:NEP}, defined by
\begin{eqnarray}
  \boldsymbol{\mathcal{C}}^{\rm c}[\mathbf{H}(q)] = &  \big [\mathbf{S}^{\rm c}(q)-
  \mathbf{F}^{\rm c}(q) \big ] \big [\boldsymbol{\mathcal{F}}^{\rm c}
  [\mathbf{F}^{\rm c},\mathbf{H}](q) \nonumber \\
  & + \boldsymbol{\mathcal{F}}^{\rm c}[\mathbf{H},\mathbf{F}^{\rm c}](q)\big ]
  \big [\mathbf{S}^{\rm c}(q)- \mathbf{F}^{\rm c}(q) \big ] ,
  \label{eq:defstabmat}
\end{eqnarray}
has a unique right eigenvector $\mathbf{H}(q)$ with eigenvalue $E_0=1$. Here the superscript ``c'' means that all static input is evaluated at $\Tc$. The normalization factors of $\mathbf{H}(q)$ are determined by the convention 
\begin{equation}
  \eqalign{\widehat{\mathbf{H}}(q):\mathbf{H}(q) = 1 , \\
  \widehat{\mathbf{H}}(q):\big \{\mathbf{H}(q)
  [\mathbf{S}^{\rm c}(q)-\mathbf{F}^{\rm c}(q)]^{-1}\mathbf{H}(q)\big \} = 1 ,}
  \label{eq:convention}
\end{equation}
where $\widehat{\mathbf{H}}(q)$ is the left eigenvector of $\boldsymbol{\mathcal{C}}^{\rm c}$ with $E_0=1$ and the double-dot operator includes integration over $q$.

Close to $\Tc$ the solutions of \eref{eq:eomMCT} and \eref{eq:eomiisf} show that $\mathbf{S}(q,t)$ and $\phi^{{\rm s},\alpha}(q,t)$ stay close to a plateau given by $\mathbf{F}^{\rm c}(q)$ and $f^{{\rm sc},\alpha}(q)$ for an intermediate time interval. This time interval is called ``$\beta$ relaxation regime'' in MCT. Whereas the $\beta$ relaxation exists both above and below $\Tc$ (below $\Tc$, $\mathbf{F}^{\rm c}$ and $f^{{\rm sc},\alpha}$ are replaced by the $T$ dependent long-time limits \eref{eq:deflongtlimits}), a decay of $\mathbf{S}(q,t)$ from the plateau to zero---i.e.\ the $\alpha$ relaxation---can only occur in the liquid phase for $T > \Tc$. The $\beta$ and the $\alpha$ process are characterized by two time scales: the $\beta$ relaxation time $t_{\sigma}$,
\begin{eqnarray}
  t_{\sigma} = \frac{t_0}{|\sigma|^{1/2a}} \quad
  \mbox{(for $T \rightarrow \Tc^\pm$)},
  \label{tps_beta}
\end{eqnarray}
and the $\alpha$ relaxation time $t'_{\sigma}$, 
\begin{eqnarray}
  t'_{\sigma} = \frac{t_0}{(-\sigma)^{\gamma}}, 
  \quad \gamma=\frac{1}{2a} + \frac{1}{2b} 
  \quad \mbox{(for $T \rightarrow \Tc^+$)}.
  \label{tps_alpha}
\end{eqnarray}
Here $t_0$ represents a system-specific microscopic time scale and $\sigma$ is the ``separation parameter'' quantifying the distance to the critical point where the bifurcation occurs. Close to $\Tc$ the separation parameter can be expressed as
\begin{eqnarray}
  \sigma=C \varepsilon,  \quad \varepsilon = \frac{\Tc - T }{\Tc}
  \label{eq:sigma2Tc}
\end{eqnarray}
with $C$ being a constant. MCT refers to $a$ as ``critical exponent'' and to $b$ as ``von Schweidler exponent''. They are connected to one another by the ``exponent parameter'' $\lambda$,
\begin{eqnarray}
  \lambda=\frac{\mathit{\Gamma}(1-a)^2}{\mathit{\Gamma}(1-2a)} =
  \frac{\mathit{\Gamma}(1+b)^2}{\mathit{\Gamma}(1+2b)} 
  \quad (1/2 \le \lambda < 1) ,
\label{lambda_parameter}
\end{eqnarray}
where $\mathit{\Gamma}$ is the Gamma function. The parameter $\lambda$ is a static quantity that can be calculated from the equilibrium structure of the glass former at $\Tc$ by 
\begin{eqnarray}
  \lambda = \widehat{\mathbf{H}}(q) : \big \{ \big [ & \mathbf{S}^{\rm c}(q)-
  \mathbf{F}^{\rm c}(q) \big ] \nonumber \\
  & \boldsymbol{\mathcal{F}}^{\rm c}[\mathbf{H},\mathbf{H}](q)
  \big [\mathbf{S}^{\rm c}(q)-\mathbf{F}^{\rm c}(q) \big ] \big \} .
  \label{eq:lambdafromMCT}
\end{eqnarray}
Since $1/2 \leq \lambda < 1$ for the $\mathcal{A}_2$ bifurcation \cite{GoetzeBook2009}, \eref{lambda_parameter} gives $0 < a < 0.3953$ and $0 < b \leq 1$, and so $\gamma > 1.7649$ due to \eref{tps_alpha}. 

On cooling the liquid toward $\Tc$, the ratio $t'_{\sigma} / t_{\sigma}$ increases. The smaller $T - \Tc$, the more separated are the $\beta$ and $\alpha$ relaxation regimes. MCT therefore predicts a two-step relaxation. The intermediate time interval of the $\beta$ regime is defined by $t_0 \ll t \leq t'_\sigma$. This interval comprises $t \sim t_{\sigma}$ where $\mathbf{S}(q,t) \sim \mathbf{F}^{\rm c}$ or $\phi^{{\rm s},\alpha}(q,t) \sim f^{{\rm sc},\alpha}(q)$. The $\alpha$ regime begins for $t > t_\sigma$ and leads to $\mathbf{S}(q,t) \rightarrow \mathbf{0}$ or $\phi^{{\rm s},\alpha}(q,t) \rightarrow 0$ for $t \gg t'_{\sigma}$. Both regimes overlap for $t_\sigma \leq t \leq t'_\sigma$. The latter time interval is called late $\beta$ or early $\alpha$ process in MCT.

For both the $\alpha$ and $\beta$ process, MCT makes detailed predictions \cite{GoetzeBook2009,Goetze_LesHouches}, many of which have been tested in fits to experimental and simulation data (for reviews of these tests see e.g.\ \cite{GoetzeBook2009,Janssen2018,Colmenero:JPCM2015,BaschnagelVarnik:JPCM2005,Goetze:JPCM1999,GoetzeSjoegren1992_RPP,Kob_LesHouches2003,%
kobreview1999}). In the following, we will also perform such fits for the binary Voronoi mixture. This analysis will be carried out for the coherent intermediate scattering function,
\begin{eqnarray}
  \phi(q,t) = \frac{S_\iAA(q,t) + S_\iBB(q,t)
   + 2 S_\iAB(q,t)}{S(q)}, 
  \label{eq:defisf}
\end{eqnarray}
the incoherent scattering functions $\phi^{{\rm s},\alpha}(q,t)$ and the mean-square displacements (MSDs),
\begin{equation}
  g_{0,\alpha}(t) = \frac{1}{N_\alpha} \sum_{j = 1}^{N_\alpha}
  \left \langle \left [ \bi{r}_{j}(t) - \bi{r}_{j}(0) 
  \right ]^2 \right \rangle ,
  \label{eq:defmsd}
\end{equation}
of species $\alpha = \lA, \lB$. The MSD is related to $\phi^{{\rm s},\alpha}(q,t)$ by $g_{0,\alpha}(t) = \lim_{q \rightarrow 0} 6[ 1 - \phi^{{\rm s},\alpha}(q,t)]/q^2$. Therefore, we summarize below the MCT predictions pertinent for this analysis.

\subsubsection*{Predictions for the $\beta$ regime}
In the $\beta$ regime MCT predicts a ``factorization theorem'' according to which all density correlators (and all quantities coupling to them) can be expressed as a sum of the nonergodicity parameter and a correction term that exhibits a factorization into a wavevector-dependent and a time-dependent part \cite{GoetzeBook2009,Goetze_LesHouches}:
\begin{eqnarray}
  \phi(q,t)   & = f^{\rm c}(q) + h(q) G(t) , 
  \label{coherent_approximation} \\
  \phi^{{\rm s},\alpha}(q,t) & = f^{{\rm sc},\alpha}(q) + h^{{\rm s},\alpha}(q) G(t) .
  \label{incoherent_approximation}
\end{eqnarray}
The nonergodicity parameters, $f^{\rm c}(q) = \lim_{t \rightarrow \infty} \phi(q,t)$ and $f^{{\rm sc},\alpha}(q) = \lim_{t \rightarrow \infty} \phi^{{\rm s},\alpha}(q,t)$, and the ``critical amplitudes'', $h(q)$ and $h^{{\rm s},\alpha}(q)$, are evaluated at $\Tc$ and are thus independent of $T$. The temperature dependence resides in the ``$\beta$ correlator'' $G(t)$ which, for $T \rightarrow \Tc^+$, is given by
\begin{eqnarray}
G(t) = \sqrt{|\sigma|}g(t/t_{\sigma}) \stackrel{t \gg t_{\sigma}}{\longrightarrow} -B(\lambda) \left( \frac{t}{t'_{\sigma}} \right)^b .
\label{beta_correlator}
\end{eqnarray}
Here $B(\lambda)$ is a $T$-independent constant and $G(t) \sim - (t/t'_{\sigma})^b$ is the so-called von Schweidler law which holds for $t_{\sigma} \ll t \ll t'_{\sigma}$. Both the factorization theorem and the von Schweidler law are MCT results in leading order of $|\sigma|$. 
Second order corrections to \eref{coherent_approximation} and \eref{incoherent_approximation} are also known \cite{FranoschFuchsGoetze1997,FuchsGoetzeMayr1998}:
\begin{eqnarray}
  \phi(q,t)
    = f^{\rm c}(q) & - h(q) B(\lambda) \left( \frac{t}{t'_{\sigma}} \right)^b
      \nonumber \\
  & + h(q) B^2(\lambda) B(q) \left(\frac{t}{t'_{\sigma}}\right)^{2b} ,
  \label{coherent_vonschweidler}
\end{eqnarray}
\begin{eqnarray}
  \phi^{{\rm s},\alpha}(q,t)
    = f^{{\rm sc},\alpha} & (q) - h^{{\rm s},\alpha}(q) B(\lambda)
  \left( \frac{t}{t'_{\sigma}} \right)^b \nonumber \\
  & + 
  h^{{\rm s},\alpha}(q) B^2(\lambda) B^{{\rm s},\alpha}(q) 
  \left(\frac{t}{t'_{\sigma}}\right)^{2b} .
  \label{incoherent_vonschweidler}
\end{eqnarray}
The $q$ dependence of the correction amplitudes $B(q)$ and $B^{{\rm s},\alpha}(q)$ implies a violation of the factorization theorem. Both amplitudes are again evaluated at $\Tc$; in the $\beta$ regime the $T$ dependence therefore solely results from the time scale $t^\prime_\sigma$.

\subsubsection*{Predictions for the $\alpha$ regime}
In the $\alpha$ regime MCT predicts that the density correlators are described by $T$-independent master curves (for $T \rightarrow \Tc^{+})$:
\begin{eqnarray}
  \phi(q,t) = \widetilde{\phi}(q,t/t'_{\sigma}), \quad
  \phi^{{\rm s},\alpha}(q,t) = \widetilde{\phi}^{{\rm s},\alpha}(q,t/t'_{\sigma}),
  \label{TTSP_eq}
\end{eqnarray}
which have the following limits for $t \rightarrow 0$:
\begin{eqnarray}
  \phi(q,t \rightarrow 0) = f^{\rm c}(q), \quad 
  \phi^{{\rm s},\alpha}(q,t\rightarrow 0) = f^{{\rm sc},\alpha}(q) .
  \label{alpha_limit_nonergo}
\end{eqnarray}
\Eref{TTSP_eq} implies a time-temperature superposition principle (TTSP): For fixed $q$, $\phi(q,t)$ and $\phi^{{\rm s},\alpha}(q,t)$ collapse for different $T$ onto master curves when rescaling $t$ by some relaxation time that is proportional to $t'_{\sigma}$. For instance, we can choose $\phi(q,t)$ at the peak position $q^*$ of $S(q)$ to define the relaxation time $\tau_{q^*}$ by the criterion $\phi(q^*,\tau_{q^*})={\rm const}$. Then, we have
\begin{eqnarray}
\tau_{q^*}=C_{q^*} t'_{\sigma} ,
\label{tauTTSP}
\end{eqnarray}
where the $T$-independent prefactor $C_{q^*}$ is determined by the constant used in the definition $\phi(q^*,\tau_{q^*})={\rm const}$.

For $t \ll t'_{\sigma}$, \eref{TTSP_eq} recovers the von Schweidler law. This justifies the statement made above that the late $\beta$ and early $\alpha$ process overlap for $t_{\sigma} \ll t \ll t'_{\sigma}$. Moreover, model calculations within MCT reveal that the $\alpha$ master curves are stretched. As for experimental or simulation data, this stretched relaxation can be fitted well by a Kohlrausch--Williams--Watts (KWW) function. For $\phi(q,t)$ the KWW function reads
\begin{equation}
  \label{eq:KWW-function}
  \phi(q,t) \simeq A(q) \exp \left[-\left (\frac{t}{\tau^{\rm K}(q)
  }\right )^{\beta^\mathrm{K}(q)}\right] \quad (t \geq t_\sigma) ,
\end{equation}
where $A(q)$ is an amplitude, $\tau^{\rm K}(q)$ the relaxation time and $\beta^\mathrm{K}(q) \leq 1$ the stretching exponent. Although the KWW function is a well suited fit function, it is in general not a solution of the MCT $\alpha$ process, except in the special limit of large $q$.  In this limit, it was proved \cite{Fuchs1994_kww} that there is a time interval $t/t'_\sigma \ll t^\mathrm{\,K}_q/t'_\sigma \leq 1$ in which the $\alpha$ process obeys
\begin{equation}
  \label{eq:kww_limit}
  \lim_{q \rightarrow \infty} \phi(q,t) = 
  f^{\rm c}(q) \exp \left[-\Gamma(q) \left (\frac{t}{t'_\sigma}\right )^{b}
  \right] ,
\end{equation}
with $\Gamma(q) \propto q$. This implies
\begin{equation}
  \eqalign{
  \lim_{q\rightarrow \infty} A(q) = f^{\rm c}(q), \quad
  \lim_{q \rightarrow \infty}\beta^{\rm K}(q) = b, \cr
  \lim_{q \rightarrow \infty} \tau^{\rm K}(q) \propto 
  \frac{t'_\sigma}{q^{1/b}} .}
  \label{eq:MCT2KWW}
\end{equation}
Equations analogous to \eref{eq:kww_limit} and \eref{eq:MCT2KWW} also hold for the incoherent scattering functions $\phi^{{\rm s},\alpha}(q,t)$.

\section{Results}
\label{sec:results}

\subsection{Factorization theorem, time-temperature superposition principle}
\label{subsec:factheo+ttsp}
\begin{figure}
\begin{center}
  \includegraphics*[scale=0.38]{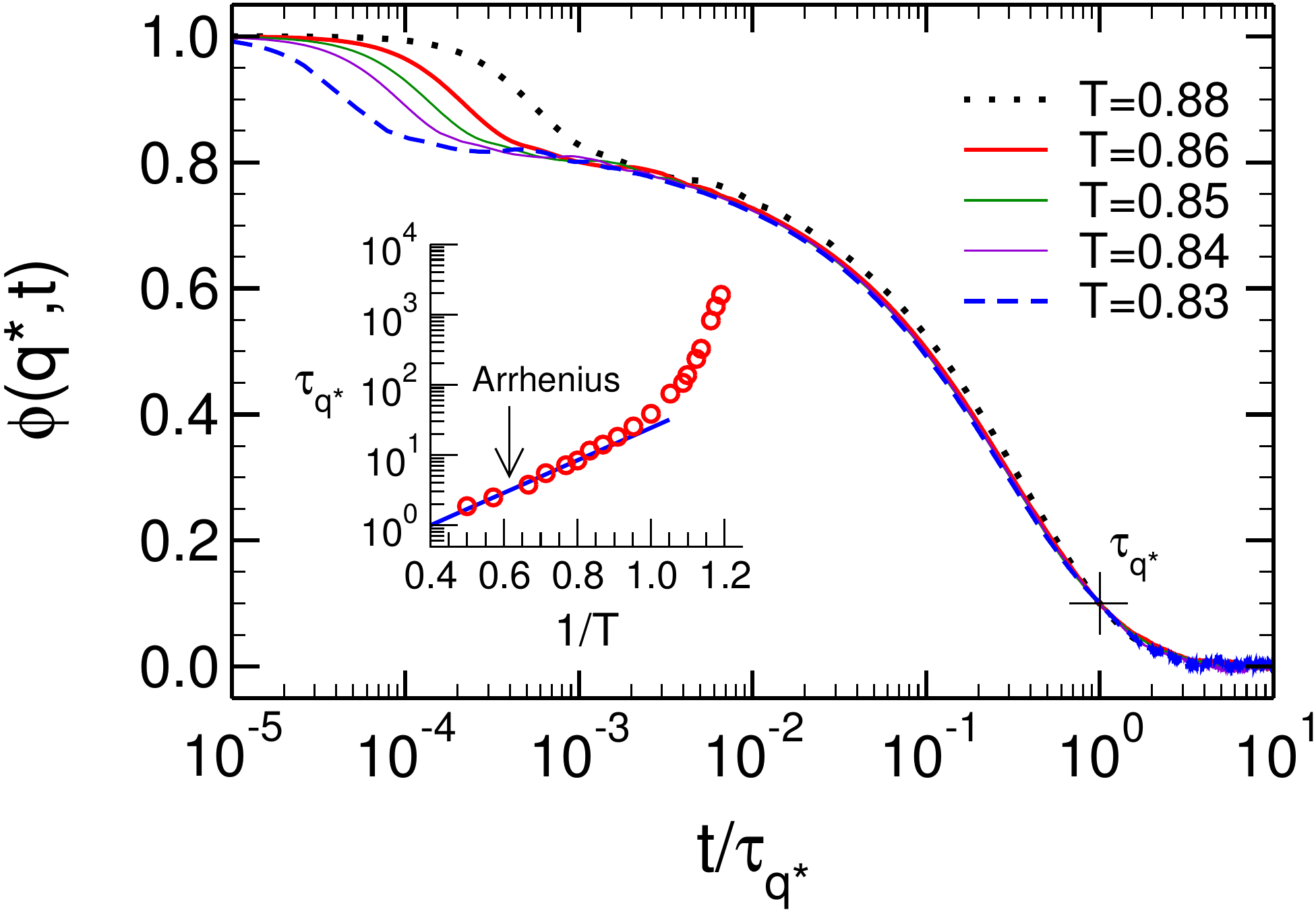}
\end{center}
\caption{Test of the TTSP for $0.83 \le T \le 0.88$: $\phi(q^*,t)$ as a function of $t/\tau_{q^*}$ where $q^*=6.85$ is the peak position of the first maximum of $S(q)$. $\tau_{q^*}$ is defined by the condition $\phi(q^*,t=\tau_{q^*})=0.1$ (indicated by a cross in the figure). The dotted black line represents $T=0.88$, the temperature above which the TTSP is violated. Inset: Arrhenius plot of $\tau_{q^*}$ versus $1/T$. The circles represent the simulation data. The solid line shows a fit to $\tau_{q^*}(T) = \tau_{q^*}^\infty \exp (E_{\rm A}/k_{\rm B} T)$ with $\tau_{q^*}^\infty = 0.1184$ and $E_{\rm A}/k_{\rm B} = 5.3259$.}
\label{fig_TTSP}
\end{figure}

Our MCT analysis of the Voronoi mixture starts with a test of the TTSP. To rescale the time axis we follow \eref{tauTTSP} and define the $\alpha$ relaxation time as the time when $\phi(q^*,t)$ has decayed to 10\% of its initial value, i.e.\ $\phi(q^*,t=\tau_{q^*})=0.1$. The threshold of 0.1 is arbitrary, but convenient: The choice ensures that the density correlator is small enough to be well in the $\alpha$ regime, but still sufficiently above the noise level so that the statistical accuracy of the data remains satisfactory. \Fref{fig_TTSP} shows $\phi(q^*,t)$ as a function of $t/\tau_{q^*}$ for $0.83 \le T \le 0.88$. This interval corresponds to the regime of the supercooled liquid where a super-Arrhenius increase of $\tau_{q^*}$ with decreasing $T$ is observed (cf inset of \fref{fig_TTSP}). For these temperatures we find that $\phi(q^*,t)$ decays in two steps, developing an intermediate time interval where $\phi(q^*,t)$ plateaus. This time interval extends upon cooling, and the second relaxation step away from the plateau toward zero obeys the TTSP for $T \lesssim 0.88$. These observations are in qualitative agreement with MCT, suggesting to focus on $T \lesssim 0.88$ for further analysis. 

The factorization theorem, \eref{coherent_approximation} and \eref{incoherent_approximation}, provides an additional means to determine whether an analysis of the observed two-step relaxation in terms of MCT is justified or not. A simple test of the theorem works directly with the simulation data without invoking any fit procedure \cite{weysser2011,WeysserEtal:PRE2010,FreyEtal:dscf2013,KobAndersen_LJ_I_1995,GleimKob2000,VoigtmannEtal:2004,horbach2002,HorbachKob:PRE2001,BernabeiEtal:JCP2009,ColmeneroEtal:JPCM2007,KhairyEtal:PRE2013,HelfferichEtal:EPJE2018}.
To this end, we fix two times $t_1$ and $t_2$ ($t_2>t_1$) in the $\beta$ regime and calculate the ratio
\begin{eqnarray}
  \nonumber
  R(q,t) & = \frac{\phi(q,t) - \phi(q,t_2)}{\phi(q,t_1)-\phi(q,t_2)} 
           = \frac{G(t)-G(t_2)}{G(t_1)-G(t_2)} \\ 
         & = \frac{\phi^{{\rm s},\alpha}(q,t) - \phi^{{\rm s},\alpha}(q,t_2)}
             {\phi^{{\rm s},\alpha}(q,t_1)-\phi^{{\rm s},\alpha}(q,t_2)}
           = R^{{\rm s},\alpha}(q,t) ,
  \label{factorization}
\end{eqnarray}
where $\alpha = \lA,\lB$. This equation shows that $R(q,t)$ and $R^{{\rm s},\alpha}(q,t)$ are independent of $q$ and superimpose on the same curve in the time window where the factorization theorem holds. Using furthermore \eref{beta_correlator}, $R(q,t)$ and $R^{{\rm s},\alpha}(q,t)$ are given by
\begin{eqnarray}
  R(q,t) = R^{{\rm s},\alpha}(q,t)=\frac{t^b-t_2^b}{t_1^b-t_2^b} .
  \label{facto_b_expo}
\end{eqnarray}
Equations~\eref{factorization} and \eref{facto_b_expo} are predicted to hold close to $\Tc$. In the following, we therefore focus on a low temperature, $T=0.84$.

\begin{figure}
\begin{center}
  \includegraphics*[scale=0.38]{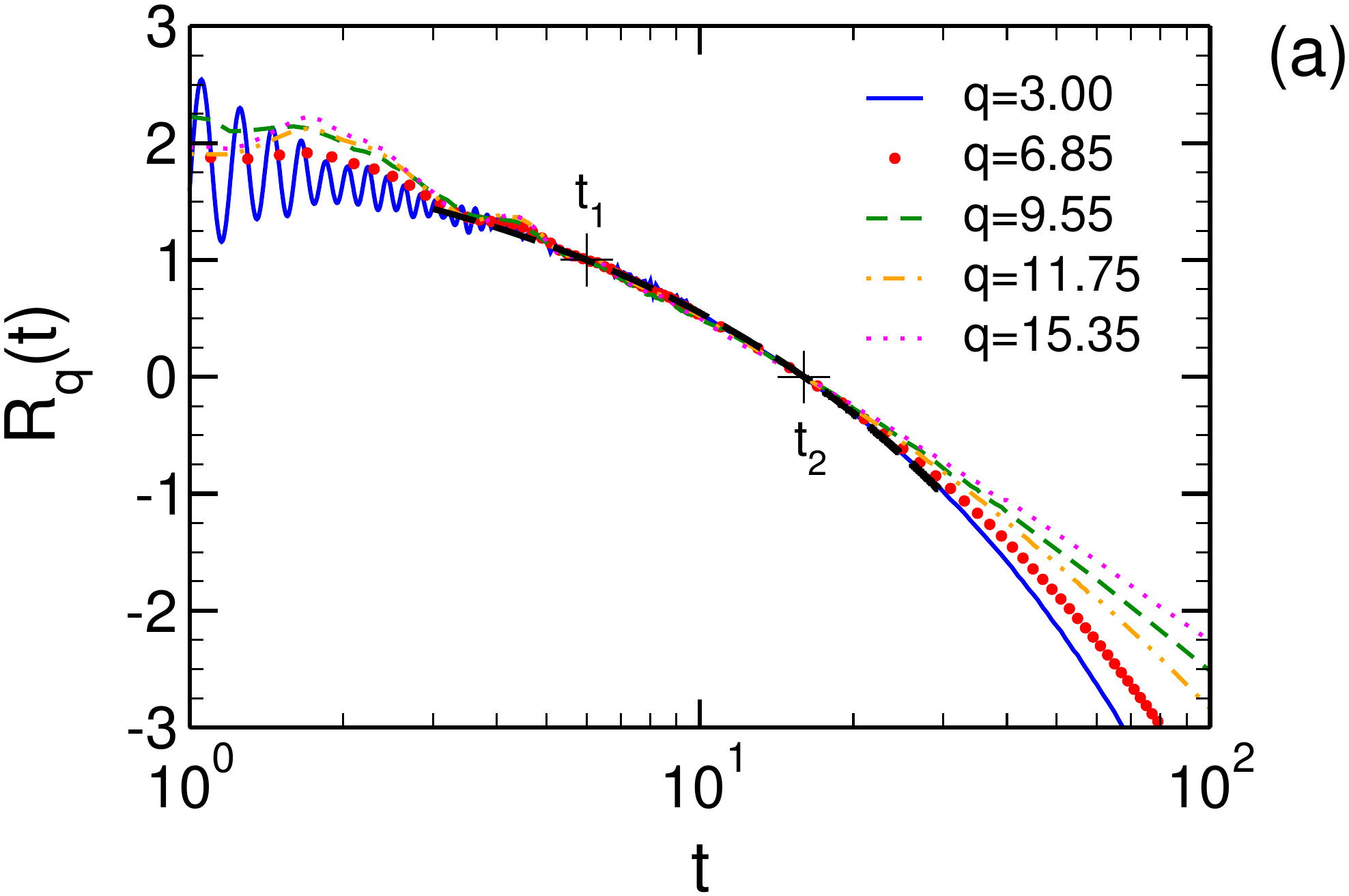}
  \includegraphics*[scale=0.38]{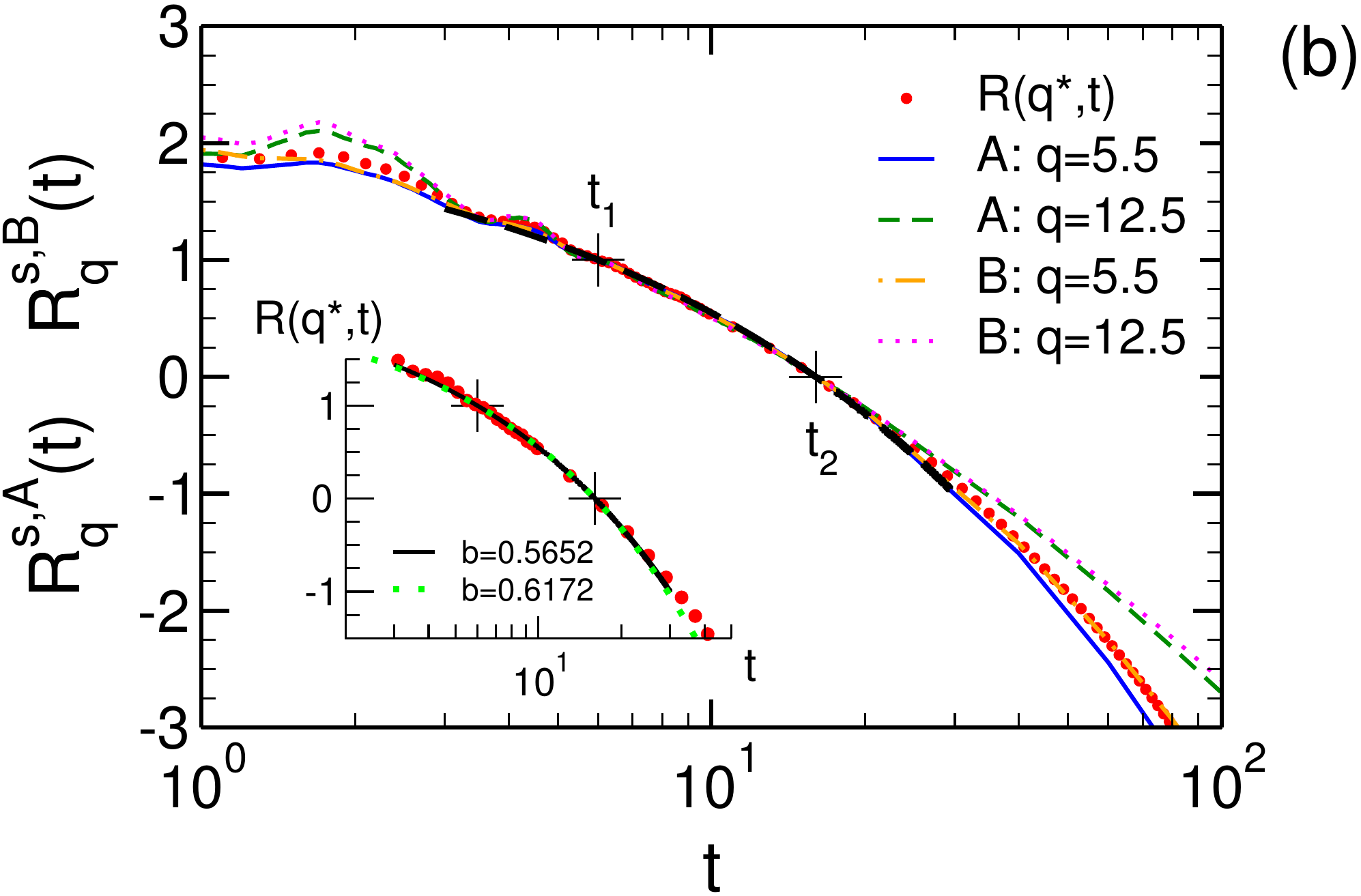}
\end{center}
\caption{Test of the factorization theorem at $T=0.84$ according to \eref{factorization} with the choice $t_1=6$ and $t_2=16$. Panel (a): $R(q,t)$ for $3 \leq q \leq 15.35$. Panel (b): $R^{{\rm s},\alpha}(q,t)$ for $q=5.5$ and $12.5$ for the $\lA$ and $\lB$ particles. By definition, $R(q,t_1=6)=R^{{\rm s},\alpha}(q,t_1=6)=1$ and $R(q,t_2=16)=R^{{\rm s},\alpha}(t_2=16)=0$. The coordinate positions of $t_1$ and $t_2$ are indicated by a plus sign. The red filled circles in panel (b) reproduce $R(q^*,t)$ from panel (a) to illustrate that $R(q,t)$ and $R^{{\rm s},\alpha}(q,t)$ collapse onto the same master curve. In both panels, the dashed black line presents \eref{facto_b_expo} with $b=0.5652$ obtained from fits to the MD data (cf \tref{tablefitparam}). The inset in panel (b) shows a zoom for $-1.5 \leq R(q^*,t) \leq 1.5$ to compare \eref{facto_b_expo} for $b=0.5652$ with $b=0.6172$ obtained from the MCT calculations based on static input (cf \tref{tabMCT}).}
\label{fig_factorization}
\end{figure}

\Fref{fig_factorization} applies \eref{factorization} to the simulation data at $T=0.84$ with the choice $t_1 = 6$ and $t_2 = 16$. We see that there is a time interval comprising $t_1$ and $t_2$ where $R(q,t)$ and $R^{{\rm s},\alpha}(q,t)$ are indeed independent of $q$ (cf top and bottom panel of \fref{fig_factorization}) and collapse onto the same master curve (bottom panel). The master curve tends to persist for $t < t_1$ in the case of $R^{{\rm s},\alpha}(q,t)$ and also for $R(q,t)$ if $q \ge 6.85$, while for $q \le 3$ strong oscillations at early times prevent the test of the data collapse for the coherent scattering. Moreover, \fref{fig_factorization} shows that the data separate at early and late times in a $q$-dependent way. This finding is expected from MCT which predicts an ordering rule \cite{FranoschFuchsGoetze1997,FuchsGoetzeMayr1998}: Since the second-order corrections to the factorization theorem have the same $q$ dependent amplitudes both for the early-time and long-time corrections, correlators that lie, for example, above the factorization theorem for short times must also lie above it for long times. Therefore, if we number the correlators in the order in which they enter the collapse regime, this numbering is preserved when the correlators leave the regime \cite{FranoschFuchsGoetze1997}. This prediction has been observed in many simulations \cite{weysser2011,WeysserEtal:PRE2010,
GleimKob2000,BernabeiEtal:JCP2009,ColmeneroEtal:JPCM2007,KhairyEtal:PRE2013,FreyEtal:dscf2013}. \Fref{fig_factorization} suggests that it also holds for our Voronoi mixture.

Finally, the black dashed lines in \fref{fig_factorization}(a) and \fref{fig_factorization}(b) indicate that the master curve is well described by \eref{facto_b_expo} with $b=0.5652$, the von Schweidler exponent found from the fits to \eref{coherent_vonschweidler} in \sref{subsec:fit}. However, there is a caveat. The inset in \fref{fig_factorization}(b) demonstrates that a description of similar quality is obtained with $b=0.6172$, the exponent from the MCT calculations based on the static input (we comment on this difference between the $b$ values in \sref{subsec:MDvsMCT}). Therefore, in the present case, we see that \eref{facto_b_expo} does not allow to determine $b$ precisely: \Eref{facto_b_expo} is an asymptotic result for $T$ close to $\Tc$. Apparently, $T=0.84$ is still too far above $\Tc$ so that the time interval over which the factorization holds---about a decade in \fref{fig_factorization}---is too narrow. From this analysis we conclude that, albeit a precise determination of $b$ via \eref{facto_b_expo} may be difficult in practice, it certainly allows to obtain bounds for $b$ that can serve as valuable input to guide the fits via \eref{coherent_vonschweidler}. We turn to such fits in the next section.

\subsection{Description of the fit procedure using the MCT predictions for the $\beta$ regime}
\label{subsec:fit}
We examine the dynamics in the supercooled regime by fitting the asymptotic MCT predictions, \eref{coherent_vonschweidler} and \eref{incoherent_vonschweidler}, to our simulation data. To this end, we write \eref{coherent_vonschweidler} in the following form: 
\begin{eqnarray}
  \phi(q,t) = f^{\rm c}(q) & - h^{{\rm fit}}(q) \left (\frac{t}{t'_{\sigma}}
  \right)^{b} \nonumber \\
  & + h^{{\rm fit}}(q) B^{{\rm fit}}(q) \left (\frac{t}{t'_{\sigma}}
  \right)^{2b} .
  \label{numerical_fit}
\end{eqnarray}
The fit constants $h^{{\rm fit}}(q)$ and $B^{{\rm fit}}(q)$ are related to the amplitudes $h(q)$ and $B(q)$ of \eref{coherent_vonschweidler} by
\begin{eqnarray}
  h^{{\rm fit}}(q)= B(\lambda) h(q) , \quad B^{{\rm fit}}(q)=B(\lambda) B(q) .
  \label{eq:fitconscoh}
\end{eqnarray}
The same equations are also valid for $\phi^{{\rm s},\alpha}(q,t)$ after substituting $f^{\rm c}(q) \rightarrow f^{\rm sc}(q)$, $h(q) \rightarrow h^{{\rm s},\alpha}(q)$ and $B(q) \rightarrow B^{{\rm s},\alpha}(q)$. 

Five fit parameters are involved in \eref{numerical_fit}. Four of them are independent of $T$, namely $f^{\rm c}(q)$, $h^{{\rm fit}}(q)$, $B^{{\rm fit}}(q)$ and $b$. One parameter, the $\alpha$ time $t'_{\sigma}$, depends on $T$. To carry out the fits it is judicious to work at low temperature. Guided by the tests of the TTSP and of the factorization theorem, we begin the analysis with the coherent scattering function at $q=q^*$ and $T=0.84$ because inspection of \fref{fig_TTSP} suggests that the plateau, i.e.\ $f^{\rm c}(q^*)$, is large and so the late $\beta$ process is pronounced. This allows us to determine $b$. Fixing $b$ and performing the fits for different $T$ gives $t'_{\sigma}(T)$. Keeping then $b$ and $t'_{\sigma}(T)$ constant, the wavevector dependence of $f^{\rm c}(q)$, $h^{{\rm fit}}(q)$ and $B^{{\rm fit}}(q)$ can finally be determined. In practice, we utilize again $T=0.84$ for the latter fits.

It is known that information from the $\alpha$ relaxation is crucial to guide the fit in the $\beta$ regime \cite{SciortinoTartaglia:JPCM1999}. The five-parameter fit is thus subjected to two constraints:
 
(i) The nonergodicity parameter $f^{\rm c}(q)$ is the initial value of the $\alpha$ master curve [cf \eref{alpha_limit_nonergo}], implying that $\widetilde{\phi}(q,t/t'_{\sigma}) < f^{\rm c}(q)$ for $t/t'_{\sigma} > 0$. This imposes a lower bound on $f^{\rm c}(q)$. The fit result for $f^{\rm c}(q)$ cannot be smaller than the value of $\phi(q,t)$ at the shortest time where the TTSP still holds. To verify this constraint \fref{fig_TTSP} serves as a guideline.
 
(ii) \Eref{coherent_vonschweidler} is invariant under the rescaling $h(q) \rightarrow \ell^b h(q)$, $B(q) \rightarrow \ell^b B(q)$ and $t'_{\sigma} \rightarrow \ell t'_{\sigma}$ where $\ell$ is a constant scale factor \cite{FoffiEtal:PRE2004}. Thus, the same fit result can be obtained for a small (small $\ell$) or large (large $\ell$) $\alpha$ time $t'_{\sigma}$, provided the amplitudes are rescaled accordingly. To guide the fit, we make use of the fully microscopic MCT calculations based on static input. Early work on binary soft-sphere mixtures showed that $0.2 < h(q) < 0.8$ for $q^*/2 \lesssim q \lesssim 2 q^*$ \cite{FuchsLatz:PhysicaA1993}. When fitting the data one can constrain $h(q)$ to lie within these bounds. In the present case, we take advantage of the MCT calculations using the static structure factors of our simulations. These calculations provide $h(q)$ and we adjust the constant $\ell$ such that the fit result matches the theoretical $h(q)$.  

A final technical aspect is to choose the time interval $[ t_{\rm min}, t_{\rm max}]$ where the fit is carried out because the latter can have a significant influence on the fit \cite{Comment_Kivelson1994,Reply_Cummins1994,goetzevoigtmann2000,SciortinoTartaglia:JPCM1999}. Certainly, $t_{\rm min}$ should be larger than the time associated with  the initial relaxation ($t_{\rm min} \gtrsim 1$, cf \fref{fig_fitvS}), whereas $t_{\rm max}$ ($>t_{\rm min}$) may not be taken too large to ensure that the second order correction in \eref{numerical_fit} remains small in comparison to the von Schweidler law. Preliminary tests at $T=0.84$ showed that the choice $[t_{\rm min}=10, t_{\rm max}=500]$ satisfies these requirements (e.g.\ we find that $|B^{\rm fit}(q^*)| (t/t'_{\sigma})^{b} \sim 0.2$ in this interval). We fix this time interval for the $\beta$ analysis in the following.

\subsection{Asymptotic analysis and MCT calculations based on static input: Exponents and critical temperature}
\label{subsec:MDvsMCT}
\begin{table}
  \begin{center}
    \begin{tabular}{c|c|c|c|c|c}
	    \hline
	    \textbf{$\Tc$} & \textbf{$\lambda$} & \textbf{$a$} & \textbf{$b$} & \textbf{$\gamma$} & \textbf{$B$}\\
      \hline
      \hline
      0.798 & 0.7457 & 0.3067 & 0.5652 & 2.5149 & 0.8918 \\
      \hline
    \end{tabular}
    \caption{\label{tablefitparam}MCT parameters obtained from fits of the MD data to the asymptotic MCT predictions. $\Tc$ is the average value of the critical temperature determined in \fref{plot_Tc}. The fit of $\phi(q=q^*,t)$ to \eref{numerical_fit} gives the von Schweidler exponent $b$ from which $\lambda$, $a$ and $\gamma$ are calculated by \eref{tps_alpha} and \eref{lambda_parameter}. The constant $B = B(\lambda)$ appearing in \eref{eq:fitconscoh} is obtained by interpolation of the data in Table~3 of \cite{Goetze:JPCM1990}.}
  \end{center}
\end{table}

\begin{figure}
\begin{center}
  \includegraphics*[scale=0.38]{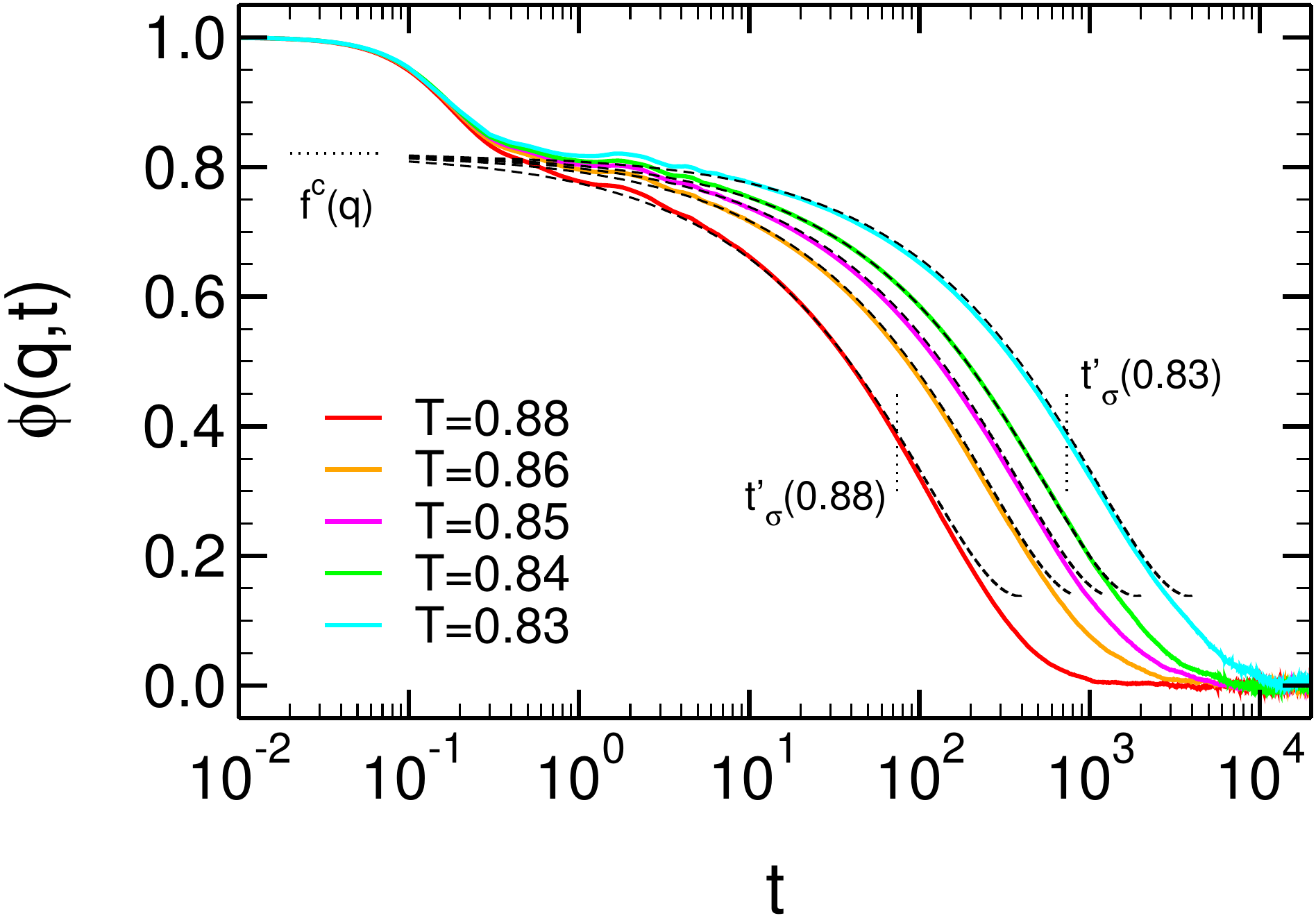}
\end{center}
\caption{Time evolution of the coherent scattering function $\phi(q,t)$ for $q=q^*=6.85$ and $0.83 \leq T \leq 0.88$ (full lines). The black dashed lines correspond to the result of the fit to \eref{numerical_fit} carried out for the interval $10 \leq t \leq 500$. The nonergodicity parameter, $f^{\rm c}(q) = 0.8212$, is represented by horizontal dotted line. Two vertical dotted lines indicate the $\alpha$ relaxation time ($t'_{\sigma}$) at $T=0.83$ and $T=0.88$, respectively.}
\label{fig_fitvS}
\end{figure}

\begin{figure}
\begin{center}
  \includegraphics*[scale=0.38]{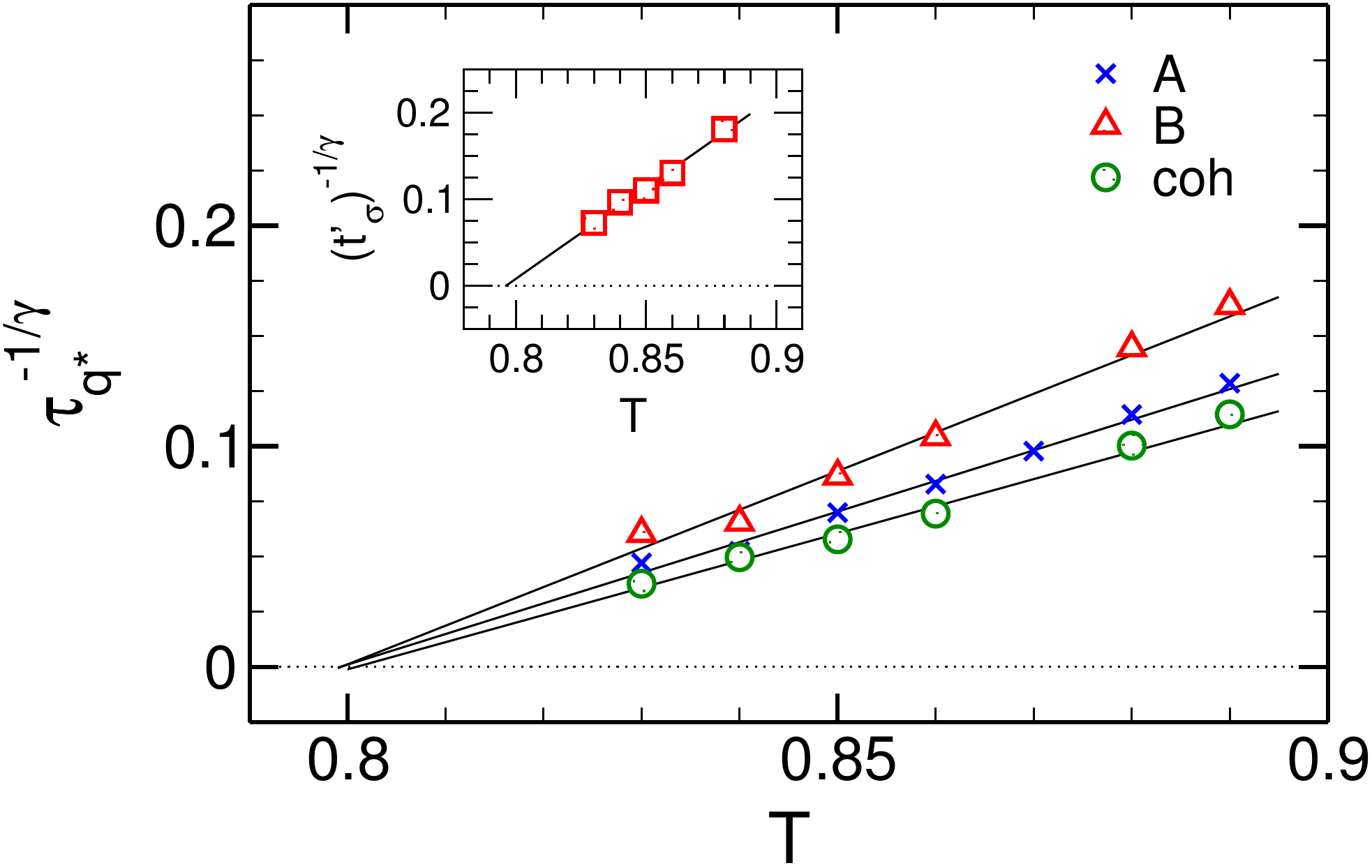}
\end{center}
\caption{Rectification plot of the $\alpha$ relaxation times for the $\lA$ particles $1/(\tau^{{\rm s},\lA}_{q^*})^{1/\gamma}$ (crosses), the $\lB$ particles $1/(\tau^{{\rm s},\lB}_{q^*})^{1/\gamma}$ (triangles) and the coherent dynamics $1/(\tau_{q^*})^{1/\gamma}$ (circles). The relaxation times are defined as the time when the incoherent or coherent intermediate scattering functions take a value of 0.1. $\gamma$ is given in \tref{tablefitparam}. The black full lines are linear extrapolations to zero giving $\Tc = 0.799$ ($\lA$ particles), $\Tc = 0.799$ ($\lB$ particles) and $\Tc = 0.801$ (coherent dynamcis). Inset: Rectification plot of the MCT $\alpha$ relaxation time $t'_\sigma$ (squares). The black full line is an extrapolation to zero giving $\Tc = 0.796$.}
\label{plot_Tc}
\end{figure}

\Fref{fig_fitvS} depicts the simulation results for $\phi(q^*,t)$ in the temperature interval $0.83 \leq T \leq 0.88$ (full lines). The dashed lines present the fits to \eref{numerical_fit}. The fits yield a good description of the MD data, over about two decades in time at $T=0.88$ and extending to about three decades at $T=0.83$. The fits extend to fairly short times; they begin to describe the MD data after the first relaxation step for $t \gtrsim 4$. The shape of this first step depends on the microscopic dynamics of the simulation method---e.g.\ Newtonian or Langevin-based \cite{GleimKob2000,GleimKobBinder1998,BerthierKob:JPCM2007}. For Newtonian MD simulations, as in our case, the first step masks the early $\beta$ relaxation toward the nonergodicity parameter \cite{WeysserEtal:PRE2010,Kob_LesHouches2003}. Due to this reason, we based our analysis on the MCT predictions for the late $\beta$ process, as many other works have done as well \cite{FoffiEtal:PRE2004,WeysserEtal:PRE2010,ChongSciortino:PRE2004,HorbachKob:PRE2001,ColmeneroEtal:JPCM2007,KhairyEtal:PRE2013,SciortinoFabbianChen1997}. 

From the fits to \eref{numerical_fit} we find $b = 0.5652$. \Eref{lambda_parameter} then leads to $\lambda = 0.7457$ (cf \tref{tablefitparam}). This is a typical value. Similar results for $\lambda$ are found for hard spheres \cite{FranoschFuchsGoetze1997,FuchsGoetzeMayr1998,WeysserEtal:PRE2010} and various binary mixtures \cite{DasEtal:PRB2008,NaurothKob1997,FoffiEtal:PRE2004,FlennerSzamel:PRE2005_2} (see also \cite{ColmeneroEtal:JPCM2007} for an overview of $\lambda$ and further MCT parameters for simple and polymeric liquids). In this respect, our binary Voronoi mixture is comparable to other glass-forming systems.

The fit also provides $t'_\sigma(T)$. Following \eref{tps_alpha} a plot of $1/(t'_\sigma)^{1/\gamma}$ against $T$, with $\gamma$ calculated from $a$ and $b$ via \eref{tps_alpha}, should give a straight line that extrapolates to 0 at $\Tc$. By virtue of \eref{tauTTSP}, the same behavior is expected for the $\alpha$ relaxation times defined by $\phi(q^*,\tau_{q^*}) = 0.1$ and $\phi^{{\rm s},\alpha}(q^*,\tau^{{\rm s},\alpha}_{q^*}) =0.1$. \Fref{plot_Tc} tests these expectations. For all relaxation times we find straight lines extrapolating to almost the same value of $\Tc$. From these results we calculate the average $\Tc = 0.798$ given in \tref{tablefitparam}. This $\Tc$ is in excellent agreement with the independent estimate $\Tc = 0.804$ determined from the vanishing of the negative directions associated with saddle points of the potential energy landscape \cite{PhDthesis:Celine}.

\begin{table*}
  \begin{center}
    \begin{tabular}{l|c|c|c|c|c} 
	    \hline
	    \textbf{$\Tc$} & \textbf{$E_0$} & \textbf{$\lambda$} & \textbf{$a$} & \textbf{$b$} & \textbf{$\gamma$} \\
      \hline
      \hline
      0.979\,245(60546875) & 0.9998 & 0.7142 & 0.3209 & 0.6172 & 2.3682\\
      0.979\,24(31640625) & 0.9989 &0.7123 & 0.3217 & 0.6203 & 2.3603\\
      0.979\,23(828125) & 0.9982 & 0.7106 & 0.3225 & 0.6231 & 2.3528\\
      0.979\,21(875) & 0.9966 & 0.7070 & 0.3240 & 0.6291 & 2.3380\\
      0.970\,6(25) & 0.9910 & 0.6948 & 0.3291 & 0.6493 & 2.2894\\
      0.970 & 0.9354 & 0.5833 & 0.3699 & 0.8419 & 1.9456\\
      \hline
    \end{tabular}
    \caption{\label{tabMCT}Impact of the precision of the MCT critical point location ($\Tc$) on the value of $\lambda$ and the exponents $a$, $b$ and $\gamma$, going from the most precise (first row, with an eigenvalue $E_0$ of the stability matrix $\boldsymbol{\mathcal{C}}$ close to 1) to the least precise (last row). The parameters have been obtained from binary MCT using the (interpolated) static structure factors as input.}
  \end{center}
\end{table*}

As described in \sref{sec:mct_universal}, the critical temperature and the MCT parameters can also be predicted by MCT calculations in a fit-parameter-free manner based on the static input of the system. More specifically, once the critical temperature is determined, the long time limit of density correlation functions $\mathbf{F}^{\rm c}(q)$ and the matrix form of the critical amplitude $\mathbf{H}(q)$ can be obtained by solving (\ref{eq:NEP}), (\ref{eq:defstabmat}), and (\ref{eq:convention}). The nonergodicity parameters and the critical amplitude which will be used to compare to the simulation results are related to the components of  $\mathbf{F}^{\rm c}(q)$ and $\mathbf{H}(q)$ via
\begin{equation}
  \eqalign{f^{\rm c}(q)
  = \frac{F^{\rm c}_{\iAA}(q)+F^{\rm c}_{\iBB}(q)+2F^{\rm c}_{\iAB}(q)}{S(q)} ,\\ 
  h(q)=\frac{H_{\iAA}(q)+H_{\iBB}(q)+2H_{\iAB}(q)}{S(q)} ,
  }
  \label{eq:fcqhq2components}
\end{equation}
with $S(q)$ being given by \eref{eq:defSnn}. The exponent parameter $\lambda$ and the exponents $a$, $b$ can be obtained via \eref{eq:lambdafromMCT} and \eref{lambda_parameter}. Since all these quantities are determined at $\Tc$, it is vital to accurately predict the critical temperature.
To determine $\Tc$, we used linear interpolations for the partial static structure factors between $T=0.97$ and $0.98$. The results are summarized in \tref{tabMCT}. From the bottom to the top the precision of $\Tc$ increases. The best estimate for $\Tc$ is $\Tc=0.979\,245$, leading to $\lambda=0.7142$. 

\Tref{tabMCT} illustrates the high sensitivity of the MCT parameters on the precise location of the critical point. This sensitivity is documented in the literature. The original work on the monodisperse hard-sphere system reported a critical packing fraction of $\varphi_{\rm c} = 0.52$ and $\lambda = 0.758$ \cite{GoetzeSjoegren1991_colloids}. Later, the estimate of the critical point was refined to $\varphi_{\rm c} = 0.515\,912\,13(1)$, leading to $\lambda = 0.723$ \cite{FranoschFuchsGoetze1997}. Reference~\cite{FranoschFuchsGoetze1997} points out that this high accuracy of $\varphi_{\rm c}$ is necessary to reproduce the slow dynamics over many orders of magnitude within MCT. A similar sensitivity of $\lambda$ on $\Tc$ is also reported for the Kob--Andersen binary mixture. Using static input from simulations the first predictions were $\Tc = 0.922$ and $\lambda = 0.708$ \cite{NaurothKob1997}, whereas later work suggested a more precise estimate of $\Tc = 0.951\,5$ and along with that a different value of ($\gamma = 2.46$ corresponding to) $\lambda = 0.735$ \cite{FlennerSzamel:PRE2005_2}. 

When comparing the results of \tref{tablefitparam} and \tref{tabMCT} two differences can be noted. First, $\Tc^{\rm MD} =0.798 < \Tc^{\rm MCT} \approx 0.979$. Qualitatively, this difference is in line with previous findings. Indeed, for many systems, including hard spheres and binary mixtures \cite{DasEtal:PRB2008,NaurothKob1997,FlennerSzamel:PRE2005_2,GoetzeSjoegren1991_colloids}
(but not simple polymer models \cite{Ciarella2019,ChongEtal:Polymer1,FreyEtal:dscf2013}), the factorization of the memory kernel \eref{eq:defMCTMF} tends to overestimate the glassiness. Here we find $\Tc^{\rm MCT} \approx 1.2 \Tc^{\rm MD}$. This overestimation by a factor of 1.2 is smaller than for the Kob--Andersen mixture \cite{NaurothKob1997,FlennerSzamel:PRE2005_2}, where a factor of about 2 is reported, and also for a metallic alloy where a factor of about 1.5 is found \cite{DasEtal:PRB2008}. 


\begin{figure}
\begin{center}
	\includegraphics*[scale=0.38]{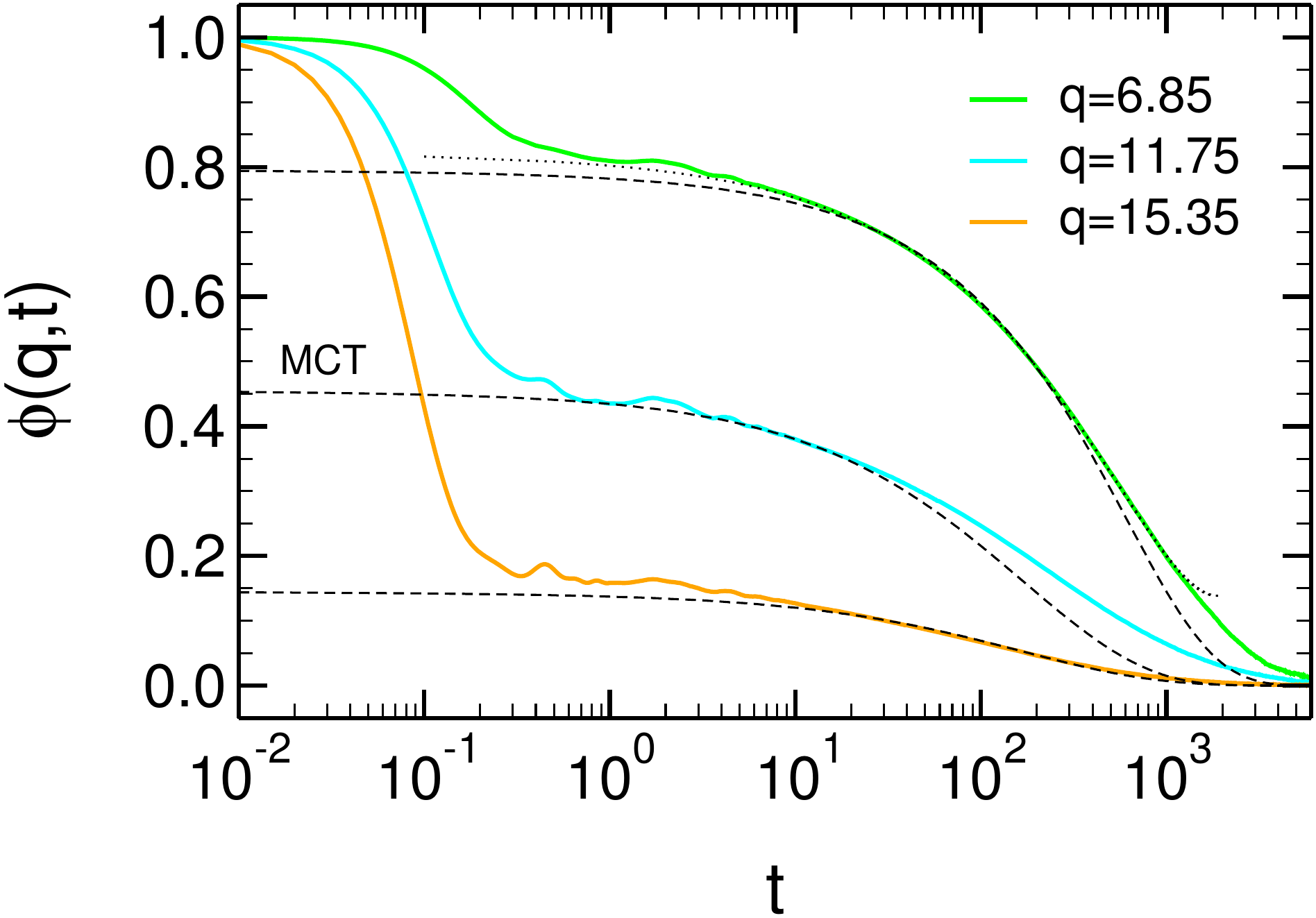}
\caption{Time dependence of $\phi(q,t)$ for various $q$. The full lines depict the MD data for $q=6.85, 11.75, 15.35$ at $T=0.84$, i.e.\ at $\varepsilon = (\Tc - T)/\Tc = - 5.26 \times 10^{-2}$. For $q=6.85$ the dotted line reproduces the fit result to \eref{numerical_fit} from \fref{fig_fitvS}. The dashed lines present the results of the MCT calculations based on the static input for $q=6.89, 11.69, 15.29$ at $T=0.979258$, i.e.\ at $\varepsilon = -1.27 \times 10^{-5}$. The MCT results are shifted along the $t$ axis so as to optimize the overlap with the MD data for $5 \lesssim t \lesssim 50$. The shift factors are $1.5 \times 10^9$ for $q=6.89$, $1.4 \times 10^9$ for $q=11.69$ and $5 \times 10^8$ for $q=15.29$.}
	\label{comparison_phiq_MD_theo}
\end{center}
\end{figure}

A second difference concerns the value of $\lambda$ and the associated von Schweidler exponent $b$. We see that $b^{\rm MCT}$ decreases with increasing precision of $\Tc$, but always stays larger than $b^{\rm MD}$ obtained from the fits. A smaller value of $b$ implies more stretching of the $\alpha$ relaxation. This is illustrated in \fref{comparison_phiq_MD_theo} which compares the MD results for $\phi(q,t)$ at $T=0.84$ and different $q$ (full lines) with the MCT calculations (dashed lines). The MCT calculations correspond to a temperature very close to $\Tc^{\rm MCT}$ and are therefore good proxies for the $\alpha$ master curve at the wave vectors considered. The MCT curves are shifted along the time axis so as to optimize the overlap with the MD data for $t \sim 10$, that is in the time window shown in the inset of \fref{fig_factorization} where a distinction between $b^{\rm MCT} = 0.6172$ and $b^{\rm MD} = 0.5652$ is not possible. This is highlighted again in \fref{comparison_phiq_MD_theo} where the fit result to \eref{numerical_fit} from \fref{fig_fitvS} is reproduced for $q=6.85$ (dotted line). \Fref{comparison_phiq_MD_theo} also shows that the MD data at long times lie above the MCT calculations and are thus more stretched (for $q =15.35$ this is not visible on the scale of the figure). The fit based on the MD data models this enhanced stretching by a smaller value of the von Schweidler exponent. 

Although $b^{\rm MCT}$ decreases with increasing precision of $\Tc^{\rm MCT}$, \tref{tabMCT} indicates that the decrease is fairly weak. It is thus unlikely that further improvement of $\Tc$ might make $b^{\rm MCT}$ converge to $b^{\rm MD}$. Does this mean that MCT cannot account for the enhanced stretching of the $\alpha$ relaxation? Not necessarily, albeit (presumably) not within the idealized MCT in the present case. Extensions of the theory need to round off the ideal glass transition and account for activated processes. Recent efforts in this direction involve the inclusion of activated events at the single particle level \cite{Chong:PRE2008,ChongEtal:JPCM2009,MirigianSchweizerI:JCP2014,MirigianSchweizerII:JCP2014},  the implementation of spatially heterogeneous relaxation by considering the distance to $\Tc$ as a spatially fluctuating variable \cite{RizzoVoigtmann:EPL2015}, or a hierarchical framework systematically shifting the factorization approximation to high-order dynamic multi-point correlations \cite{Szamel:PRL2003,JanssenReichman:PRL2015,JanssenEtal:JSM2016,LuoJanssen:GMCTPY2019}. The latter approach, referred to as generalized mode-coupling theory (GMCT), has recently been examined numerically for Percus-Yevick (PY) hard spheres by performing explicitly wavenumber- and time-dependent calculations up to sixth order \cite{LuoJanssen:GMCTPY2019}. The results indicate that the inclusion of more levels in the GMCT hierarchy leads to a systematic increase of $\gamma$ (and of the predicted critical packing fraction $\varphi_{\rm c}$). Due to \eref{tps_alpha} an increase of $\gamma$ implies a decrease of $b$ and so more stretching (cf table~1 in \cite{LuoJanssen:GMCTPY2019}).
Qualitatively, we can thus expect that the inclusion of more levels in the GMCT hierarchy would probably lead to a better approximation of the activated dynamics.
\subsection{Coherent and incoherent dynamics: Nonergodicity parameters, critical and long-time correction amplitudes}
\label{subsec:cohincdyn}
\begin{figure}
\begin{center}
  \includegraphics*[scale=0.38]{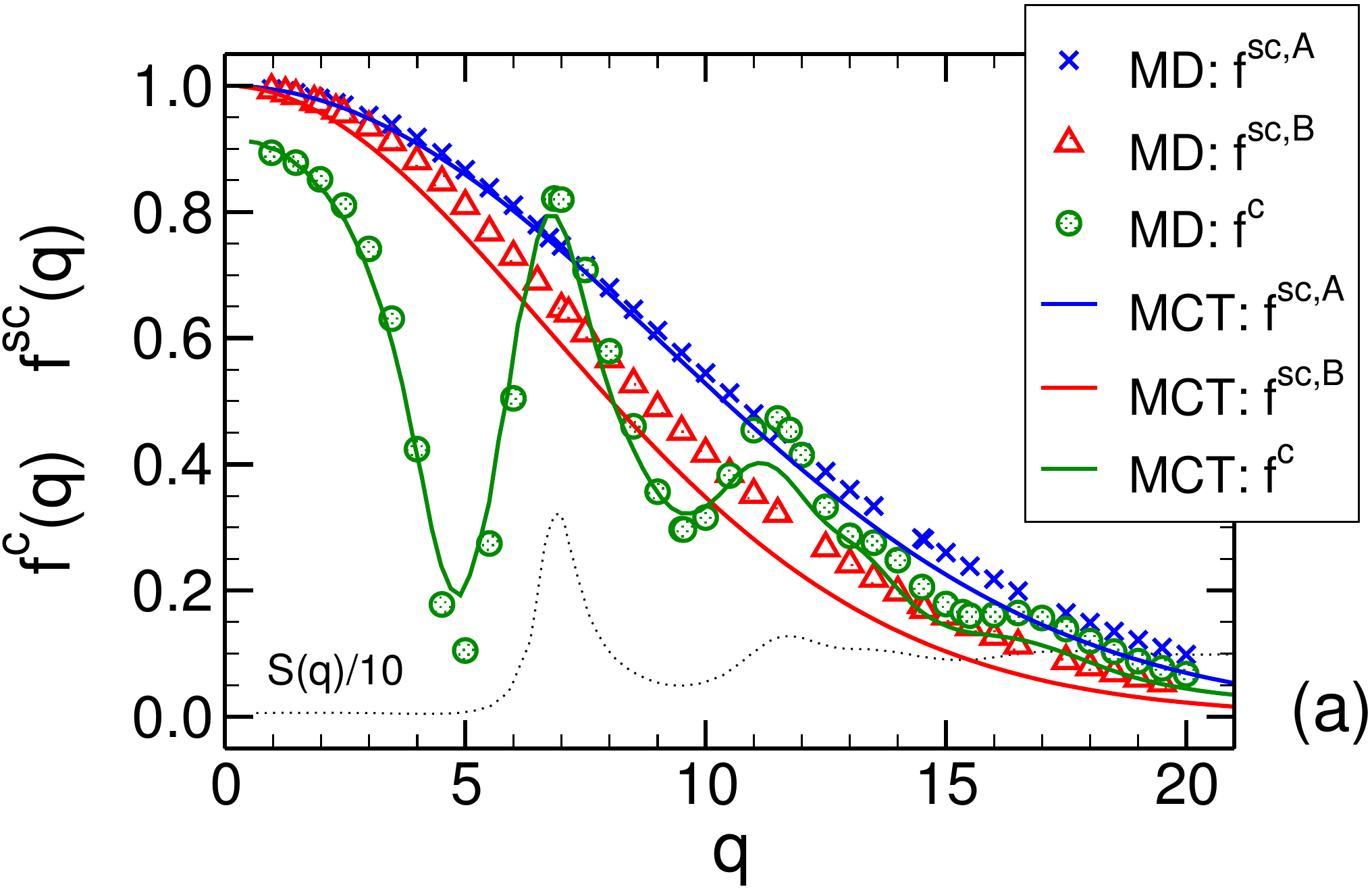}\\[2mm]
  \includegraphics*[scale=0.38]{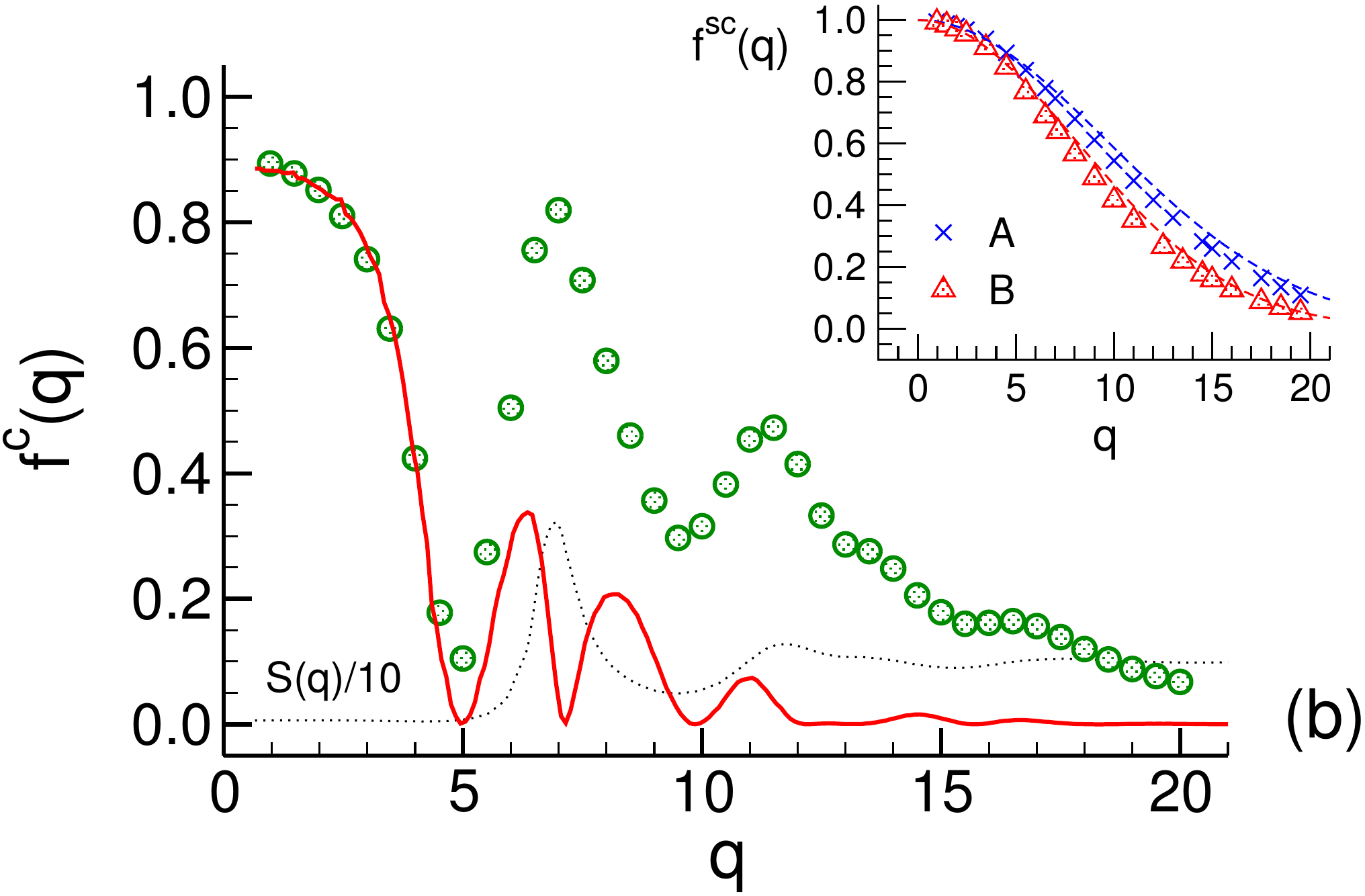}
\end{center}
\caption{Panel (a): $q$ dependence of $f^{{\rm sc},\lA}(q)$, $f^{{\rm sc},\lB}(q)$ and $f^{\rm c}(q)$. The symbols (labeled ``MD'') depict the results of the fit to \eref{numerical_fit}. The full lines (labeled ``MCT'') correspond to the results of the MCT calculations based on static input. 
Panel(b): The symbols in the main figure and in the inset reproduce the fit results from panel (a). The full line in the main figure shows $S_{\rm nc}^2(q)/[S_{\rm cc}(q) S(q)]$ related to composition fluctuations [cf \eref{eq:fccompfluc}]. $S_{\rm nc}(q)$, $S_{\rm cc}(q)$ and $S(q)$ are obtained from the partial structure factors at $T=0.85$, cf \sref{sec:statics}. The dashed lines in the inset present the Gaussian approximation \eref{gaussian_nonergo} with $r_{{\rm sc},\lA} = 0.0731$ and $r_{{\rm sc},\lB} = 0.0873$. The values of the Lindemann localization lengths were obtained by extrapolating the fit results for $f^{{\rm sc},\lA}(q)$ and $f^{{\rm sc},\lB}(q)$ to $q \rightarrow 0$.
In both panels the dotted line shows $S(q)$ divided by 10 for comparison.}
\label{nonergo_various_q}
\end{figure}

\begin{figure}
\begin{center}
  \includegraphics*[scale=0.38]{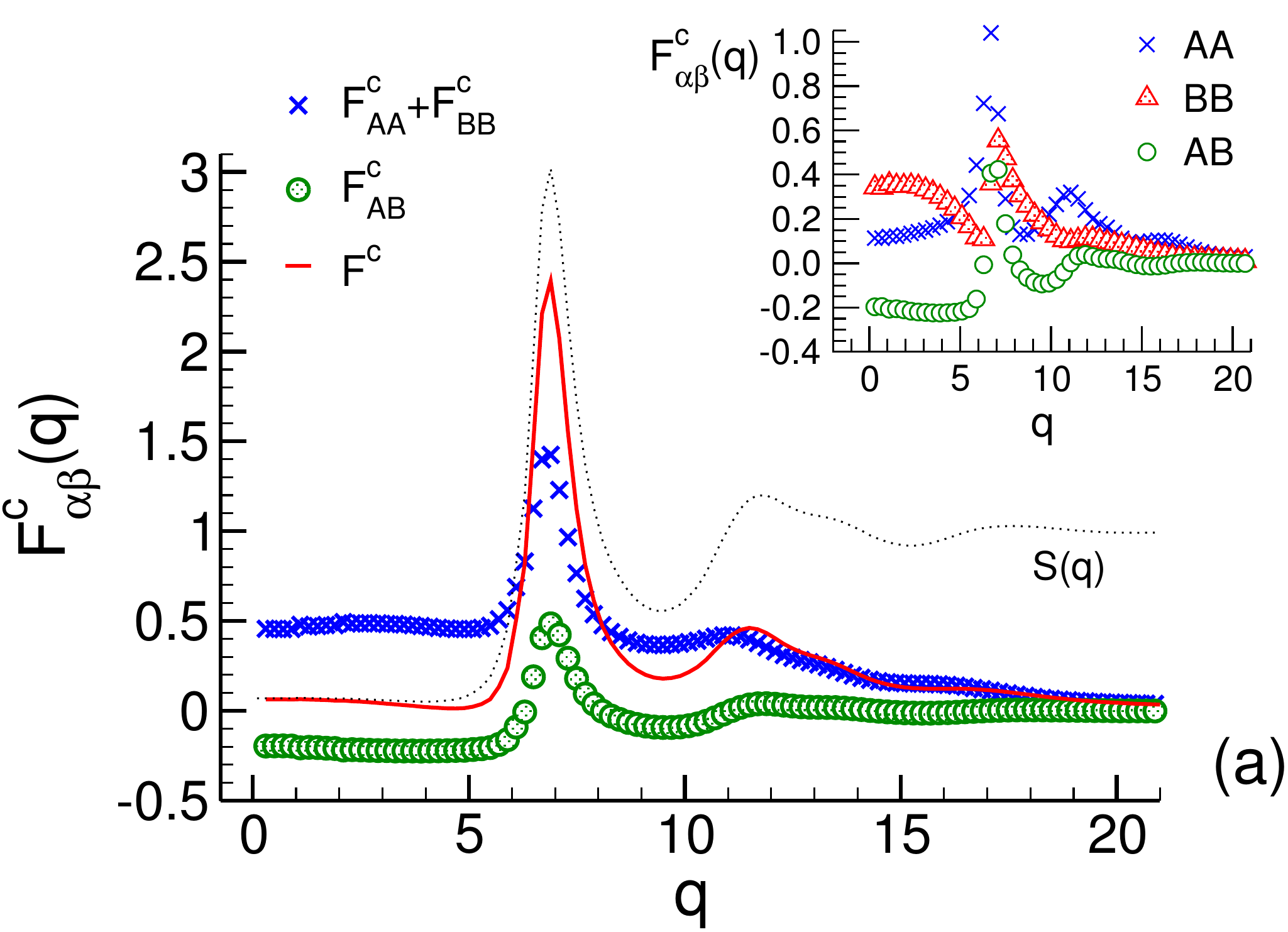}\\[2mm]
  \includegraphics*[scale=0.38]{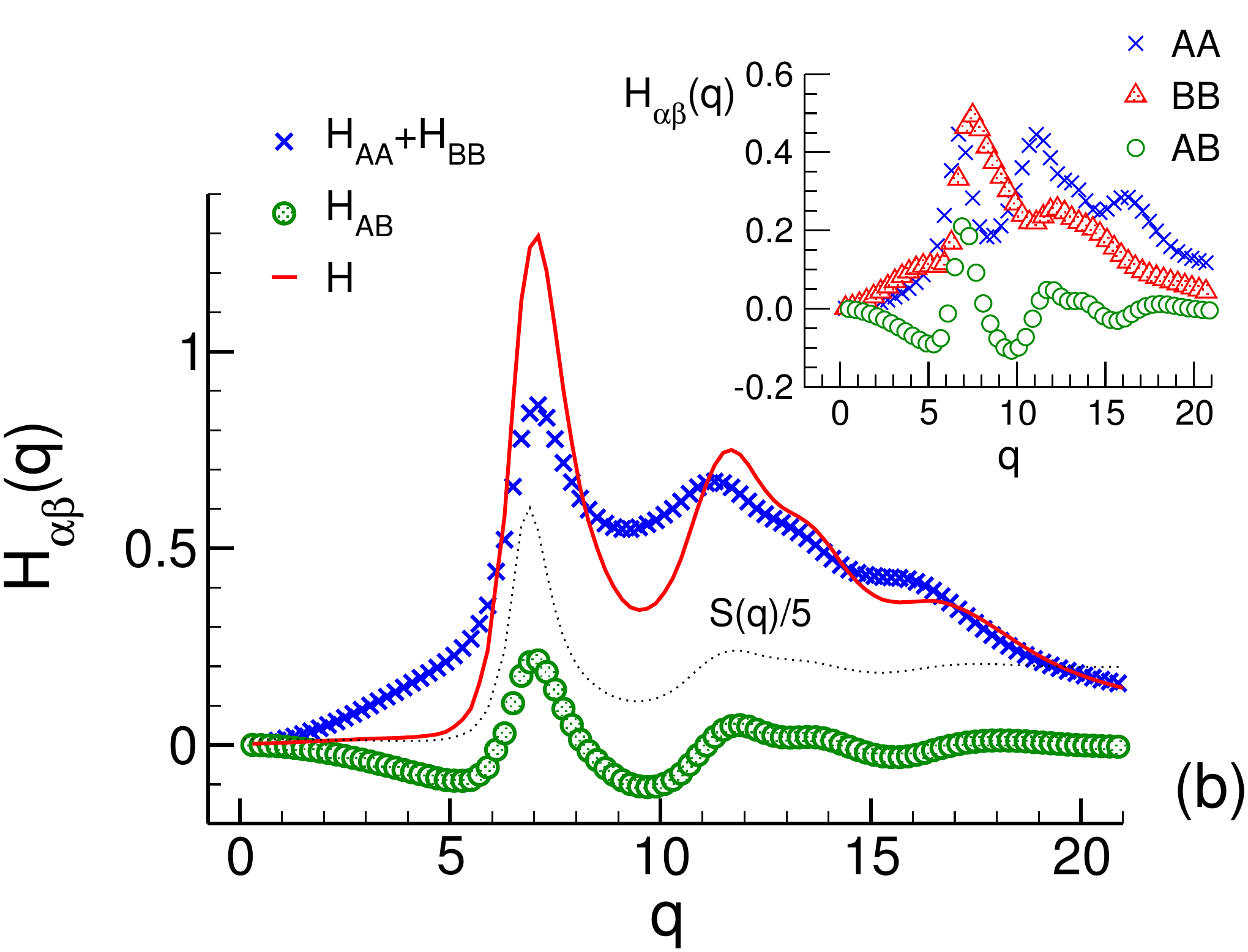}
\end{center}
\caption{Results from the fully microscopic MCT calculations for the nonergodicity parameters $F^{\rm c}_{\alpha\beta}(q)$ (panel (a)) and the critical amplitudes $H_{\alpha\beta}(q)$ (panel (b)) at $\Tc = 0.979\,245$ (cf \tref{tablefitparam}). Both the nonergodicity parameters and critical amplitudes are not normalized by $S(q)$. The insets in panel (a) and panel (b) show the partial components of like ($\iAA$, $\iBB$) and unlike ($\iAB$) particles. The main figure in panel (a) presents $F^{\rm c}_{\iAA}(q) + F^{\rm c}_{\iBB}(q)$ (crosses), $F^{\rm c}_{\iAB}(q)$ (circles), and $F^{\rm c}(q) = F^{\rm c}_{\iAA}(q) + F^{\rm c}_{\iBB}(q) + 2 F^{\rm c}_{\iAB}(q)$ (full line). The dotted line indicates $S(q)$ at $\Tc$. The main figure in panel (b) depicts $H_{\iAA}(q) + H_{\iBB}(q)$ (crosses), $H_{\iAB}(q)$ (circles), and $H(q) = H_{\iAA}(q) + H_{\iBB}(q) + 2 H_{\iAB}(q)$ (full line). The dotted line indicates $S(q)/5$ at $\Tc$ for comparison.}
\label{fig:unnorm_nonergo_critampl}
\end{figure}

\begin{figure}
\begin{center}
  \includegraphics*[scale=0.38]{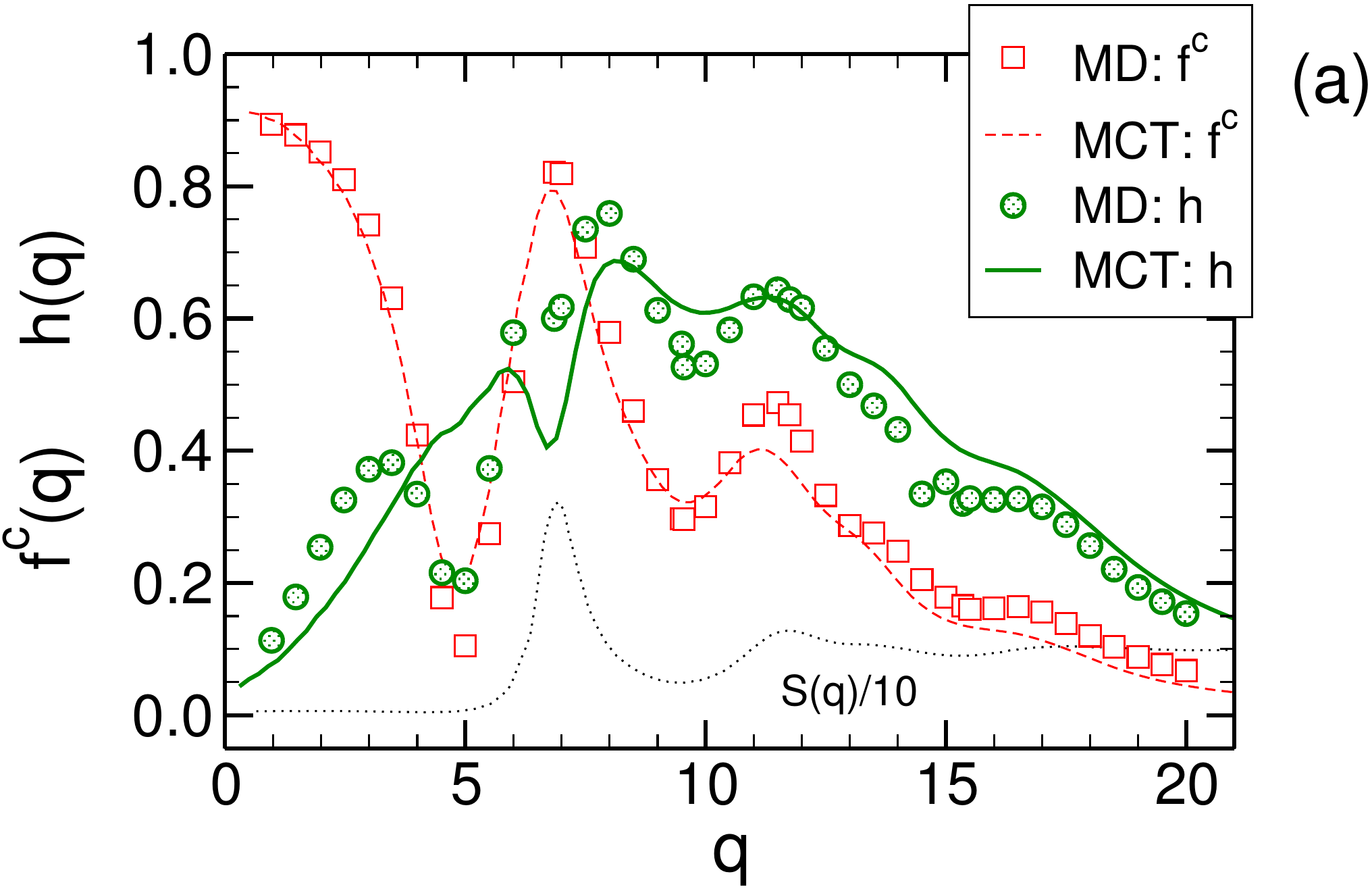}
  \includegraphics*[scale=0.38]{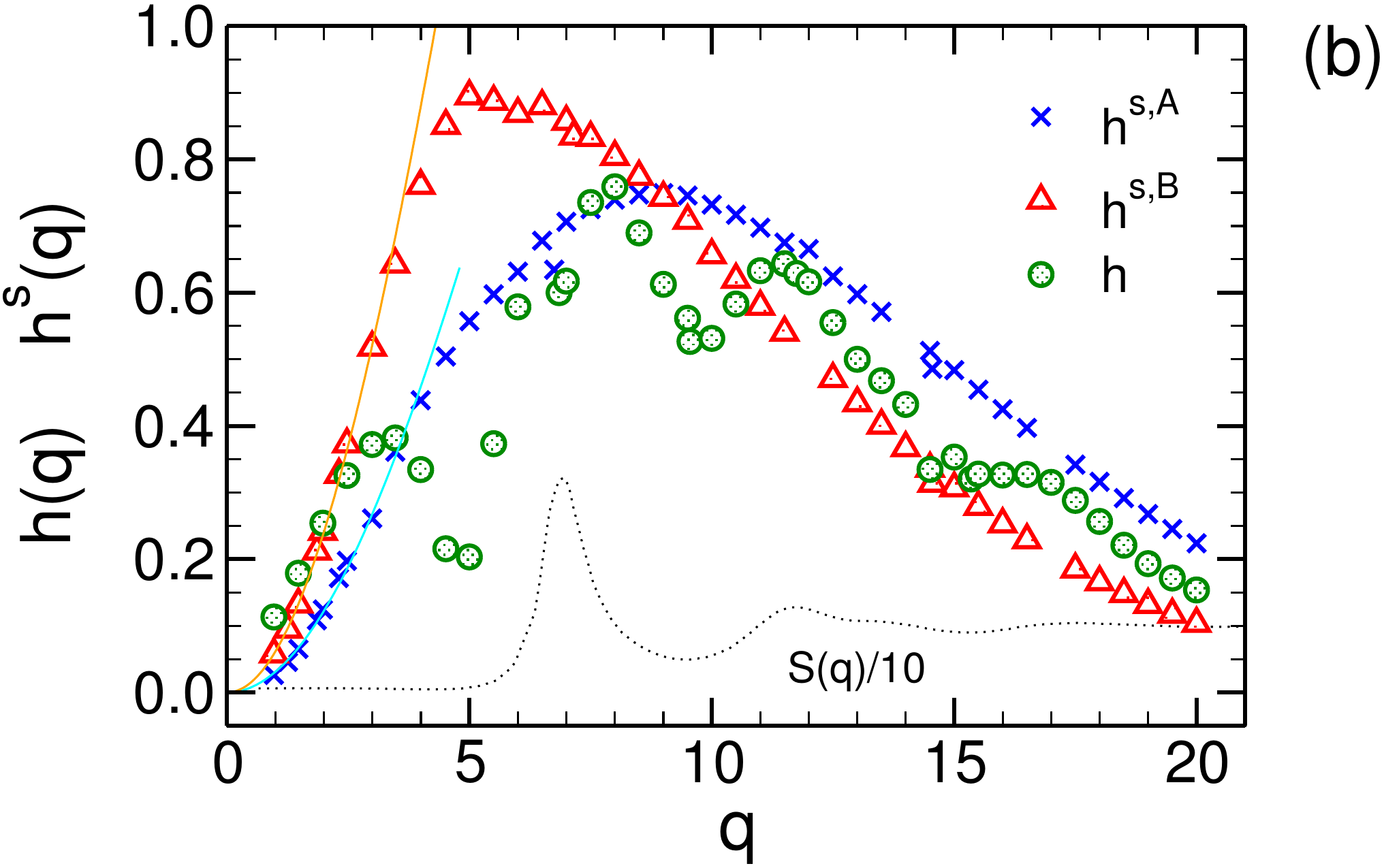}
\end{center}
\caption{
Panel (a): Critical amplitude $h(q)$ versus $q$. The circles (labeled ``MD'') depict the results of the fit to \eref{numerical_fit}. The full line (labeled ``MCT'') corresponds to the result for $h(q)$ from the MCT calculations based on static input. For comparison $f^{\rm c}(q)$ from \fref{nonergo_various_q} is reproduced (squares: fit results to \eref{numerical_fit}, dashed line: MCT calculations). The dotted line shows $S(q)$ divided by 10.
Panel (b):  $q$ dependence of $h^{{\rm s},\lA}(q)$, $h^{{\rm s},\lB}(q)$ and $h(q)$. The symbols depict the results of the fit to \eref{numerical_fit}.  The full lines present the Gaussian approximation \eref{gaussian_h} with $h^{\lA}_{\rm msd} = 0.0313$ and $h^{\lB}_{\rm msd} = 0.0622$. These values were obtained by fitting \eref{gaussian_h} to $h^{{\rm s},\alpha}(q)$ to $q \lesssim 3$ while keeping $r^2_{{\rm sc},\alpha}$ fixed at the values given in \fref{nonergo_various_q}. The dotted line shows $S(q)$ divided by 10 for comparison.}
\label{h_various_q}
\end{figure}

\begin{figure}
\begin{center}
  \includegraphics*[scale=0.38]{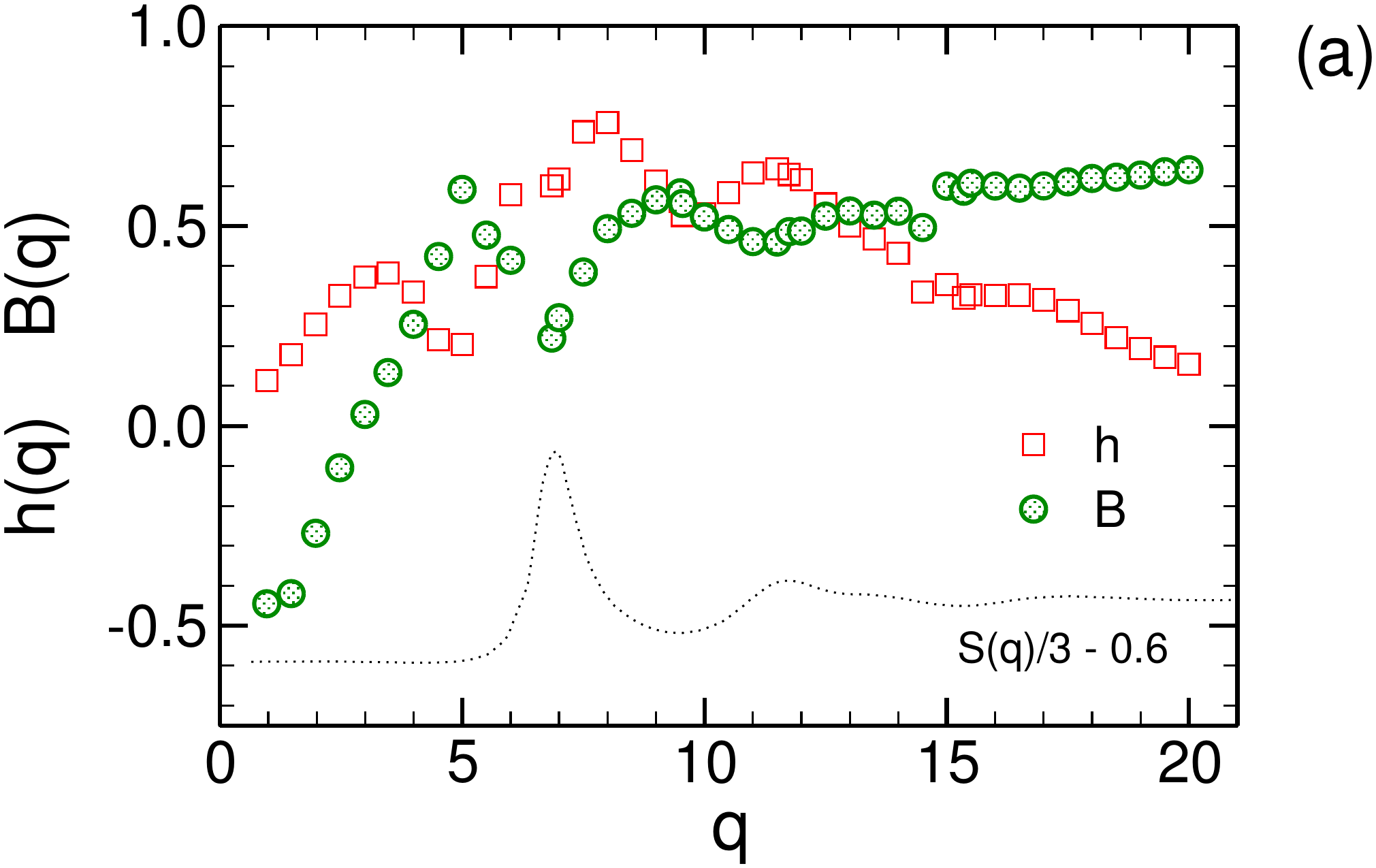}
  \includegraphics*[scale=0.38]{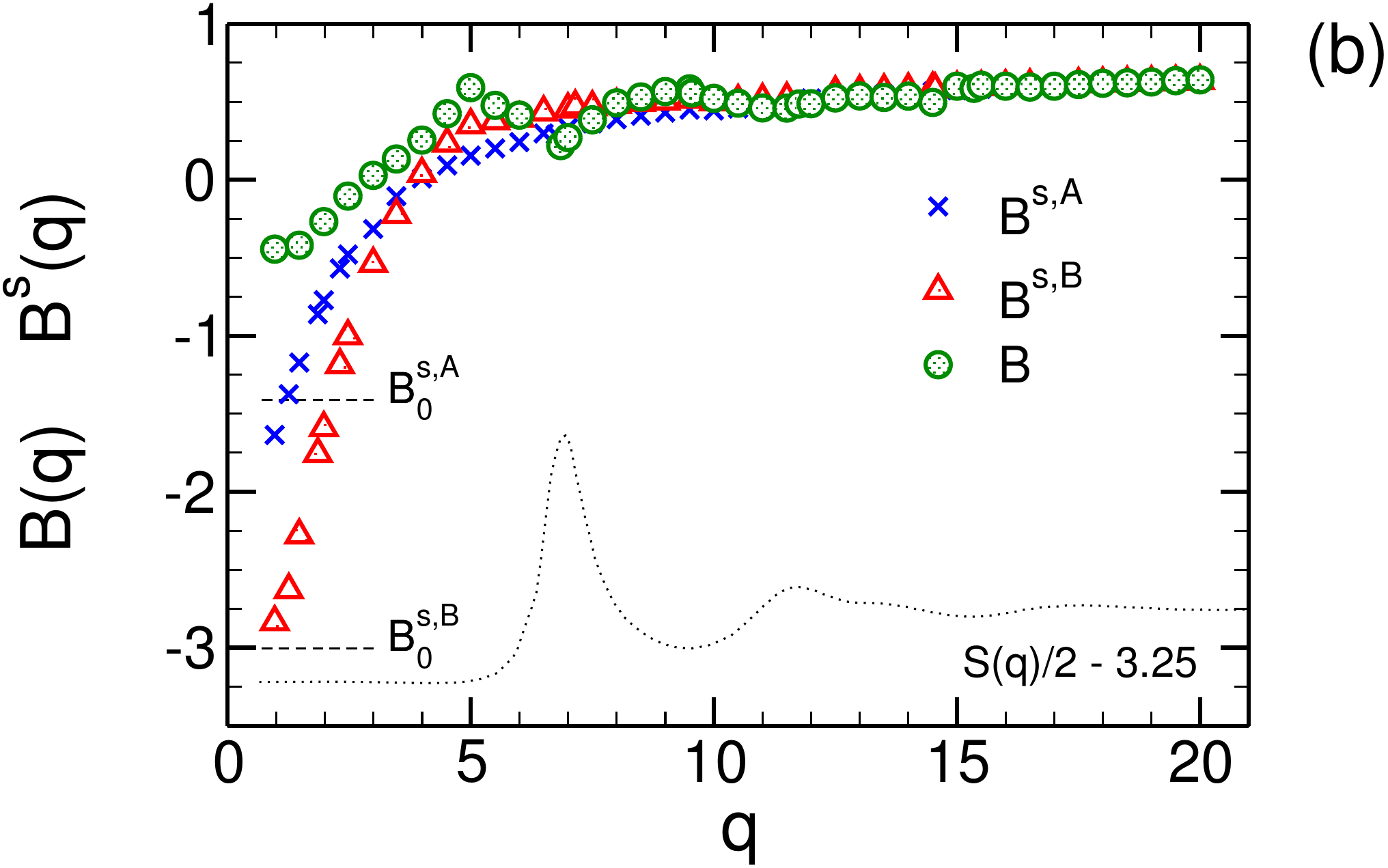}
\end{center}
\caption{Panel (a): Long-time correction amplitude $B(q)$ versus $q$. The circles depict the results of the fit to \eref{numerical_fit}. The squares reproduce the fit results for the critical amplitude $h(q)$ from \fref{h_various_q}. The dotted line shows $S(q)/3 - 0.6$ for comparison.
Panel (b): $q$ dependence of $B^{{\rm s},\lA}(q)$, $B^{{\rm s},\lB}(q)$ and $B(q)$. The horizontal dotted lines indicate the limit $B^{{\rm s},\alpha}_0 = \lim_{q\rightarrow 0}B^{{\rm s},\alpha}(q)$ obtained from fits of \eref{msd_fit_equation} to the MSD of species $\alpha$, $B_0^{{\rm s},\lA} = -1.41093$ and $B_0^{{\rm s},\lB} = -3.00259$ (cf \fref{fig_msd}). The dotted line shows $S(q)/5 - 3.25$ for comparison.}
\label{hB_various_q}
\end{figure}

From the fits to \eref{numerical_fit} we obtain the $q$ dependence of $f^{\rm c}(q)$, $h(q)$, $B(q)$ and of their incoherent counterparts. The nonergodicity parameters, $f^{\rm c}(q)$ and $f^{{\rm sc},\alpha}(q)$, and the critical amplitude, $h(q)$, were also calculated by binary MCT based on the simulated static input. \Fref{nonergo_various_q} to \fref{hB_various_q} show the results.

As seen in \fref{nonergo_various_q}(a), the nonergodicity parameters from the fits (symbols) and the MCT calculations (lines) are in semiquantitative agreement. For $q \gtrsim q^*$ the MCT calculations tend to lie below the fit results. This trend is evident for the incoherent scattering and also visible for $q \gtrsim 10$ in the coherent scattering. A similar underestimation was observed for polydisperse hard spheres and rationalized as follows \cite{WeysserEtal:PRE2010}: MCT predicts structural arrest at $\Tc^{\rm MCT} > \Tc^{\rm MD}$. As the glass stiffens with decreasing $T$, one can expect $f^{\rm c}(q)$ from the fits to be larger than from the MCT calculations. This argument is corroborated by the GMCT analysis of the PY hard sphere system, which finds $f^{\rm c}(q)$ to increase with increasing $\varphi_{\rm c}$ \cite{LuoJanssen:GMCTPY2019}.

For $q < q^*$ the agreement between the fit results and MCT calculations improves with decreasing wave vector. For small $q$, $f^{\rm c}(q)$ strongly increases and tends to a value of about 0.9 in the $q \rightarrow 0$ limit. This behavior is unusual compared to the one-component PY hard-sphere system for which one rather finds a weak $q$ dependence for small $q$ and $f^{\rm c}(q \rightarrow 0) \approx 0.4$ \cite{WeysserEtal:PRE2010,FranoschFuchsGoetze1997,LuoJanssen:GMCTPY2019}. In \cite{WeysserEtal:PRE2010} it has been argued that this difference between the one-component and polydisperse system is a consequence of composition fluctuations, in reference to an analysis of the hydrodynamic limit of the MCT equations \eref{eq:eomMCT} to \eref{eq:dcf} for binary mixtures \cite{FuchsLatz:PhysicaA1993}. For the binary Voronoi mixture we can test these predictions. For $q \rightarrow 0$ one expects (cf (10b) in \cite{FuchsLatz:PhysicaA1993})
\begin{equation}
  f^{\rm c}(q \rightarrow 0)
  = \lim_{q \rightarrow 0} \left [\frac{S_{\rm nc}^2(q)}{S_{\rm cc}(q)S(q)}
  \right ] .
  \label{eq:fccompfluc}
\end{equation}
The ratio $S_{\rm nc}^2(q\rightarrow 0)/S_{\rm cc}(q\rightarrow 0)$ corresponds to the second term, $\delta^2 S_{\rm cc} (q \rightarrow 0)$, of \eref{eq:Snnq0} that represents the contribution to the static structure factor due to composition fluctuations. Using the Bhatia--Thornton structure factors at $T=0.85$ (cf \fref{sdek}(c)) we can estimate the term in the square brackets of \eref{eq:fccompfluc}. The full line in \fref{nonergo_various_q}(b) presents the result. We find good agreement with $f^{\rm c}(q)$ from the fits for $q < 5$, thereby confirming \eref{eq:fccompfluc}. For $q \gtrsim q^*$, on the other hand, $f^{\rm c}(q)$ is in phase with $S(q)$ (cf dotted line in \fref{nonergo_various_q}(b)) and the contribution due to composition fluctuations decreases in amplitude with increasing $q$. This suggests that composition fluctuations do not play a prominent role for $q \gtrsim q^*$, a conclusion that resonates with the findings of \cite{WeysserEtal:PRE2010} and the MCT predictions in \cite{FuchsLatz:PhysicaA1993}. \Fref{nonergo_various_q}(b) thus indicates that a crossover between a composition-fluctuation dominated small-$q$ regime and a packing dominated large-$q$ regime occurs at $q \approx 5$, leading to a minimum in $f^{\rm c}(q)$ at $q \approx 5$ for our Voronoi mixture.

The minimum at $q \approx 5$ can also be understood from binary MCT in terms of the partial components $F^{\rm c}_{\alpha\beta}(q)$ determining $f^{\rm c}(q)$ via \eref{eq:fcqhq2components}. From \fref{fig:unnorm_nonergo_critampl}(a) we see that the components of like particles, $F^{\rm c}_{\iAA}(q)$ and $F^{\rm c}_{\iBB}(q)$, are always positive (cf inset), and so is their sum (crosses). By contrast, the component of unlike particles, $F^{\rm c}_{\iAB}(q)$, becomes negative for $q < q^*$, leading to a shallow minimum at $q \approx 5$ when the numerator $F^{\rm c}(q) = F^{\rm c}_{\iAA}(q) + F^{\rm c}_{\iBB}(q) + 2 F^{\rm c}_{\iAB}(q)$ of \eref{eq:fcqhq2components} is calculated (full line). The depth of the minimum is amplified after division by $S(q)$ (dotted line). As seen in \fref{fig:unnorm_nonergo_critampl}(a), $S(q)$ is similar in magnitude to $F^{\rm c}(q)$ for $q \rightarrow 0$ and $q \approx q^*$, while $S(q) > F^{\rm c}(q)$ for $q \approx 5$. This gives rise to values near 1 for the normalized nonergodicity parameter $f^{\rm c}(q)$ for $q \rightarrow 0$ and $q \approx q^*$, and explains the pronounced minimum at $q \approx 5$.

For the incoherent scattering MCT predicts that $f^{{\rm sc},\alpha}(q) = 1 - (qr_{{\rm sc},\alpha})^2$ for $q \rightarrow 0$ \cite{FuchsGoetzeMayr1998}, where $r_{{\rm sc},\alpha}$ is the ``Lindemann localization length'' of species $\alpha$. Fitting this relation for $q \lesssim 3$ to the data in \fref{nonergo_various_q}(a) gives $r_{{\rm sc},\lA} = 0.0731 \approx 0.054 \times (2 R_\lA)$ and $r_{{\rm sc},\lB} = 0.0873 \approx 0.078 \times (2 R_\lB)$ where $R_\lA$ and $R_\lB$ are the natural radii of the Voronoi mixture (cf \sref{sec:voronoiliquid}). If we take $R_\lA$ and $R_\lB$ as approximations for the particle radii, we see that the localization lengths are on the order of $10\%$ of the particle diameters, as suggested by MCT \cite{GoetzeVoigtmann:PRE2003,FuchsGoetzeMayr1998,FuchsLatz:PhysicaA1993}. Moreover, MCT predicts that the Gaussian approximation,
\begin{eqnarray}
  f^{{\rm sc},\alpha}(q) = \exp(-q^2 r^2_{{\rm sc},\alpha}) 
  \quad \mbox{($\alpha,\beta = \lA, \lB$)} ,
  \label{gaussian_nonergo}
\end{eqnarray}
gives a reasonable description of the $q$ dependence of the nonergodicity parameter. The inset in \fref{nonergo_various_q}(b) confirms this expectation.

\Fref{h_various_q}(a) displays the critical amplitude $h(q)$ for the coherent scattering. The circles correspond to the fit results, the full line to the MCT calculations. Recall from \sref{subsec:fit} that the fits involve a constant, but arbitrary, scale factor $\ell$: To fix this factor we adjust $\ell$ so that the fitted $h(q)$ closely matches the $h(q)$ from the MCT calculations (here we took $\ell = 0.4$). Then, the found $q$ dependence can be better compared. For $q > q^*$ \fref{h_various_q}(a) shows that fits and MCT agree well with each other, albeit the agreement is a bit worse than for $f^{\rm c}(q)$ (cf squares and dashed line). The MCT calculations indicate that $h(q)$ oscillates in phase with $f^{\rm c}(q)$ for $q > q^*$, whereas it is in antiphase with $f^{\rm c}(q)$ for $q \leq q^*$. For $q > q^*$ the fitted $h(q)$ has the same $q$ dependence as the MCT calculations. On the other hand, for $2 \lesssim q \lesssim q^*$---that is, in the regime where composition fluctuations become important---qualitative differences occur. The fit results exhibit an oscillation, while the calculations rather predict a weak shoulder at $q \approx 4$ followed by monotonic decrease with $q$, in qualitative agreement with other MCT studies \cite{FuchsLatz:PhysicaA1993}. The presence of the shoulder can be understood from the interplay of the partial components $H_{\alpha\beta}(q)$ determining $h(q)$ via \eref{eq:fcqhq2components}. \Fref{fig:unnorm_nonergo_critampl}(b) shows that, similar to the nonergodicity parameters, the components of like particles, $H_{\iAA}(q)$ and $H_{\iBB}(q)$, are always positive, whereas the component of unlike particles, $H_{\iAB}(q)$, becomes negative for $q < q^*$ (cf inset). However, contrary to the nonergodicity parameters, the sum $H_{\iAA}(q)+ H_{\iBB}(q)$ (crosses) increases steeply in the range $q \sim 4$. This increase cannot be outweighed by $H_{\iAB}(q)$ so that the numerator $H(q) = H_{\iAA}(q) + H_{\iBB}(q) + 2 H_{\iAB}(q)$ of \eref{eq:fcqhq2components} plateaus for $q \approx 4$ (full line). This gives rise to a shoulder after division by $S(q)$.

To verify our fit results for $2 \lesssim q \lesssim q^*$ we attempted to impose in \eref{numerical_fit} the value of $h(q)$ from the MCT calculations, whereas $f^{\rm c}(q)$ and $b$ were fixed at the values found before from the fits. A fit of comparable quality is only obtained, if we allow $t'_\sigma$ to depend on $q$, which is not acceptable within MCT. Therefore, it seems we cannot achieve better agreement between fits and MCT calculations for $2 \lesssim q \lesssim q^*$. For smaller $q$, however, the fit and MCT results appear to converge again toward one another. Both approaches suggest that $h(q) \lesssim 0.1$ for $q < 1$. This value is much smaller than in the one-component PY hard-sphere system \cite{FranoschFuchsGoetze1997} and may be attributed to a composition-fluctuation effect \cite{WeysserEtal:PRE2010,FuchsLatz:PhysicaA1993}.   

For the critical amplitudes $h^{{\rm s},\alpha}(q)$ of the species $\alpha = \lA, \lB$ no MCT calculations are currently available for the Voronoi mixture. The corresponding fit results are shown in \fref{h_various_q}(b). For $q \gtrsim 10$ we find that $h^{{\rm s},\lA}(q)$ and $h^{{\rm s},\lB}(q)$ bracket $h(q)$, as it is also observed for the nonergodicity parameters in \fref{nonergo_various_q}(a). MCT calculations for the one-component PY hard-sphere system suggest that $h^{\rm s}(q)$ vanishes in the limits $q \rightarrow 0$ and $q \rightarrow \infty$ and has a maximum near the second peak of $S(q)$ \cite{FuchsGoetzeMayr1998}. Similar behavior is found here for the $\lA$ and $\lB$ particles. In particular, \fref{h_various_q}(b) shows that $h^{{\rm s},\alpha}(q) \rightarrow 0$ for $q \rightarrow 0$. In this limit, the critical amplitude is supposed to be well described by the Gaussian approximation \cite{FuchsGoetzeMayr1998},
\begin{eqnarray}
  h^{{\rm s},\alpha}(q)=h_{{\rm msd},\alpha} q^2 
  \exp(-q^2 r^2_{{\rm sc},\alpha}) ,
  \label{gaussian_h}
\end{eqnarray}
where $h_{{\rm msd},\alpha}$ are constants. We fix $r^2_{{\rm sc},\alpha}$ to the values from \fref{nonergo_various_q}(b) and fit \eref{gaussian_h} for $q \lesssim 3$ to the data in \fref{hB_various_q}(a) to determine $h_{{\rm msd},\alpha}$. This gives $h_{{\rm msd},\lA} = 0.0313$ and $h_{{\rm msd},\lB} = 0.0622$. Using these results the dashed lines in \fref{hB_various_q}(a) show \eref{gaussian_h} for both particle species. \Eref{gaussian_h} provides a good description for $q < 4$. 

By contrast to the critical amplitudes, \fref{hB_various_q}(a) and \fref{hB_various_q}(b) show that the long-time correction coefficients, $B(q)$ and $B^{{\rm s},\alpha}(q)$, change sign. This is expected from the literature on MCT \cite{FranoschFuchsGoetze1997,FuchsGoetzeMayr1998,FuchsLatz:PhysicaA1993}. However, comparison of these literature results and the data in \fref{hB_various_q} also reveals some differences, for $q \lesssim q^*$. From MCT calculations for binary mixtures \cite{FuchsLatz:PhysicaA1993} one expects $B(q)$ to be in phase with $h(q)$ for $q \lesssim q^*$, to be negative at $q^*$ and to tend to a small positive value for $q \rightarrow 0$. \Fref{hB_various_q}(a) does not support this expectation. Moreover, for the tagged-particle dynamics the Gaussian approximation should become valid in the limit $q \rightarrow 0$, predicting that $B^{{\rm s},\alpha}(q)$ is larger than the constant $B_0^{{\rm s},\alpha} = \lim_{q \rightarrow 0} B_q^{{\rm s},\alpha}$ \cite{FuchsGoetzeMayr1998}. The constant $B_0^{{\rm s},\alpha}$ enters the long-time correction to the von Schweidler law for the mean-square displacement (MSD) of species $\alpha$, cf \eref{msd_fit_equation}. We determined $B_0^{{\rm s},\alpha}$ from the MSD and the results are shown as horizontal dashed lines in \fref{hB_various_q}(b). While $B_q^{{\rm s},\lB} > B_0^{{\rm s},\lB}$, this is not the case for the $\lA$ particles. As pointed out in \cite{FoffiEtal:PRE2004}, the determination of the correction amplitudes is impeded for data which cannot be chosen close enough to $\Tc$ \cite{FoffiEtal:PRE2004}. In part, the here described differences may be attributed to such uncertainties. 

\subsection{Mean-square displacements}
\label{subsec:msd}
\begin{figure}
  \begin{center}
  \includegraphics*[scale=0.38]{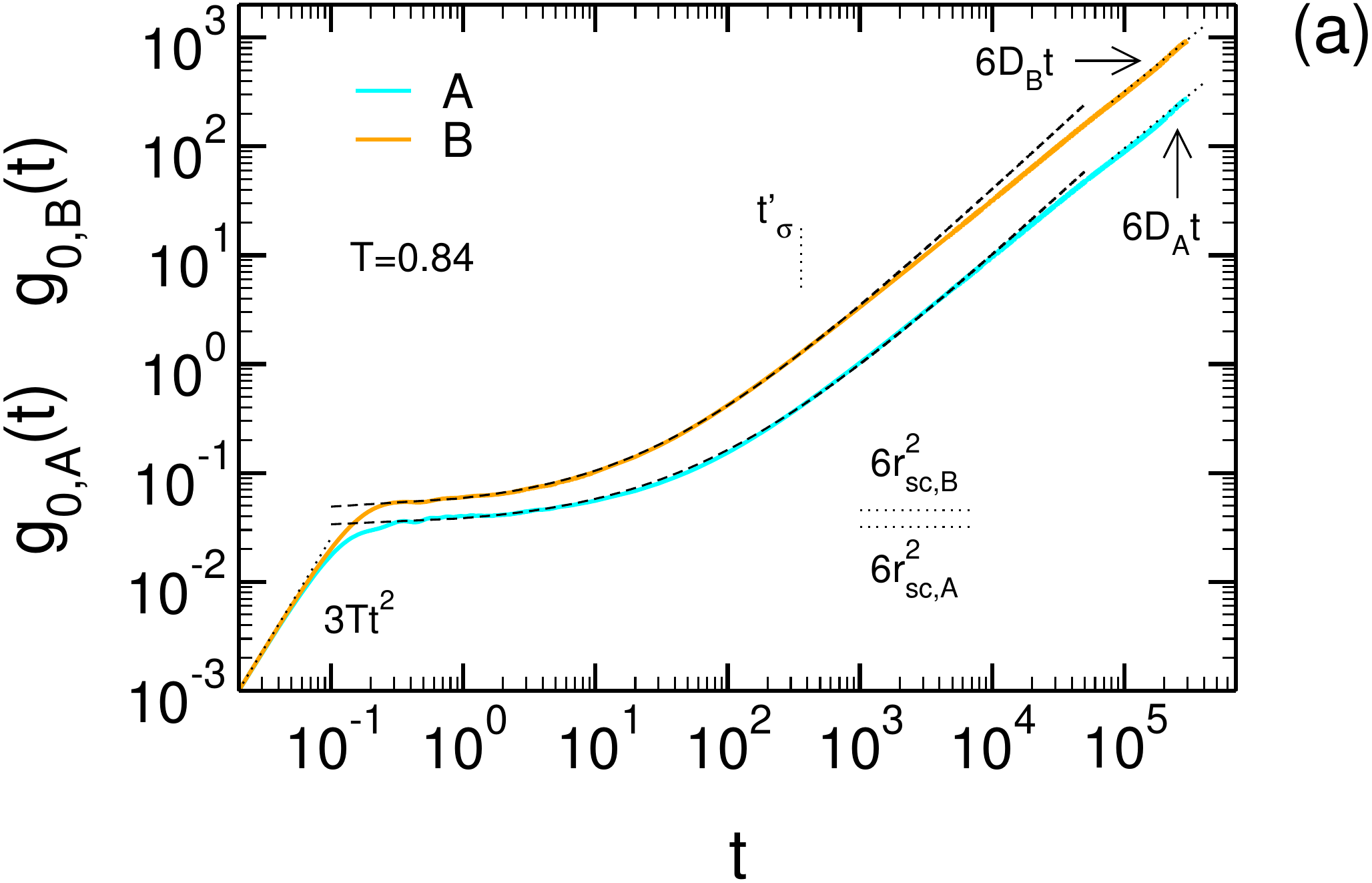}\\[2mm]
  \includegraphics*[scale=0.38]{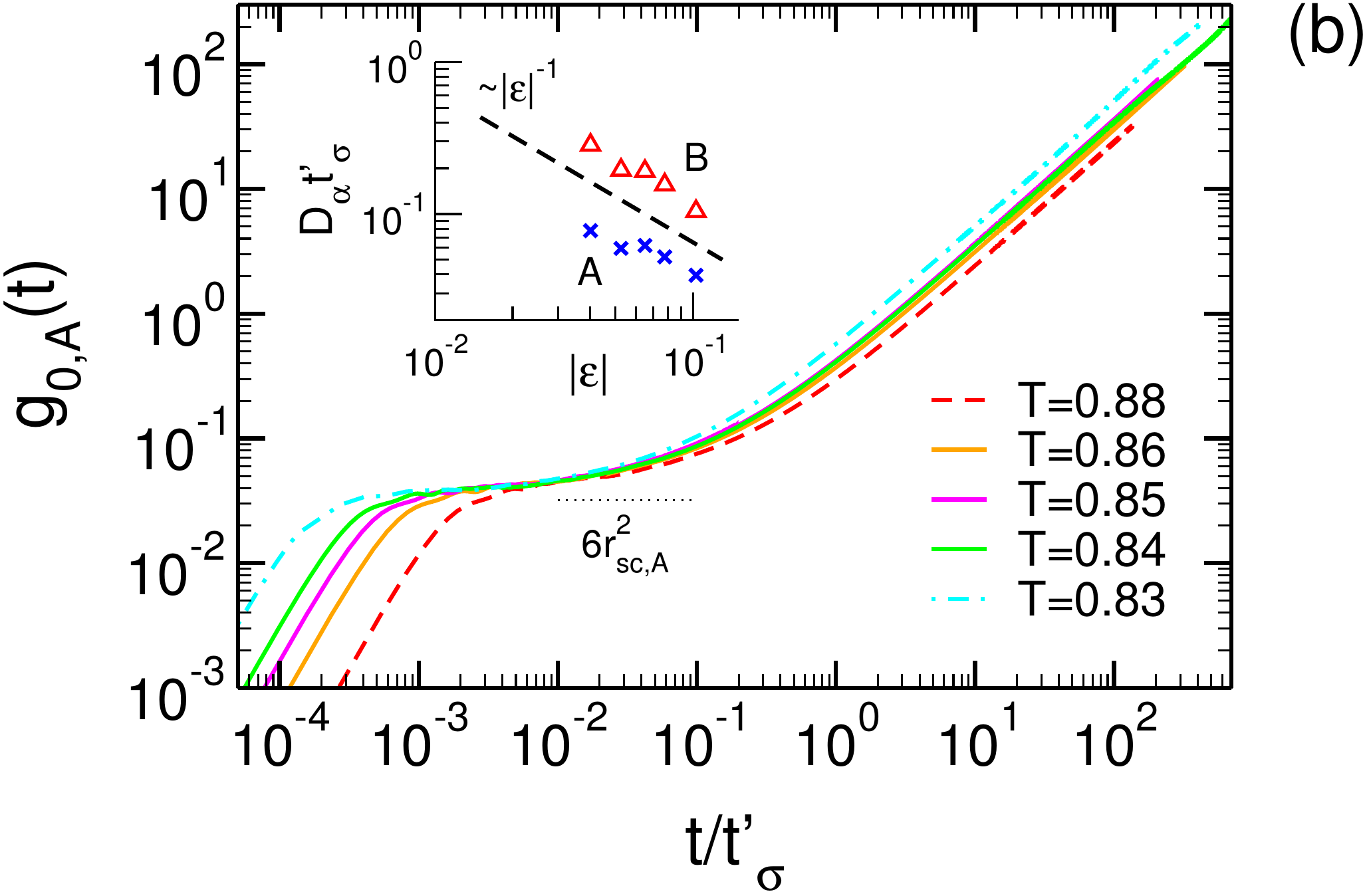}
  \end{center}
  \caption{Panel (a): Time dependence of the mean-square displacements (MSDs) at $T=0.84$ for the $\lA$ particles $g_{0,\lA}(t)$ (blue full line) and the $\lB$ particles $g_{0,\lB}(t)$ (orange full line). The ballistic motion ($3T t^2$) at short times and the diffusive motion ($6 D_\alpha t$) at long times are indicated by dotted lines. The diffusion coefficients are $D_\lA = 1.652 \times 10^{-4}$ and $D_\lB = 5.418 \times 10^{-4}$. The dashed lines present the fit results to \eref{msd_fit_equation} where only the long-time correction coefficients $B_0^{{\rm s},\alpha}$ were adjusted, yielding $B_0^{{\rm s},\lA} = -1.41093$  and $B_0^{{\rm s},\lB} = -3.00259$. The Lindemann localization lengths ($r_{{\rm sc},\lA} = 0.0731$, $r_{{\rm sc},\lB} = 0.0873$) were taken from \fref{nonergo_various_q}, the critical amplitudes ($h^{\lA}_{\rm msd} = 0.0313$, $h^{\lB}_{\rm msd} = 0.0622$) from \fref{hB_various_q}. The nonergodicity parameters of the MSDs, $6 r^2_{{\rm sc},\alpha}$, are indicated by horizontal dotted lines. The vertical dotted line shows the value of the MCT $\alpha$ time scale $t'_\sigma$ ($=359$) at $T=0.84$.
  Panel(b): Test of the TTSP for the A particles by plotting $g_{0,\lA}(t)$ versus $t/t'_\sigma$. Deviations are visible for $T=0.83$ (dash-dotted line) and $T=0.88$ (dashed line). The inset shows $D_\alpha t'_\sigma$ as a function of $|\varepsilon| = |(\Tc - T)/\Tc|$ with $\Tc =0.798$. The dashed line indicates the power law $1/|\varepsilon|$.}
  \label{fig_msd}
\end{figure}

\Fref{fig_msd}(a) shows the MSD of the $\lA$ particles, $g_{0,\lA}(t)$, and of the $\lB$ particles, $g_{0,\lB}(t)$, at $T=0.84$. For both species the MSD starts from the ballistic regime ($3Tt^2$). Outside this regime, the small ($\lB$) particles always move much farther than the large ($\lA$) particles in a given time. For $t > 0.1$ the MSD crosses over to a species-specific plateau, the height of which is comparable to the respective Lindemann localization length (see horizontal dotted lines) and thus much smaller than the particle diameter. This illustrates the temporary localization of the particles in their nearest-neighbor cages. For the increase of the MSD beyond the plateau MCT predicts the following relation \cite{FuchsGoetzeMayr1998} 
\begin{eqnarray}
  g_{0,\alpha}(t) = 6 r^2_{{\rm sc},\alpha} + &
  6 h_{\rm msd}^\alpha B \left( \frac{t}{t'_{\sigma}} \right )^b 
  \nonumber \\
  & - 6 h_{\rm msd}^\alpha B^2 B_0^{{\rm s},\alpha} 
  \left( \frac{t}{t'_{\sigma}} \right )^{2b} ,
  \label{msd_fit_equation}
\end{eqnarray}
with the localization lengths $r_{{\rm sc},\alpha}$ [\eref{gaussian_nonergo}], the critical amplitudes $h_{\rm msd}^\alpha$ [\eref{gaussian_h}], and the long-time correction coefficients $B_0^{{\rm s},\alpha}$. \Eref{msd_fit_equation} is a consequence of \eref{incoherent_vonschweidler}, since $g_{0,\alpha}(t) = \lim_{q \rightarrow 0} 6[ 1 - \phi^{{\rm s},\alpha}(q,t)]/q^2$. When comparing \eref{msd_fit_equation} only the long-time corrections need to be fitted; all other parameters are taken from the previous analysis. \Fref{fig_msd}(a) shows that \eref{msd_fit_equation} describes the MSD over approximately four decades in time for both species before the crossover to diffusion occurs at late times. In this long-time regime, $g_{0,\alpha}(t)=6 D_\alpha t$ with $D_\alpha$ being the self-diffusion coefficient of species $\alpha$. 

From \eref{TTSP_eq} it follows that the MSD should obey the TTSP when plotting $g_{0,\alpha}(t)$ against $t/t'_\sigma$. \Fref{fig_msd}(b) tests this prediction for the $\lA$ particles in the $T$ interval where $\phi(q^*,t)$ obeys the TTSP (cf \fref{fig_TTSP}). We see that the TTSP holds for the MSD only in a narrower temperature interval (for $T=0.84$, 0.85 and 0.86), whereas deviations occur for higher and lower $T$. This is highlighted in the inset which plots $D_\alpha t'_\sigma$ against $| \varepsilon | = (T - \Tc)/\Tc$. The product $D_\alpha t'_\sigma$ is not constant over the whole interval $0.83 \leq T \leq 0.88$, but appears to increase as $1/|\varepsilon |$. With \eref{tps_alpha} and \eref{eq:sigma2Tc} this would imply a fractional Stokes-Einstein relation \cite{ParmarEtal:PRL2017} $D \sim 1/(t'_\alpha)^\xi$ with exponent $\xi = (\gamma -1)/\gamma \approx 0.6$.


\subsection{Kohlrausch--Williams--Watts analysis of the $\alpha$ relaxation}
\label{subsec:kww}
\begin{figure}
  \begin{center}
  \includegraphics*[scale=0.38]{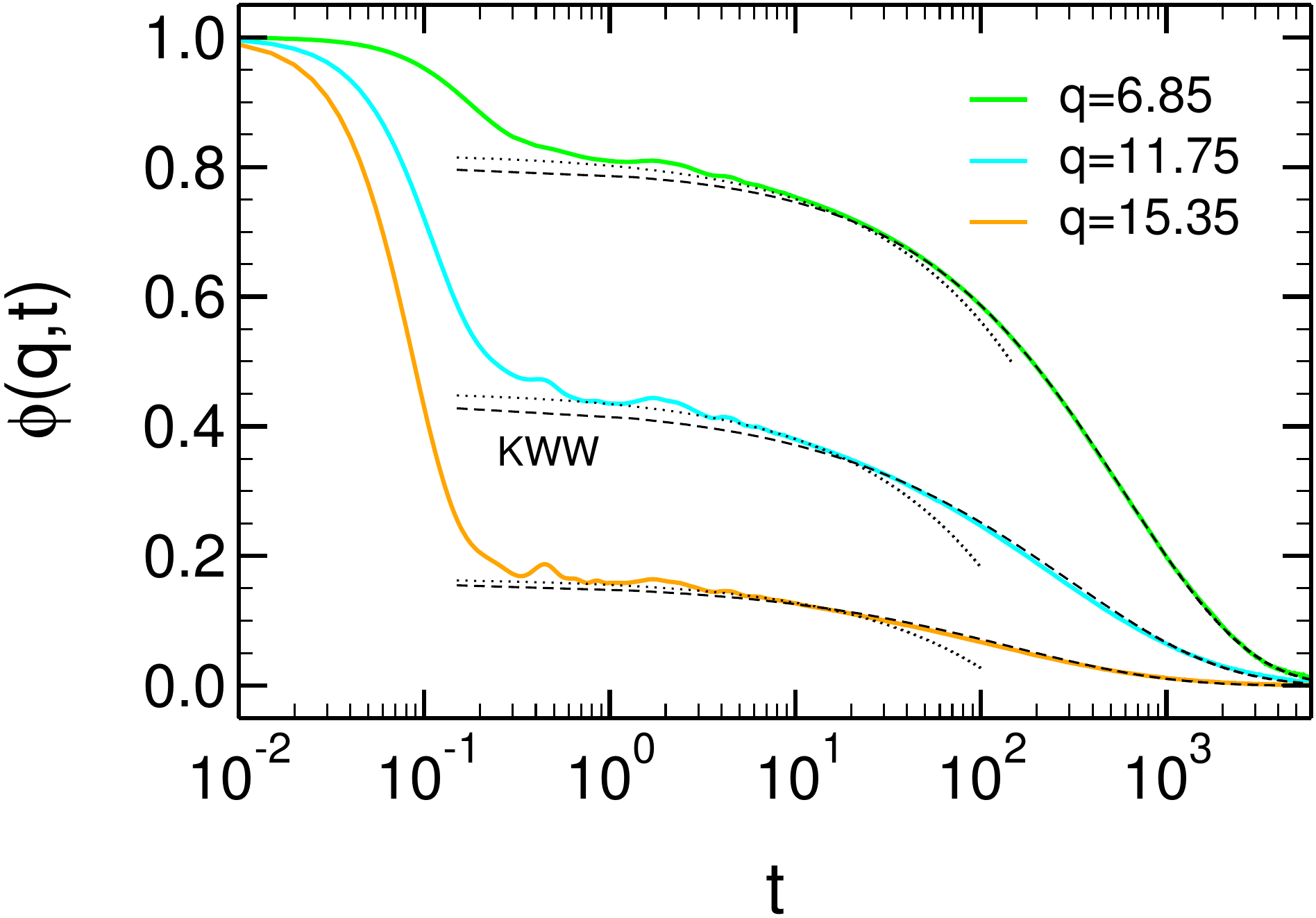}
  \end{center}
  \caption{Plot of $\phi(q,t)$ for $q=6.85, 11.75, 15.35$ at $T=0.84$ (full lines). The dashed lines show the KWW fits, the dotted lines the von Schweidler law, i.e.\ the first line of \eref{coherent_vonschweidler}.}
  \label{fig:comp_beta_KWW}
\end{figure}

The KWW function \eref{eq:KWW-function} is often used as a convenient parameterization of the $\alpha$ process in experiments and simulations \cite{Goetze:JPCM1999,GoetzeSjoegren1992_RPP,AngellEtal_JAP2000}. When fitting the $\alpha$ relaxation with \eref{eq:KWW-function} similar caveats as discussed for the late $\beta$ analysis (cf \sref{subsec:fit}) apply: The parameters $A(q)$, $\tau^{\rm K}(q)$ and $\beta^{\rm K}(q)$ are sensitive to the choice of the time interval employed for the fit \cite{WeysserEtal:PRE2010,FuchsHofackerLatz1992,CDFGHLLTerwMCT1993}, in particular the stretching exponent appears to be plagued by this effect \cite{BaschnagelVarnik:JPCM2005,alphaDynamics}. To guide the KWW fits we here draw upon the asymptotic MCT results from \sref{sec:mct_universal} and subject the fits to two constraints. First, since \eref{eq:KWW-function} is a model for the $\alpha$ process, we require $A(q) \leq f^{\rm c}(q)$. Second, the early $\alpha$ process should be excluded from the fit because $\beta^{\rm K}(q) \neq b$ for finite $q$ and so the short-time expansion of \eref{eq:KWW-function} cannot agree with the von Schweidler law \eref{beta_correlator} \cite{VoigtmannEtal:2004}. Different strategies to cope with this problem have been proposed (see \cite{WeysserEtal:PRE2010,BaschnagelVarnik:JPCM2005,alphaDynamics} and references therein). One possibility is to focus on the late $\alpha$ process only \cite{KaemmererKobSchilling1998,KobAndersen_LJ_II_1995} by restricting the fit to times for which $\phi(q,t)$ is smaller than $f^{\rm c}(q)$ by some factor $x_{\rm cut} < 1$. We varied $x_{\rm cut}$ in the interval $[0.3, 0.9]$ \cite{alphaDynamics} and found that $x_{\rm cut}=0.9$ is the most appropriate choice.

\Fref{fig:comp_beta_KWW} exemplifies the results of the KWW fits for $\phi(q,t)$ at $T=0.84$ and three wave vectors. As desired, the KWW function (dotted lines) provides a good description of the final relaxation and barely overlaps with the early $\beta$ process (von Schweidler law, dotted lines) for $q = 6.85$ and 11.75. For $q=15.35$, however, the KWW function is at short times close to the von Schweidler law. This suggests that the regime $q \gtrsim 15$ corresponds to the asymptotic large-$q$ regime where we may expect \eref{eq:kww_limit} and \eref{eq:MCT2KWW} to hold. Analysis of the $q$ dependence of the stretching exponents and relaxation times can test this expectation.

\begin{figure}
  \begin{center}
  \includegraphics*[scale=0.38]{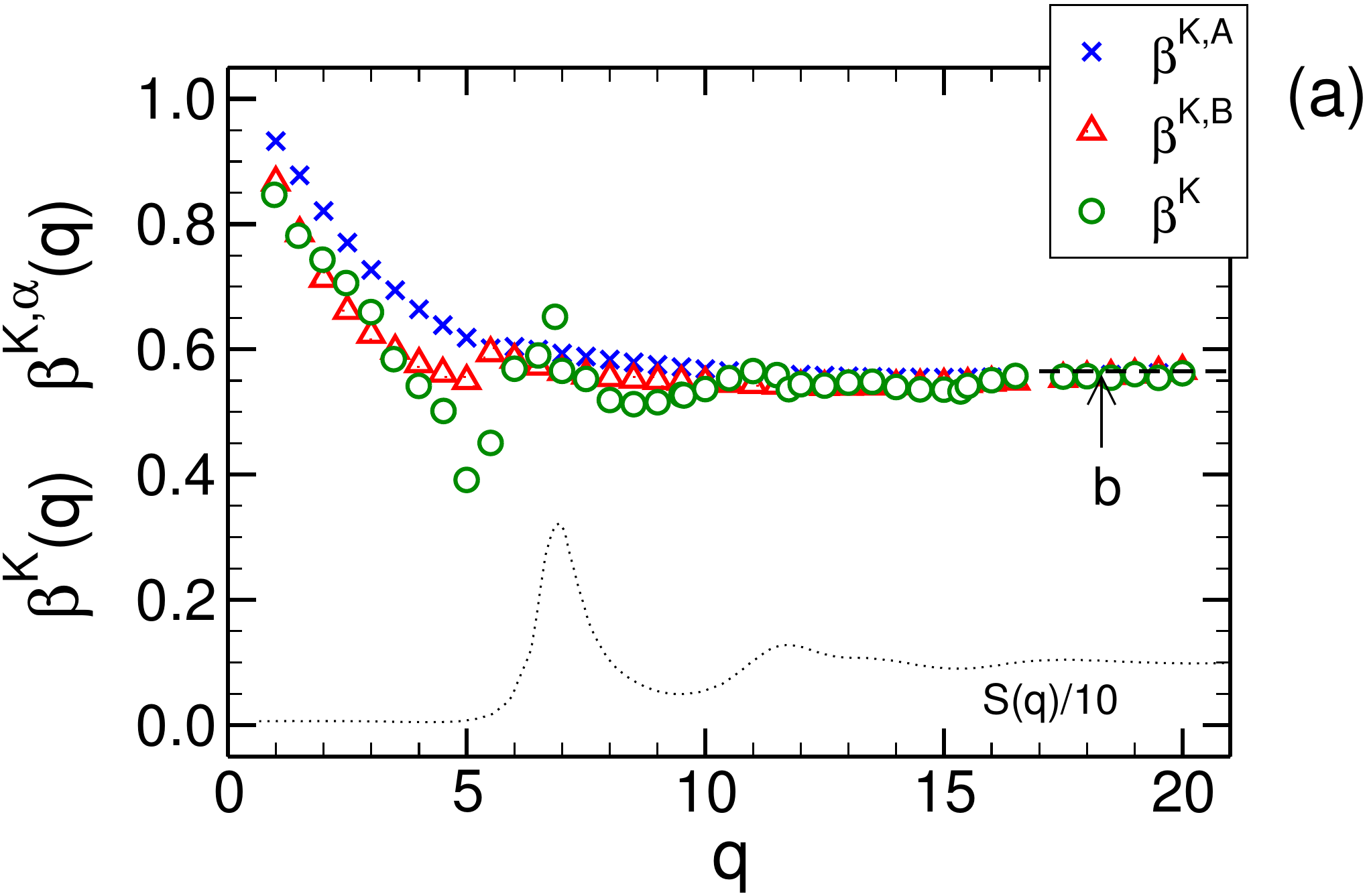}\\[2mm]
  \includegraphics*[scale=0.38]{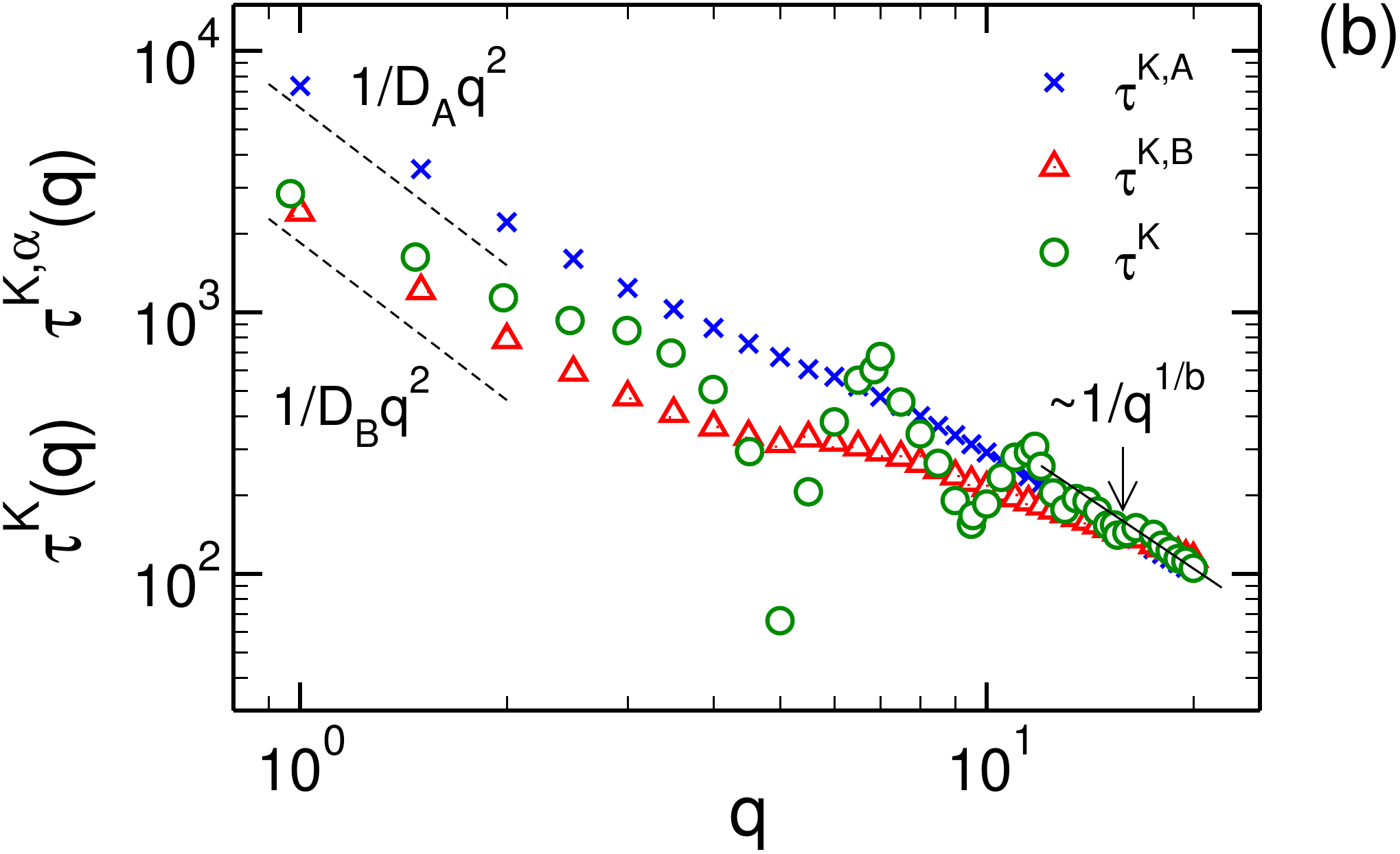}
  \end{center}
  \caption{Panel (a): $q$ dependence of the KWW stretching exponents $\beta^{{\rm K},\lA}(q)$ (crosses), $\beta^{{\rm K},\lB}(q)$  (triangles) and $\beta^{{\rm K}}(q)$ (circles). The horizontal dashed line indicates the value of von Schweidler exponent $b = 0.5652$ from the fits in the $\beta$ regime. The dotted line shows $S(q)$ divided by 10 for comparison.
  Panel (b): Log-log plot of the KWW relaxation times at $T=0.84$ versus $q$: $\tau^{{\rm K},\lA}(q)$ (crosses), $\tau^{{\rm K},\lB}(q)$  (triangles) and $\tau^{{\rm K}}(q)$ (circles). For the tagged-particle dynamics the dashed lines present the behavior $1/(D_\alpha q^2)$ expected for $q \rightarrow 0$ with the self-diffusion coefficients $D_\alpha$ taken from \fref{fig_msd}. The full line indicates the MCT prediction $\sim 1/q^{1/b}$ for large $q$ with $b = 0.5652$ [cf \eref{eq:MCT2KWW}].}
  \label{KWW_beta_tau}
\end{figure}

\Fref{KWW_beta_tau}(a) shows the results for the stretching exponents and \fref{KWW_beta_tau}(b) for the relaxation times. For $q \gtrsim q^*$ the stretching exponent $\beta^{\rm K}(q)$, obtained from $\phi(q,t)$, is roughly in phase with $S(q)$ and tends to the von Schweidler exponent $b$ for large $q$. The same large-$q$ asymptote is also found for $\beta^{{\rm K},\alpha}$, the stretching exponents of $\phi^{{\rm s},\alpha}(t)$. Along with that, the  relaxation times $\tau^{\rm K}(q)$ and $\tau^{{\rm K},\alpha}(q)$ for coherent and incoherent scattering also converge to the same large-$q$ asymptote which is proportional to $1/q^{1/b}$. These findings agree with the MCT predictions \eref{eq:kww_limit} and \eref{eq:MCT2KWW}. However, a reservation has to be mentioned: From \fref{KWW_beta_tau}(a) it seems as if the limit $\lim_{q\rightarrow \infty}\beta^{\rm K}(q)=b$ is approached from below. However, according to theory \cite{FuchsHofackerLatz1992,LuoJanssen:GMCTPY2019}, the limit should be approached from above. Such an approach has been seen in several simulations \cite{FoffiEtal:PRE2004,ChongSciortino:PRE2004,Colmenero:JPCM2015,VoigtmannEtal:2004,SciortinoFabbianChen1997,StarrSciortino1999}. Certainly, data with high accuracy at long times are needed to verify \eref{eq:MCT2KWW}, since the amplitude of the $\alpha$ process becomes small at large $q$ (cf \fref{nonergo_various_q}). This may be a prime source of uncertainty in the present analysis. 
  
\begin{figure}
  \begin{center}
  \includegraphics*[scale=0.38]{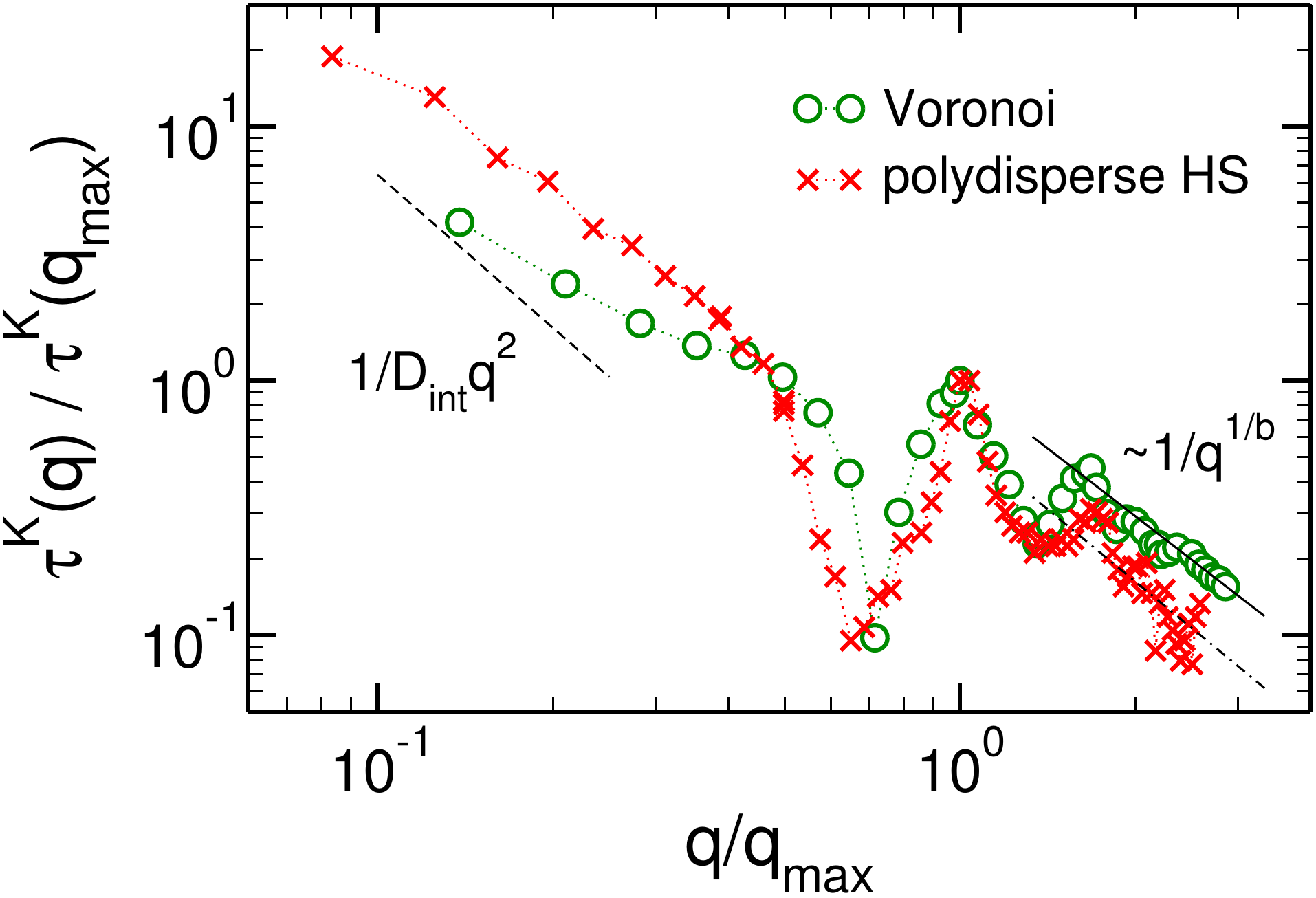}
  \end{center}
  \caption{Comparison of the binary Voronoi mixture (circles) and the polydisperse hard-sphere-like model (crosses) of \cite{WeysserEtal:PRE2010}: The figure shows KWW relaxation times $\tau^{\rm K}(q)$. The data of the hard-sphere model were digitized from the upper panel of figure~7 in \cite{WeysserEtal:PRE2010}. The axes are scaled by $q_{\rm max} \approx q^*$ (Voronoi: $q_{\rm max} = 7$, Hard spheres: $q_{\rm max} = 7.1$) and $\tau^{\rm K}(q_{\rm max})$ (Voronoi: $\tau^{\rm K}(q_{\rm max}) = 677.74$, Hard spheres: $\tau^{\rm K}(q_{\rm max}) = 0.50$). The full line indicates the MCT prediction $\sim 1/q^{1/b}$ with $b=0.5652$ for the Voronoi model, the dash-dotted line for the hard-sphere model with $b = 0.53$ from \cite{WeysserEtal:PRE2010}. The dashed line shows the hydrodynamic behavior $1/D_{\rm int} q^2$ where the interdiffusion coefficient $D_{\rm int}$ was estimated for the Voronoi mixture from the Darken equation \eref{eq:Darken}.}
  \label{fig:comp_voro_pHS}
\end{figure}

In the hydrodynamic limit we expect all scattering functions to decay as single exponentials: $\phi^{{\rm s},\alpha}(q,t) \propto \exp(-q^2 D_\alpha t)$ due to self diffusion and $\phi(q,t) \propto \exp(-q^2 D_{\rm int} t)$ due to interdiffusion, with $D_{\rm int}$ being the interdiffusion coefficient \cite{FuchsLatz:PhysicaA1993}. Therefore, $\beta^{\rm K}(q\rightarrow 0) = \beta^{{\rm K},\alpha}(q\rightarrow 0) =1$ and $\tau^{\rm K}(q) \sim \tau^{{\rm K},\alpha}(q) \sim 1/q^2$ for $q\rightarrow 0$. For $q < q^*$ we see from \fref{KWW_beta_tau}(a) that the stretching exponents increase toward 1 with decreasing $q$, but clearly the linear dimension of the simulation box is still too small so that the hydrodynamic limit is not reached for the smallest accessible $q$ values. By the same token, we cannot expect $\tau^{\rm K}(q)$ or $\tau^{{\rm K},\alpha}(q)$ to attain the hydrodynamic limit. Still, \fref{KWW_beta_tau}(b) shows that $\tau^{{\rm K},\alpha}(q)$ tend to the expected behavior, $\tau^{{\rm K},\alpha}(q) = 1/D_\alpha q^2$, for $q \rightarrow 0$. 

For the collective dynamics we have not determined the interdiffusion coefficient $D_{\rm int}$ (this would be possible via an Einstein relation similar to the one for the self-diffusion coefficients \cite{HorbachEtal:PRB2007}). However, \cite{HorbachEtal:PRB2007} suggests that the following linear combination of the self-diffusion coefficients, known as the ``Darken equation'',
\begin{equation}
  D_{\rm int} = \frac{\xA\xB}{S_{\rm cc}(q \rightarrow 0)} \,
  \big (\xA D_\lB + \xB D_\lA \big ) ,
  \label{eq:Darken}
\end{equation}
represents a good approximation even in the supercooled regime. We estimate $D_{\rm int}$ from the data shown in \fref{sdek} and \fref{fig_msd}. The result ($1/D_{\rm int}q^2$) is included as a dashed line in \fref{fig:comp_voro_pHS}. This figure compares the Voronoi mixture to the polydisperse hard-sphere-like model studied in \cite{WeysserEtal:PRE2010} in order to assess to what extent the $q$ dependence of $\tau^{\rm K}(q)$ is model specific. For a better comparison we superimpose the data at one point, $q_{\rm max}$ and $\tau^{\rm K}(q_{\rm max})$, where $q_{\rm max} \approx q^*$ for both models. We see that the relaxation times for both models are in good qualitative agreement. For large $q$ they are compatible with the scaling $\sim 1/q^{1/b}$ with a model-specific von Schweidler exponent and for small $q$ they tend to the hydrodynamic behavior. For the $q$ regime near $q_{\rm max} \approx q^*$ the agreement is even semiquantitative. In particular, the drop of $\tau^{\rm K}(q \approx 0.7 q_{\rm max})$ by an order of magnitude relative to $\tau^{\rm K}(q_{\rm max})$ is present for both models. This drop is accompanied by a low amplitude of the $\alpha$ process (cf \fref{nonergo_various_q} and figure~5 in \cite{WeysserEtal:PRE2010}) and a pronounced stretching of the KWW function (cf \fref{KWW_beta_tau} and figure~8 in \cite{WeysserEtal:PRE2010}). These features therefore appear to be independent of the model and rather characteristic of the collective dynamics in multicomponent systems on length scales where the crossover between large-scale composition fluctuations and local-scale liquid-like packing constraints occurs. 

\Fref{fig:comp_voro_pHS} also suggests that the hard-sphere-like model reaches the hydrodynamic limit ($\sim 1/q^2$) earlier than the Voronoi mixture. A slow convergence to the hydrodynamic limit was also observed for the sound attenuation in the monodisperse Voronoi liquid and could be traced back to the fact that the product of the infinite frequency shear modulus ($G_\infty$) and the isothermal compressibility ($\chi_T$) is exceptionally small (compared Lennard-Jones systems) \cite{RuscherEtal:JCP2017}. It would be worthwhile to explore whether a similar mechanism also protracts the crossover to the hydrodynamic limit for the interdiffusion process in the binary Voronoi mixture.

\section{Summary and discussion}
\label{sec:sum}
The Voronoi liquid is a fluid model whose interactions are local, many-body and soft \cite{RuscherEtal:EPL2015,RuscherEtal:JCP2017}. Here we study a generalization of the Voronoi liquid to binary mixtures. Our mixture is equimolar, weakly polydisperse and additive. This binary Voronoi mixture is a relatively new model. Up to now, only its thermodynamic and structural properties, from the normal liquid to the supercooled state, have been investigated \cite{RuscherEtal:PRE2018}. With the present work we extend the analysis to dynamic properties. The focus of our analysis is a comparison of MD results for the incoherent and coherent scattering functions with the idealized MCT. Overall, we find that the glassy dynamics of the binary Voronoi fluid conforms to the same qualitative phenomenology as that of simple liquids, albeit with a few subtleties. 

As in every multicomponent system, the binary Voronoi mixture exhibits transport processes related to composition fluctuations. In the hydrodynamic limit, these processes are described by the interdiffusion of the two particle species. The idealized MCT obeys this hydrodynamic limit and makes a number of predictions \cite{FuchsLatz:PhysicaA1993}. For $q \rightarrow 0$ the nonergodicity parameter of $\phi(q,t)$ is determined by the ratio $S_{\rm nc}^2(q\rightarrow 0)/S_{\rm cc}(q\rightarrow 0)$ of the Bhatia--Thornton structure factors, $\phi(q,t)$ decays exponentially and the corresponding relaxation time is given by $1/D_{\rm int} q^2$. Although the systems simulated are still too small to fully realize the hydrodynamic limit, figures~\ref{nonergo_various_q}, \ref{KWW_beta_tau} and \ref{fig:comp_voro_pHS} reveal that our simulation results approach the predicted behavior with decreasing $q$. In this small-$q$ regime the $\alpha$ process of $\phi(q,t)$ is dominated by transport processes due to composition fluctuations.

A hallmark of glassy slowing down is the super-Arrhenius increase of the local relaxation times with decreasing $T$. \Fref{fig_TTSP} provides an example for $\tau_{q^*}$. MCT attributes this slowing down to the nonlinear coupling between dynamic density fluctuations, which amplifies weak structural changes of the dense packing in the neighbor shells of the liquid (``cage effect''). As a consequence, the $\alpha$ process of $\phi(q,t)$ exhibits the fingerprint of $S(q)$ for $q \gtrsim q^*$. We find evidence for this in-phase modulation with $S(q)$ for $f^{\rm c}(q)$ (\fref{nonergo_various_q}), $\beta^{\rm K}(q)$ (\fref{KWW_beta_tau}) and $\tau^{\rm K}(q)$ (\fref{fig:comp_voro_pHS}). Therefore, at intermediate $q$ a crossover exists between the composition-fluctuation dominated small-$q$ regime and the cage-effect dominated large-$q$ regime. This crossover occurs in the range $q \approx 0.7 q^*$, not only for the Voronoi mixture but also for polydisperse hard spheres (\fref{fig:comp_voro_pHS}). Here the amplitude of the $\alpha$ process is weak and $\tau^{\rm K}(q)$ is about an order of magnitude smaller than $\tau^{\rm K}(q^*)$, while the decay of $\phi(q,t)$ is strongly stretched. 

We compare our MD simulations with two MCT approaches, with fits to the asymptotic predictions valid for $T \gtrsim \Tc$ and with MCT calculations using the partial static structure factors from the simulations as input to compute the dynamics. Fits to the asymptotic predictions have been carried out for many experimental and simulated systems in the past \cite{GoetzeBook2009,Goetze:JPCM1999}, including binary Lennard-Jones \cite{KobAndersen_LJ_I_1995,GleimKob2000,KobAndersen_LJ_II_1995} and hard-sphere mixtures \cite{FoffiEtal:PRE2004} or metallic alloys \cite{DasEtal:PRB2008}. Compared to these studies, we get similar results for the Voronoi mixture, despite its more complicated many-body potential. The MCT $\alpha$ time ($t'_\sigma$) is strongly coupled to the $\alpha$ relaxation times of the coherent and incoherent scattering functions at $q^*$ (cf \fref{plot_Tc}), allowing for a consistent extrapolation from all of these relaxation times to estimate $\Tc$ ($= 0.798$). For $T \gtrsim \Tc$ we find evidence for the space-time factorization in the $\beta$ regime (\fref{fig_factorization}) and the TTSP in the $\alpha$ regime (\fref{fig_TTSP}) from the scattering functions at finite wave vectors. On the other hand, time-temperature superposition by scaling time with $t'_\sigma$ appears to become violated for $q \rightarrow 0$, as shown for the MSD in \fref{fig_msd}, implying a decoupling of the $\alpha$ relaxation time and self-diffusion. It could be that single-particle hopping processes are responsible for this decoupling \cite{FlennerSzamel:PRE2005_2,FlennerSzamel:PRE2005_1,Chong:PRE2008,ChongEtal:JPCM2009,CharbonneauEtal:PNAS2014}. Investigations in this direction for the Voronoi mixture, following e.g.\ the lines of \cite{PastoreEtal:SoftMatter2014,PastoreEtal:JSTM2016,KeysEtal:PRX2011}, would be interesting. 

The binary MCT calculations based on static input give very good agreement for $f^{\rm c}(q)$ (\fref{nonergo_various_q}), whereas the agreement is worse for $h(q)$, in particular in the regime of the crossover between composition fluctuations and cage effect (\fref{h_various_q}). We note that our MCT calculations have used only the partial static structure factors, i.e.\ two-point correlation functions, as input, even though the fluid itself contains many-body interactions by construction. In this regard, it may be considered striking that some of the MCT predictions are in such good agreement with simulation. Indeed, our work suggests that even for a complex fluid such as the Voronoi mixture, one of the simplest measures of structure (i.e.\ $S_{\alpha \beta}(q)$) already constitutes a major portion of the relevant structural information needed to predict the dynamics.
Nonetheless, discrepancies in e.g.\ the prediction for $h(q)$ highlight the need for more refined theory. Currently, the origin of these discrepancies is unclear. To resolve this issue, it would be worthwhile to carry out the comparison between MCT and simulation for the partial dynamic structure factors $S_{\alpha\beta}(q,t)$ because they are the primary correlators calculated by the theory (cf \sref{sec:mct}). Such a comparison would allow one to identify whether the observed differences in $h(q)$ stem from one particle species (A or B), or from the interplay between them. Unfortunately, $S_{\alpha\beta}(q,t)$ was not determined in the present simulations, but work in this direction is planned for the future.

The MCT calculations also illustrate the very high precision required of $\Tc$ to get convergent results for $\lambda$ (cf \tref{tabMCT}). $\lambda$ only settles if $\Tc$ is accurate to the fifth or sixth digit after the decimal point. Still, the final value is not so satisfying when compared to the results from the asymptotic analysis (cf \tref{tablefitparam}). The $\alpha$ process is more stretched than predicted by MCT (\fref{comparison_phiq_MD_theo}). This difference could be related to the overestimation of $\Tc$ ($\simeq 0.979$) by the idealized theory. Extensions of MCT, developed by some of us \cite{JanssenReichman:PRL2015,JanssenEtal:JSM2016,LuoJanssen:GMCTPY2019,JanssenEtal:PRE2014}, allow to delay the factorization approximation of the memory kernel to higher order. Application of this generalized mode-coupling theory (GMCT) to simulated hard spheres \cite{JanssenReichman:PRL2015} and Percus-Yevick hard spheres \cite{LuoJanssen:GMCTPY2019} suggests that the critical packing fraction improves and shifts to larger values compared to the idealized MCT and along with that, the stretching of the $\alpha$ process increases. It might therefore be worthwhile to extend the GMCT to binary mixtures, as studied here.

\ack
Financial support by the ANR LatexDry project grant ANR-18-CE06-0001 of the French Agence Nationale de la Recherche, the Canada First Research Excellence Fund, Quantum Materials and Future Technologies Program  and by the Dutch Research Council (NWO) through a START-UP grant is gratefully acknowledged. We are indebted to M Fuchs (Konstanz), H Meyer, A N Semenov (both Strasbourg) for very helpful discussions and to O Benzerara (Strasbourg) for valuable technical support with the MD simulations. The simulations were made possible by a generous grant of computer time on the HPC cluster of the University of Strasbourg.

\vspace*{2ex}

\bibliography{references_Glass_MCT_and_LiquidTheory.bib,references_CTRW.bib,references_Simulation.bib}
\end{document}